\documentclass[12pt,letterpaper]{article}

\usepackage[T1]{fontenc} %
\usepackage[utf8]{inputenc} %
\usepackage{lmodern} %
\usepackage{textcase} %
\usepackage{amsmath,amssymb,amsthm}%
\usepackage[nodisplayskipstretch]{setspace} %
\usepackage{booktabs}%
\usepackage{multirow}%
\usepackage{pdflscape, afterpage} %
\usepackage[tracking=true, letterspace=50, expansion=false]{microtype} %
\usepackage[authoryear]{natbib}%
\usepackage{fullpage,enumitem}%
\usepackage[pass]{geometry}%
\usepackage[dvips]{graphicx}
\usepackage{hyperref,comment}

\newcommand{\EHWse}{\widehat{\sigma}_{\textnormal{EHW}}}
\newcommand{\CRV}{CRV}
\newcommand{\CRVse}{\widehat{\sigma}_{\textnormal{\CRV}}}
\newcommand{\limCRVse}{\widehat{\sigma}_{\textnormal{\CRV},\infty}}
\newcommand{\SC}{K}%

\DeclareMathOperator{\cv}{cv} 
\DeclareMathOperator{\trace}{trace} 
\DeclareMathOperator{\sign}{sign} 
\newcommand{\indist}{\stackrel{d}{\rightarrow}}%
\newcommand{\tauh}{\tau_{h}}%
\newcommand{\thetah}{\theta_{h}}%

\RequirePackage{mathtools}
\DeclarePairedDelimiter\abs{\lvert}{\rvert}
\DeclarePairedDelimiter\hol{(}{]}
\DeclarePairedDelimiter\hor{[}{)}

\newcommand{\norm}[1]{\lVert#1\rVert}%
\newcommand{\E}{\mathbb{E}}%
\newcommand{\1}[1]{\mathbb{I}\left\{#1\right\}}%
\newcommand{\Var}{\mathbb{V}}%
\DeclareMathOperator*{\vvec}{vec}%
\DeclareMathOperator*{\diag}{diag}%
\newtheorem{theorem}{Theorem}%
\newtheorem{lemma}{Lemma}%
\newtheorem{assumption}{Assumption}%
\newtheorem{proposition}{Proposition}%
\theoremstyle{definition}%
\newtheorem{remark}{Remark}%

\usepackage{titlesec}
\titleformat{\section}[block]{\centering\normalfont}{\thesection.}{0.5em}{\MakeTextUppercase}
\titleformat{\subsection}[block]{\normalfont}{\thesubsection.}{0.5em}{\bfseries}
\titleformat{\subsubsection}[block]{\normalfont}{\thesubsubsection.}{0.5em}{\bfseries}
\titlespacing*\section{0pt}{18pt plus 4pt minus 2pt}{2pt plus 2pt minus 2pt}
\titlespacing*\subsection{0pt}{8pt plus 2pt minus 1pt}{2pt plus 2pt minus 2pt }
\titlespacing*\subsubsection{0pt}{8pt plus 2pt minus 1pt}{2pt plus 2pt minus 2pt }
\numberwithin{equation}{section}%

\usepackage[stable,multiple]{footmisc}

\title{Inference in Regression Discontinuity Designs with a Discrete Running
  Variable\thanks{We thank Joshua Angrist, Tim Armstrong, Guido Imbens, Philip
    Oreopoulos, and Miguel Urquiola and seminar participants at Columbia University,
    Villanova University, and the 2017 SOLE Annual Meeting for helpful
    comments and discussions.} }

\author{{Michal Kolesár}\thanks{Woodrow Wilson
    School and Department of Economics, Princeton University, \texttt{mkolesar@princeton.edu}} \and
  {Christoph Rothe}\thanks{Department of Economics,
    University of Mannheim, \texttt{rothe@vwl.uni-mannheim.de}}}

\date{\normalsize\today}

\begin{document}

\bibliographystyle{ecta}
\singlespacing

\maketitle

\begin{abstract}
  We consider inference in regression discontinuity designs when the running
  variable only takes a moderate number of distinct values. In particular, we
  study the common practice of using confidence intervals (CIs) based on
  standard errors that are clustered by the running variable as a means to make
  inference robust to model misspecification \citep{lee2008regression}. We
  derive theoretical results and present simulation and empirical evidence
  showing that these CIs do not guard against model misspecification, and that they
        have poor coverage properties. We therefore recommend against using these CIs in
  practice. We instead propose two alternative CIs with guaranteed coverage
  properties under easily interpretable restrictions on the conditional
  expectation function.
\end{abstract}

\newpage

\onehalfspacing

\section{Introduction}

The regression discontinuity design (RDD) is a popular empirical strategy that
exploits fixed cutoff rules present in many institutional settings to estimate
treatment effects. In its basic version, the sharp RDD, units are treated if and
only if an observed running variable falls above a known threshold. For example,
students may be awarded a scholarship if their test score is above some
pre-specified level. If unobserved confounders vary smoothly around the
assignment threshold, the jump in the conditional expectation function (CEF) of
the outcome given the running variable at the threshold identifies the average
treatment effect (ATE) for units at the margin for being treated
\citep{hahn2001identification}.

A standard approach to estimate the ATE is local polynomial regression. In its
simplest form, this amounts to fitting a linear specification separately on each
side of the threshold by ordinary least squares, using only observations that
fall within a prespecified window around the threshold. Since the true CEF is
typically not exactly linear, the resulting estimator generally exhibits
specification bias. If the chosen window is sufficiently narrow, however, the
bias of the estimator is negligible relative to its standard deviation. One can
then use a confidence interval (CI) based on the conventional Eicker-Huber-White
(EHW) heteroskedasticity-robust standard error for
inference.\footnote{Proceeding like this is known as ``undersmoothing'' in the
  nonparametric regression literature. See \citet{calonico2014robust} for an
  alternative approach.}

This approach can in principle be applied whether the running variable is
continuous or discrete. However, as \citet[][LC from hereon]{lee2008regression}
point out, if the running variable only takes on a moderate number of distinct
values, and the gaps between the values closest to the threshold are
sufficiently large, there may be few or no observations close to the threshold.
Researchers may then be forced to choose a window that is too wide for the bias
of the ATE estimator to be negligible, which in turn means that the EHW CI
undercovers the ATE, as it is not adequately centered. This concern applies to
many empirical settings, as a wide range of treatments are triggered when
quantities that inherently only take on a limited number of values exceed some
threshold. Examples include the test score of a student, the enrollment number
of a school, the number of employees of a company, or the year of birth of an
individual.\footnote{This setting is conceptually different from a setting with
  a continuous latent running variable, of which only a discretized or rounded
  version is recorded in the data. See \citet{dong2015regression} for an
  analysis of RDDs with this type of measurement error.}

Following LC's suggestion, it has become common practice in the empirical
literature to address these concerns by using standard errors that are clustered
by the running variable (CRV). This means defining observations with the same
realization of the running variable as members of the same ``cluster'', and then
using a cluster-robust procedure to estimate the variance of the ATE estimator.
Recent papers published in leading economics journals that follow this
methodology include \citet{oreopoulos2006estimating},
\citet{10.1257/aer.98.5.2242}, \citet{urquiola2009class},
\citet{martorell2011help}, \citet{fredriksson2013long},
\citet{10.1257/aer.103.7.2683}, \citet{10.1257/aer.103.6.2087}, and
\citet{hinnerich2014democracy}, among many others. The use of CRV standard
errors is also recommended in survey papers \citep[e.g.][]{ll10} and government
agency guidelines for carrying out RDD studies
\citep[e.g.][]{schochet2010standards}.

 In this paper, we present theoretical, empirical and simulation evidence
showing that clustering by the running variable is generally unable to
resolve bias problems in discrete RDD settings. Furthermore, using the usual
cluster-robust standard error formula can lead to CIs with substantially \emph{worse}
coverage properties than those based on the EHW standard errors.
To motivate our analysis, and to demonstrate the quantitative importance
of our findings, we first conduct a Monte Carlo study based on real data. Our
exercise mimics the process of conducting an empirical study by drawing random
samples from a large data set extracted from the Current Population Survey, and
estimating the effect of a placebo treatment ``received'' by individuals over
the age of 40 on their wages, using a discrete RDD with age in years as the
running variable. By varying the width of the estimation window, we can vary the
accuracy of the fitted specification from ``very good'' to ``obviously
incorrect''. For small and moderate window widths, CRV standard errors turn out
to be {smaller} than EHW standard errors in most of the samples, and the
actual coverage rate of CRV CIs with nominal level 95\% is as low as 58\%. At
the same time, EHW CIs perform well and have coverage much closer to 95\%. For
large window widths (paired with large sample sizes), CRV CIs perform relatively
better than EHW CIs, but both procedures undercover very severely.

To explain these findings, we derive novel results concerning the asymptotic
properties of CRV standard errors and the corresponding CIs. In our analysis,
the data are generated by sampling from a fixed population, and specification
errors in the (local) polynomial approximation to the CEF are population
quantities that do not vary across repeated samples. This is the standard
framework used throughout the theoretical and empirical RDD literature. In
contrast, LC motivate their approach by modeling the specification error as
random, with mean zero conditional on the running variable, and independent
across the ``clusters'' formed by its support points. As we explain in
Section~\ref{sec:motiv-using-clust} however, their setup is best viewed as a
heuristic device; viewing it as a literal description of the data generating
process is unrealistic and has several undesirable implications.

Our results show that in large samples, the average difference between CRV and
EHW variance estimators is the sum of two components. The first component is
generally negative, does not depend on the true CEF, decreases in magnitude
with the number of support points (that is, the number of ``clusters''), and
increases with the variance of the outcome. This component is an analog of the
usual downward bias of the cluster-robust variance estimator in settings with a
few clusters; see \citet{CaMi14} for a recent survey. The second component is
non-negative, does not depend on the variance of the outcome, and is increasing
in the average squared specification error. Heuristically, it arises because the
CRV variance estimator treats the (deterministic) specification errors as
cluster-specific random effects, and tries to estimate their variability. This
component is equal to zero under correct specification.

This decomposition shows that CRV standard errors are on average larger than EHW
standard errors if the degree of misspecification is sufficiently large relative
to the sampling uncertainty and the number of support points in the estimation
window. However, the coverage of \CRV\ CIs may still be arbitrarily far below
their nominal level, as the improvement in coverage over EHW CIs is typically
not sufficient when the coverage of EHW CIs is poor to begin with. Moreover,
when the running variable has many support points, a more effective way to
control the specification bias is to choose a narrower estimation window. In the
empirically relevant case in which the degree of misspecification is small to
moderate (so that it is unclear from the data that the model is misspecified),
clustering by the running variable generally \emph{amplifies} rather than ameliorates the
distortions of EHW CIs.

In addition to CRV CIs based on conventional clustered standard errors, we also
consider CIs based on alternative standard error formulas suggested recently in
the literature on inference with a few clusters \citep[cf.][Section VI]{CaMi14}.
We find that such CIs ameliorate the undercoverage of CRV CIs when the
misspecification is mild at the expense of a substantial loss in power relative
to EHW CIs (which already work well in such setups); and that their coverage is
still unsatisfactory in settings where EHW CIs work poorly. This is because the
second term in the decomposition does not adequately reflect the magnitude of
the estimator's bias.

These results caution against clustering standard errors by the running variable
in empirical applications, in spite of its great popularity.\footnote{Of course,
  clustering the standard errors at an appropriate level may still be justified
  if the data are not generated by independent sampling. For example, if the
  observational units are students, and one first collects a sample of schools
  and then samples students from those schools, clustering on schools would be
  appropriate.} We therefore propose two alternative CIs that have guaranteed
coverage properties under interpretable restrictions on the CEF\@. The first
method relies on the assumption recently considered in \citet{ArKo16honest} that
the second derivative of the CEF is bounded by a constant, whereas the second
method assumes that the magnitude of the approximation bias is no larger at the
left limit of the threshold than at any point in the support of the running
variable below the threshold, and similarly for the right limit. Both CIs are
``honest'' in the sense of \citet{li89}, which means they achieve asymptotically
correct coverage uniformly over all CEFs satisfying the respective assumptions.
Their implementation is straightforward using the software package
\texttt{RDHonest}, available at \texttt{https://github.com/kolesarm/RDHonest}.

We illustrate our results in two empirical applications, using data from
\citet{oreopoulos2006estimating} and \citet{lalive08}. In both applications, we
find that the \CRV\ standard errors are \emph{smaller} than EHW standard errors,
up to an order of magnitude or more, with the exact amount depending on the
specification. Clustering by the running variable thus understates the
statistical uncertainty associated with the estimates, and in the
\citet{oreopoulos2006estimating} data, it also leads to incorrect claims about
the statistical significance of the estimated effect. Honest CIs based on
bounding the second derivative of the CEF are slightly wider than those based on
EHW standard errors, while the second type of honest CI gives quite conservative
results unless one restricts the estimation window.

The rest of the paper is organized as follows. The following section illustrates
the distortion of CRV CIs using a simulation study based on data from the
Current Population Survey. Section~\ref{sec:properties-lee-card} reviews the
standard sharp RDD, explains the issues caused by discreteness of the running
variable, and discusses LC's original motivation for clustering by the running
variable. Section~\ref{sec:asymptotic-results} contains the main results of our
theoretical analysis, and Section~\ref{sec:honest-conf-interv} discusses the
construction of honest CIs. Section~\ref{sec:empir-appl} contains two empirical
illustrations. Section~\ref{sec:conclusions} concludes. Technical arguments and
proofs are collected in the appendix. The supplemental material contains some
additional theoretical results, and also additional simulations that study the
performance of CRV and EHW CIs as well as honest CIs from
Section~\ref{sec:honest-conf-interv}.

\section{Evidence on Distortions of CRV Confidence Intervals}\label{sec:dist-crv-conf}

To illustrate that clustering by the running variable can lead to substantially
distorted inference in empirically relevant settings, we investigate the
performance of the method by applying it to wage data from the Current Population
Survey.

\subsection{Setup}\label{sec:setup}

\begin{figure}[!t]
  \begin{center}
    \includegraphics[width=0.9\textwidth]{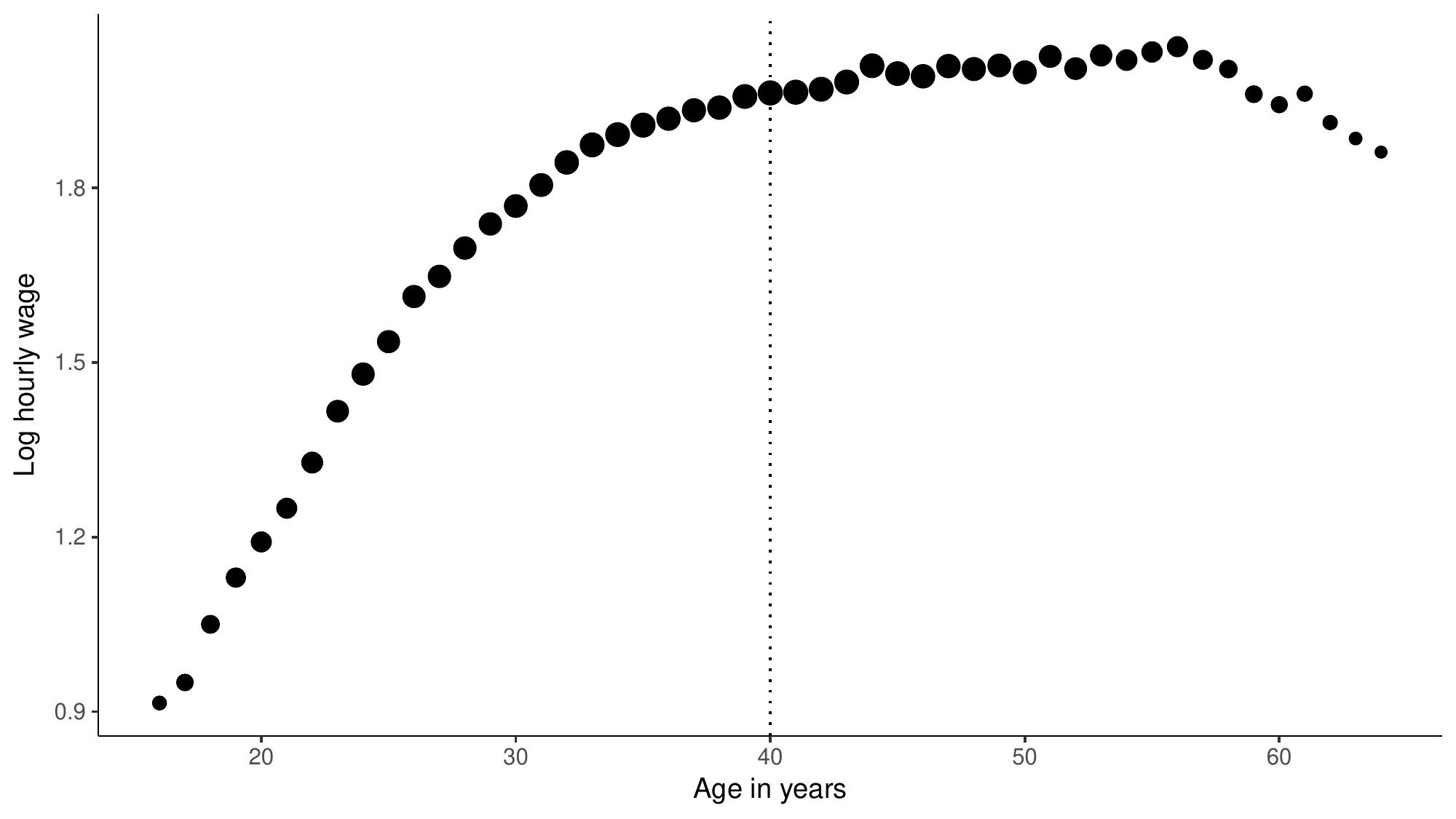}
    \caption{Average of natural logarithm of hourly wage by age of the worker.
      Vertical line indicates the cutoff value of 40 years for the placebo
      treatment. Volume of dots is proportional to share of workers in the full
      CPS data with the corresponding age.\label{fig_data1}}
  \end{center}
\end{figure}

Our exercise is based on wage data from the Outgoing Rotation Groups of the
Current Population Survey (CPS) for the years 2003--2005. The data contain
170,693 observations of the natural logarithm of hourly earnings in US\$ and the
age in years for employed men aged 16 to 64. The data are augmented with an
artificial treatment indicator that is equal to one for all men that are at
least 40 years old, and equal to zero otherwise.\footnote{See
  \citet{lemieux2006increasing}, who uses the same dataset, for details on its
  construction. The artificial threshold of 40 years corresponds to the median
  of the observed age values. Considering other artificial thresholds leads to
  similar results.} Figure~\ref{fig_data1} plots the average of the log hourly
wages in the full CPS data by worker's age, and shows that this relationship is
smooth at the treatment threshold. This indicates that our artificial treatment
is indeed a pure placebo, with the causal effect equal to zero.

For our exercise, we draw many small random samples from the full CPS data, and
estimate the effect of the artificial treatment using an RDD, with log hourly
wage as the outcome and age as the running variable, on each of them.
Figure~\ref{fig_data1} therefore plots the true CEF of log hourly wages given
age for the population from which our samples are taken.
In particular, we draw random samples of size $N_h\in\{100, 500, 2000, 10000\}$,
from the set of individuals in the full CPS data whose age differs by no more
than $h\in\{5,10,15,\infty\}$ from the placebo cutoff of 40. In each sample, we
use OLS to estimate the regression
\begin{equation}\label{eq:cps-regression}
\begin{split}
  \log(\text{WAGE}_i) &= \alpha_{h} + \tau_{h} \cdot \1{\text{AGE}_i \geq 40}
  + \sum_{j=1}^p \beta_{hj}^- \cdot (\text{AGE}_i-40)^p \\
  &\qquad + \sum_{j=1}^p \beta_{hj}^+ \cdot \1{\text{AGE}_i \geq 40}\cdot
  (\text{AGE}_i-40)^p + U_i,
\end{split}
\end{equation}
for $p=1$ (a linear model with different intercept and slope on each side of the
cutoff), and $p=2$ (a quadratic specification). The OLS estimate $\widehat\tau$
of $\tauh$ is then the empirical estimate of the causal effect of the placebo
treatment for workers at the threshold. The parameter $h$ is a window width (or
bandwidth) that determines the range of the running variable over which the
specification approximates the true CEF of log hourly wages given age. The
regression coefficients in~\eqref{eq:cps-regression} are subscripted by $h$
because their value depends on the window width (the dependence on $p$
is left implicit). Running OLS on specification~\eqref{eq:cps-regression} is equivalent
to local linear ($p=1$) or local quadratic ($p=2$) regression with a bandwidth
$h$ and a uniform kernel function on a sample with $N_h$ individuals whose age
differs by less than $h$ from the threshold.

Since we know the true CEF in the context of our exercise, we can compute the population values of the
coefficients in the model~\eqref{eq:cps-regression} for the various values of
$h$ and $p$ under consideration by estimating it on the full CPS data. Inspection
of Figure~\ref{fig_data1} shows that the
model~\eqref{eq:cps-regression} is misspecified to some degree for all values of
$h$ and $p$ under consideration, as the dots representing the true CEF of log
wages never lie \emph{exactly} on a straight or quadratic line
(Figures~\ref{fig_data1a} and~\ref{fig_data1b} in Appendix C show the resulting
approximations to the CEF). As a result, the parameter $\tauh$, reported in the
second column of Table~\ref{tab:empmc_results1}, which approximates the true
jump of the CEF at the placebo cutoff, is never exactly equal to zero. The degree
of misspecification varies with $h$ and $p$. In the linear case ($p=1$), it is
very small for $h=5$, but
increases substantially for larger values of $h$. On the other hand, the
quadratic model ($p=2$) is very close to being correctly specified for all values
 of $h$ that we consider.

To assess the accuracy of different methods for inference, in addition to the
point estimate $\widehat\tau$, we compute both the CRV and EHW standard error
(using formulas that correspond to the default settings in STATA\@; see below for
details) in each random
sample. Because the OLS estimator $\widehat\tau$ is centered around $\tauh$,
which due to misspecification of the model~\eqref{eq:cps-regression} differs
from the true effect of zero, we expect that the 95\% EHW CI, given by
$[\widehat\tau\pm 1.96 \times\text{EHW standard error}]$, contains the zero less
than 95 percent of the time. However, when $\tauh$ is close to zero relative to
the standard deviation of $\widehat\tau$ we expect this distortion to be small.
Moreover, if clustering by the running variable is a suitable way to correct the
distortion, the CRV standard error of $\widehat\tau$ should generally be larger
than the EHW standard error (to account for the incorrect centering), and the
CRV CI, given by $[\widehat{\tau}\pm 1.96\times\text{CRV standard error}]$,
should contain the true effect of zero in roughly 95 percent of the random
samples across all specifications.

\subsection{Main results}

\begin{table}[!htbp] \caption{Performance of EHW and CRV inference for placebo
    treatment.\label{tab:empmc_results1}}
  \begin{tabular*}{\textwidth}{@{\extracolsep{\fill}}lll ccc ccc@{}}
    \toprule
    \multicolumn{4}{c}{}
    &\multicolumn{2}{c}{Average SE}
    &\multirow{2}{*}{\mbox{\shortstack{Rate CRV SE \\[2pt] $>$ EHW SE}}}&\multicolumn{2}{c}{CI coverage rate}\\
    \cmidrule(rl){5-6}\cmidrule(rl){8-9}
   $h$ & $\tauh$ & $N_h$ & $SD(\widehat\tau)$ &
    CRV & EHW  & & CRV  &EHW  \\
    \midrule
    \multicolumn{9}{c}{\emph{Linear Specification ($p=1$)}}\\[1ex]
 5&$-0.008$&   100&   0.239 & 0.166 & 0.234 &     0.14 & 0.773 & 0.938\\
    &      &   500&   0.104 & 0.073 & 0.104 &     0.13 & 0.780 & 0.947\\
    &      &  2000&   0.052 & 0.036 & 0.052 &     0.13 & 0.773 & 0.949\\
    &      & 10000&   0.021 & 0.015 & 0.023 &     0.09 & 0.772 & 0.961\\[0.8ex]
10&$-0.023$&   100&   0.227 & 0.193 & 0.223 &     0.26 & 0.873 & 0.939\\
    &      &   500&   0.099 & 0.086 & 0.099 &     0.25 & 0.876 & 0.944\\
    &      &  2000&   0.049 & 0.044 & 0.050 &     0.27 & 0.860 & 0.930\\
    &      & 10000&   0.021 & 0.021 & 0.022 &     0.39 & 0.781 & 0.829\\[0.8ex]
15&$-0.063$&   100&   0.222 & 0.197 & 0.216 &     0.31 & 0.884 & 0.927\\
    &      &   500&   0.095 & 0.089 & 0.096 &     0.34 & 0.853 & 0.899\\
    &      &  2000&   0.048 & 0.047 & 0.048 &     0.45 & 0.712 & 0.730\\
    &      & 10000&   0.020 & 0.028 & 0.021 &     0.92 & 0.348 & 0.153\\[0.8ex]
$\infty$&$-0.140$&100&0.208 & 0.196 & 0.205 &     0.38 & 0.856 & 0.886\\
    &      &   500&   0.091 & 0.094 & 0.091 &     0.54 & 0.673 & 0.667\\
    &      &  2000&   0.045 & 0.058 & 0.046 &     0.93 & 0.292 & 0.134\\
    &      & 10000&   0.019 & 0.043 & 0.020 &     1.00 & 0.006 & 0.000\\
    \midrule
        \multicolumn{9}{c}{\emph{Quadratic Specification  ($p=2$)}}\\[0.8ex]
  5&$-0.010$&   100&0.438 & 0.206 & 0.427 &     0.03 & 0.607 & 0.932\\
   &        &   500&0.189 & 0.086 & 0.190 &     0.01 & 0.599 & 0.947\\
   &        &  2000&0.093 & 0.042 & 0.095 &     0.01 & 0.587 & 0.951\\
   &        & 10000&0.038 & 0.018 & 0.042 &     0.00 & 0.595 & 0.964\\[0.8ex]
 10& $0.008$&   100&0.361 & 0.258 & 0.349 &     0.15 & 0.795 & 0.933\\
   &        &   500&0.157 & 0.110 & 0.156 &     0.12 & 0.790 & 0.948\\
   &        &  2000&0.077 & 0.055 & 0.078 &     0.11 & 0.794 & 0.947\\
   &        & 10000&0.033 & 0.025 & 0.035 &     0.13 & 0.808 & 0.956\\[0.8ex]
 15& $0.014$&   100&0.349 & 0.270 & 0.329 &     0.20 & 0.836 & 0.923\\
   &        &   500&0.146 & 0.117 & 0.147 &     0.18 & 0.851 & 0.946\\
   &        &  2000&0.073 & 0.058 & 0.073 &     0.17 & 0.839 & 0.946\\
   &        & 10000&0.031 & 0.026 & 0.033 &     0.18 & 0.828 & 0.937\\[0.8ex]
$\infty$&$-0.001$&1&0.316 & 0.267 & 0.303 &     0.26 & 0.876 & 0.930\\
   &        &   500&0.134 & 0.117 & 0.135 &     0.24 & 0.890 & 0.949\\
   &        &  2000&0.068 & 0.058 & 0.067 &     0.23 & 0.887 & 0.947\\
   &        & 10000&0.029 & 0.027 & 0.030 &     0.26 & 0.910 & 0.960\\
    \bottomrule
  \end{tabular*}

  {\vspace{1ex}\small Note: Results are based on 10,000 simulation runs. }
\end{table}

Table~\ref{tab:empmc_results1} reports the empirical standard deviation of
$\widehat\tau$ across simulation runs, the average values of the EHW and CRV
standard errors, the proportion of runs in which the CRV standard error was
larger than the EHW standard error, and the empirical coverage probabilities of
the EHW and CRV CI with nominal level 95\%. These results are as expected
regarding the properties of EHW standard errors and CIs. They also show,
however, that the CRV standard error is a downward-biased estimator of the
standard deviation of $\widehat\tau$ in almost all the specifications that we
consider, and is thus typically smaller than the EHW standard error.
Correspondingly, the CRV CI has coverage properties that are mostly worse than
those of the EHW CI\@.

A closer inspection of Table~\ref{tab:empmc_results1} reveals further
interesting patterns. Consider first the quadratic specification $(p=2)$, which,
as pointed out above, is close to correct for all bandwidth values. First, we
see that when holding the value of $h$ constant, the coverage rate of the CRV CI
is hardly affected by changes in the sample size, and always below the nominal
level of 95\%. The undercoverage for $h=5$ is very severe, about 35 percentage
points, and decreases as the value of $h$, and thus the number of ``clusters''
in the data, increases. However, it still amounts to 4--6 percentage points for
$h=\infty$, in which case there are, respectively, 24 and 25 ``clusters'' below
and above the threshold. Third, the EHW CI has close to correct coverage for all
bandwidth values since the degree of misspecification is small.

Next, consider the results for the linear model ($p=1$), in which the degree of
misspecification increases with the bandwidth. First, we see that when holding
the sample size constant the value of the CRV standard error is increasing in
$h$. The CRV standard error is downward-biased for the standard deviation of
$\widehat\tau$ when either $N_h$ or $h$ are small, and upward-biased if both are
large (in the latter case, misspecification is substantial relative to sampling
uncertainty). Second, even when the CRV standard error is larger than the EHW
standard error, the corresponding CIs still undercover by large amounts.

In summary, our results suggest that if the
specification~\eqref{eq:cps-regression} is close to correct, clustering by the
running variable shrinks the estimated standard error, and thereby leads to a CI
that is too narrow. This effect is particularly severe if the number of clusters
is small. On the other hand, if misspecification is sufficiently severe relative
to the sampling uncertainty, clustering by the running variable increases the
standard error, but the resulting CIs often still remain too narrow.
In the  following sections, we derive theoretical results about the properties of CRV
 standard errors that shed further light on these findings.

\begin{remark}\label{remark:monte-carlo}
  The results of our simulation exercise illustrate the potential severity of
  the issues caused by clustering standard errors at the level running variable
  in discrete RDDs. However, one might be concerned the results are driven by
  some specific feature of the CPS data. Another possible reservation is that by
  varying the bandwidth $h$ we vary both the degree of model misspecification
  and the number of ``clusters'', and cluster-robust standard errors are
  well-known to have statistical issues in other settings with a few
  clusters. To address these potential concerns, we report the results of another simulation
  experiment in the supplemental material. In that simulation study, the data
  are fully computer generated, and we consider several data generating
  processes in which we vary the degree of misspecification and the number of
  support points of the running variable independently. These additional
  simulations confirm the qualitative results reported in this section.
\end{remark}

\begin{remark}\label{remark:few-clusters-robust}
  The clustered standard error used in our simulation exercise corresponds to
  the default setting in STATA\@. Although its formula, given in
  Section~\ref{sec:properties-lee-card} below, involves a ``few clusters''
  adjustment, this variance estimator is known to be biased when the number of
  clusters is small. One might therefore be concerned that the unsatisfactory
  performance of CRV CIs is driven by this issue, and that these CIs might work
  as envisioned by LC if one used one of the alternative variance estimators
  suggested in the literature on inference with few clusters. To address this
  concern, we repeat both the CPS placebo study and the simulation exercise
  described in Remark~\ref{remark:monte-carlo} using a bias-reduction
  modification of the STATA formula developed by \citet{bm02} that is analogous
  to the HC2 modification of the EHW standard error proposed by \citet{MaWh85}.
  The results are reported in the supplemental material. The main findings are
  as follows. When combined with an additional critical value adjustment, also
  due to \citet{bm02}, this fixes the undercoverage of CRV CIs in settings where
  the specification~\eqref{eq:cps-regression} is close to correct. However, this
  comes at a substantial loss in power relative to EHW CIs, which also have
  approximately correct coverage in this case (the CIs are more than twice as
  long in some specifications than EHW CIs). Under more severe misspecification,
  the adjustments tend to improve coverage relative to EHW CIs somewhat, but
  both types of CIs generally still perform poorly. Using bias-corrected
  standard errors does therefore not make CRV CIs robust to misspecification of
  the CEF\@.
\end{remark}

\section{Econometric Framework}\label{sec:properties-lee-card}

In this section, we first review the sharp RDD,\footnote{\label{fn:kink}While
  the issues that we study in this paper also arise in fuzzy RDDs and regression
  kink designs \citep{clpw15}, for ease of exposition we focus here on the sharp
  case. For ease of exposition, we also abstract from the presence of any
  additional covariates; their presence would not meaningfully affect our
  results.} and formally define the treatment effect estimator and the
corresponding standard errors and CIs. We then discuss LC's motivation for
clustering the standard errors by the running variable, and point out some
conceptual shortcomings.

\subsection{Model, Estimator and CIs}\label{sec:basic-model-infer}

In a sharp RDD, we observe a random sample of $N$ units from some large
population. Let $Y_i(1)$ and $Y_i(0)$ denote the potential outcome for the $i$th
unit with and without receiving treatment, respectively, and let
$T_{i}\in\{0,1\}$ be an indicator variable for the event that the unit receives
treatment. The observed outcome is given by
$Y_{i}=(1-T_{i})Y_{i}(0)+T_{i}Y_{i}(1)$. A unit is treated if and only if a
running variable $X_{i}$ crosses a known threshold, which we normalize to zero,
so that $T_{i}=\1{X_{i}\geq 0}$. Let $\mu(X_{i})=\E[Y_{i}\mid X_{i}]$ denote the
conditional expectation of the observed outcome given the running variable. If
the CEFs for the potential outcomes $\E[Y_{i}(1)\mid X_{i}]$ and
$\E[Y_{i}(0)\mid X_{i}]$ are continuous at the threshold, then the discontinuity
in $\mu(x)$ at zero is equal to the average treatment effect (ATE) for units at
the threshold,
\begin{equation*}
  \tau = \E(Y_i(1)-Y_i(0)\mid X_i=0)=
  \lim_{x\downarrow 0}\mu(x) - \lim_{x\uparrow 0}\mu(x).
\end{equation*}

A standard approach to estimate $\tau$ is to use local polynomial regression. In
its simplest form, this estimation strategy involves fixing a window width, or a
bandwidth, $h>0$, and a polynomial order $p\geq 0$, with $p=1$ and $p=2$ being
the most common choices in practice. One then discards all observations for
which the running variable is outside the estimation window $[-h,h]$, keeping
only the $N_{h}$ observations $\{Y_{i},X_{i}\}_{i=1}^{N_{h}}$ for which
$\abs{X_{i}}\leq h$, and runs an OLS regression of the outcome $Y_i$ on a vector
of covariates $M_{i}$ consisting of an intercept, polynomial terms in the
running variable, and their interactions with the treatment
indicator.\footnote{Our setup also covers the global polynomial approach to
  estimating $\tau$ by choosing $h=\infty$. While this estimation approach is
  used in a number of empirical studies, theoretical results suggest that it
  typically performs poorly relative to local linear or local quadratic
  regression \citep{gelman2014high}. This is because the method often implicitly
  assigns very large weights to observations far away from the
  threshold.}\footnote{For ease of exposition, we focus on case with uniform
  kernel. Analogous results can be obtained for more general kernel functions,
  at the expense of a slightly more cumbersome notation. Estimates based on a
  triangular kernel should be more efficient in finite samples
  \citep[e.g.][]{ArKo16optimal}.} The ATE estimator $\widehat{\tau}$ of $\tau$
is then given by the resulting OLS estimate on the treatment indicator. Without
loss of generality, we order the components of $M_i$ such that this treatment
indicator is the first one, so
\begin{align*}
  \widehat{\tau} &= e_1'\widehat\theta, \qquad\widehat{\theta}=\widehat{Q}^{-1}\frac{1}{N_{h}}\sum_{i=1}^{N_{h}}M_{i}Y_{i},
                   \qquad \widehat{Q}=\frac{1}{N_{h}}\sum_{i=1}^{N_{h}} M_{i}M_{i}', \qquad M_{i}=m(X_{i}),\\
  m(x)&=
        (\1{x\geq 0},\1{x\geq 0}x,\ldots,\1{x\geq 0}x^p,1,x,\ldots,x^{p})',
\end{align*}
where $e_1 =(1,0,\ldots,0)'$ denotes the first unit vector. When $p=1$, for
instance, $\widehat{\tau}$ is simply the difference between the intercepts from
separate linear regressions of $Y_{i}$ on $X_{i}$ to the left and to the right
of the threshold. The conventional Eicker-Huber-White (EHW) or
heteroskedasticity-robust standard error of $\widehat\tau$ is
$\EHWse/\sqrt{N_{h}}$, where $\EHWse^2$ is the top-left element of the EHW
estimator of the asymptotic variance of $\widehat\theta$. That is,
\begin{align*}
  \EHWse^2
  & = e_1'\widehat{Q}^{-1} \widehat{\Omega}_{\textnormal{EHW}}
    \widehat{Q}^{-1}e_1,
  &\widehat\Omega_{\textnormal{EHW}}
  &= \frac{1}{N_{h}}\sum_{i=1}^{N_{h}} \widehat{u}_{i}^{2}M_{i} M_i',
  &\widehat{u}_{i} &= Y_i-M_{i}'\widehat\theta.
\end{align*}
An alternative standard error, proposed by LC for RDDs with a discrete running
variable, is $\CRVse/\sqrt{N_{h}}$. Here $\CRVse^2$ is the top-left element of
the ``cluster-robust'' estimator of the asymptotic variance of $\widehat\theta$
that clusters by the running variable (CRV). That is, the estimator treats units
with the same realization of the running variable as belonging to the same
cluster \citep{liang1986longitudinal}. Denoting the $G_{h}$ support points
inside the estimation window by $\{x_{1},\dotsc,x_{G_{h}}\}$, the estimator
$\CRVse^2$ has the form\footnote{\label{fn:stata2}The STATA implementation of
  these standard errors multiplies $\CRVse^2$ by
  $G_{h}/(G_{h}-1) \times (N_{h}-1)/(N_{h}-k)$, where $k=2(p+1)$ is the number
  of parameters that is being estimated. We use this alternative formula for the
  numerical results in Sections~\ref{sec:dist-crv-conf}
  and~\ref{sec:empir-appl}.}
\begin{equation*}
  \CRVse^2 = e_1'\widehat Q^{-1} \widehat{\Omega}_{\textnormal{\CRV}}\widehat
  Q^{-1}e_1,\quad \widehat\Omega_{\textnormal{\CRV}}=
  \frac{1}{N_{h}}\sum_{g=1}^{G_{h}}
  \sum_{i=1}^{N_{h}}\1{X_{i}=x_{g}}\widehat{u}_{i}M_{i}
  \sum_{j=1}^{N_{h}}\1{X_{j}=x_{g}}\widehat{u}_{j}M_{j}'.
\end{equation*}
The EHW and CRV CIs for $\tau$ with nominal level $1-\alpha$ based on these
standard errors are then given by
\begin{align*}
  \widehat\tau \pm z_{1-\alpha/2}
  \times \widehat\sigma_\textnormal{EHW}/\sqrt{N_h},
  \quad\textrm{ and }\quad
  \widehat\tau \pm z_{1-\alpha/2} \times \widehat\sigma_\textnormal{CRV}/\sqrt{N_h},
\end{align*}
respectively, where the critical value $z_{1-\alpha/2}$ is the $1-\alpha/2$
quantile of the standard normal distribution. In particular, for a CI with 95\%
nominal coverage we use $z_{0.975}\approx 1.96$.

\subsection{Potential issues due to discreteness of the running
  variable}\label{sec:potential-issues-due}

Whether the EHW CI is valid for inference on $\tau$ largely depends on the bias
properties of the ATE estimator $\widehat\tau$. It follows from basic regression
theory that in finite samples, $\widehat{\tau}$ is approximately unbiased for
its population counterpart $\tau_{h}$, given by
\begin{align*}
  \tau_{h} &= e_1'\theta_h, &\theta_{h}&=Q^{-1}\E[M_{i}Y_{i}\mid \abs{X_{i}}\leq h],&
  Q&=\E[M_{i}M_{i}'\mid \abs{X_{i}}\leq h]
\end{align*}
(for simplicity, we leave the dependence of $\tau_{h}$ on $p$ implicit).
However, $\widehat{\tau}$ is generally a biased estimator of $\tau$. The
magnitude of the bias, $\tauh-\tau$, is determined by how well $\mu(x)$ is
approximated by a $p$th order polynomial within the estimation window $[-h,h]$.
If $\mu(x)$ was a $p$th order polynomial function, then $\tauh=\tau$, but
imposing this assumption is typically too restrictive in applications.

The EHW CI is generally valid for inference on $\tau_h$ since, if $p$ is fixed,
the $t$-statistic based on the EHW standard error has a standard normal
distribution in large samples. That is, as $N_{h}\to\infty$, under mild regularity
conditions,
\begin{equation}\label{eq:tau-normal}
\sqrt{N_{h}}(\widehat\tau - \tauh)/\EHWse\indist \mathcal{N}(0,1).
\end{equation}
Using the EHW CI for inference on $\tau$ is therefore formally justified if the
bias $\tauh-\tau$ is asymptotically negligible relative to the standard error,
in these sense that their ratio converges in probability to zero,
\begin{equation}\label{eq:smallbias}
  \frac{\tauh-\tau}{\EHWse/\sqrt{N_{h}}}\stackrel{P}{\rightarrow}0.
\end{equation}
Choosing a window width $h$ such that condition~\eqref{eq:smallbias} is
satisfied is called ``undersmoothing'' in the nonparametric regression
literature. If the distribution of the running variable $X_i$ is continuous,
then~\eqref{eq:smallbias} holds under regularity conditions provided that
$h\to 0$ sufficiently quickly as the sample size grows, and that $\mu(x)$ is
sufficiently smooth \citep[e.g.][Theorem 4]{hahn2001identification}. This result
justifies the use of the EHW standard errors for inference on $\tau$ when
$X_{i}$ has rich support and $h$ is chosen sufficiently small.

In principle, the EHW CI is also valid for inference on $\tau$ when the
distribution of $X_{i}$ is discrete, as long as there is some justification for
treating the bias $\tauh-\tau$ as negligible in the sense of
condition~\eqref{eq:smallbias}.\footnote{While most results in the recent
  literature on local polynomial regression are formulated for settings with a
  continuous running variable, early contributions such as that of
  \citet{sacks1978linear} show that it is not necessary to distinguish between
  the discrete and the continuous case in order to conduct inference.} This
justification becomes problematic, however, if the gaps between the support
points closest to the threshold are sufficiently wide. This is because, to
ensure that $N_{h}$ is large enough to control the sampling variability of
$\widehat\tau$ and to ensure that the normality
approximation~\eqref{eq:tau-normal} is accurate, a researcher may be forced to
choose a window width $h$ that is ``too large'', in the sense that the
bias-standard error ratio is likely to be large. When the running variable is
discrete, the EHW CI might therefore not be appropriately centered, and thus
undercover in finite samples.

\subsection{Motivation for clustering by the running variable}%
\label{sec:motiv-using-clust}

To address the problems associated EHW standard errors discussed above, LC
proposed using CRV standard errors when the running variable is discrete. This
approach has since found widespread use in empirical economics. The rationale
provided by LC is as follows. Let $\delta(x)=\mu(x)-m(x)'\thetah$ denote the
specification bias of the (local) polynomial approximation to the true CEF, and
write $\delta_{i}=\delta(X_{i})$. Then, for observations inside the estimation
window, we can write
\begin{equation}\label{lcmodel}
  Y_{i}=M_{i}'\thetah+u_{i},\qquad u_{i}=\delta_{i}+\varepsilon_{i},
\end{equation}
where $\varepsilon_{i}= Y_{i} - \mu(X_{i})$ is the deviation of the observed
outcome from its conditional expectation, and the value of $\delta_i$ is
identical for all units that share the same realization of the running variable.
LC then treat $\delta_i$ as a random effect, rather than a specification error
that is non-random conditional on the running variable. That is, they consider a
model that has the same form as~\eqref{lcmodel}, but where
$\delta_i = \sum_{g=1}^{G_h} D_{g}\1{X_i=x_g}$, with $D = (D_1,\dotsc, D_{G_h})$
a mean-zero random vector with elements that are mutually independent and
independent of $\{(X_i,\varepsilon_i)\}_{i=1}^{N_h}$. Under this assumption,
equation~\eqref{lcmodel} is a correctly specified regression model in which the
error term $u_{i}$ exhibits within-group correlation at the level of the running
variable. That is, for $i\neq j$, we have that $\E(u_i u_j|X_i=X_j) \neq 0$ and
$\E(u_i u_j|X_i\neq X_j)=0$. LC then argue that due to this group structure, the
variance estimator $\CRVse^2$ is appropriate.

This rationale is unusual in several respects. First, it is not compatible with
the standard assumption that the data are generated by i.i.d.\ sampling from a
fixed population, with a non-random CEF\@. Under i.i.d.\ sampling,
$\delta_{i} = \delta(X_i)$ is not random conditional on $X_i$, as the function
$\delta(x)$ depends only on the population distribution of observable
quantities, and does not change across repeated samples. Assuming that
$\delta(x)$ is random implies that the CEF $\mu(x) = m(x)'\thetah +\delta(x)$
changes across repeated samples, since in each sample one observes a different
realization of $D$, and thus different deviations $\delta(x)$ of the CEF from
its ``average value'' $m(x)'\thetah$.

Second, the randomness of the CEF in the LC model implies that the ATE $\tau$,
which is a function of the CEF, is also random. This makes it unclear what the
population object is that the RDD identifies, and the \CRV\ CIs provide
inference for. One might try to resolve this problem by defining the parameter
of interest to be $\tau_{h}$, the magnitude of the discontinuity of the
``average'' CEF
$\E(Y_i\mid X_i) = \E(\E(Y_i\mid X_i,D\mid X_i)) = M_i'\theta_h$, where the
average is taken over the specification errors. But it is unclear in what sense
$\tau_{h}$ measures a causal effect, as it does not correspond to an ATE for any
particular set of units. Furthermore, proceeding like this amounts to assuming
that the chosen polynomial specification is correct after all, which seems to
contradict the original purpose of clustering by running variable.
Alternatively, one could consider inference on the conditional expectation of
the (random) ATE given specification errors, $\E(\tau\mid D)$. But once we
condition on $D$, the data are i.i.d., the LC model becomes equivalent to the
standard setup described in Section~\ref{sec:basic-model-infer}, and all the
concerns outlined in Section~\ref{sec:potential-issues-due} apply.

Third, the LC model effectively rules out smooth CEFs. To see this, note that if
the CEF $\mu(x)$ were smooth, then units with similar values of the running
variable should also have a similar value of the specification error
$\delta(X_{i})$. In the LC model, the random specification errors are
independent for any two values of the running variable, even if the values are
very similar. Since in most empirical applications, one expects a smooth
relationship between the outcome and the running variable, this is an important
downside.

Due to these issues, we argue that the LC model should not be viewed as a
literal description of the DGP, but rather as a heuristic device to motivate
using \CRV\ standard errors. Their performance should be evaluated under the
standard framework outlined in Section~\ref{sec:basic-model-infer}, in which the
data are generated by i.i.d.\ sampling from a fixed population distribution.

\section{Theoretical properties of CRV confidence
  intervals}\label{sec:asymptotic-results}

In this section, we study the properties of $\CRVse^{2}$ under the standard
setup introduced in Section~\ref{sec:basic-model-infer}, which then directly
leads to statements about the coverage properties of the corresponding CIs. The
results reported here summarize the main insights from a more formal analysis in
Appendix~\ref{sec:proofs-results-sect}. To capture the notion that $\mu(x)$ is
``close'' to a polynomial specification, our asymptotic analysis uses a sequence
of data generating processes (DGPs) under which the CEF $\mu(x)$ changes with
the sample size, so that it lies within an $N_{h}^{-1/2}$ neighborhood of the
polynomial specification $m(x)'\theta_{h}$ uniformly over the estimation window
$[-h,h]$. Changing the DGP with the sample size is a widely used technical
device that helps preserve important finite-sample issues in the asymptotic
approximation. In our setting, it allows us capture the fact that it may be
unclear in finite samples whether the chosen polynomial specification is
correct.\footnote{If the CEF $\mu(x)$, and thus the magnitude of the
  specification errors $\delta(x)$, did not change with the sample size, the
  asymptotic coverage probability of CRV CIs would be equal to either zero or
  one, depending on the magnitude of the bias $\tau-\tau_h$. This would clearly
  not be a very useful approximation of the finite-sample properties of these
  CIs.}

This setup implies that there are constants $b$ and $d_1,\ldots,d_{G_h}$ such
that, as the sample size increases,
\begin{align*}
        \sqrt{N_{h}}(\tau_{h}-\tau)\to b\quad\textrm{ and }\quad\sqrt{N_{h}}\delta(x_{g})\to d_{g}
\end{align*}
for each support point $x_{g}$ in the estimation window.
Lemma~\ref{lemma:auxilliary-lemma} in Appendix~\ref{sec:proofs-results-sect}
then shows that in large samples, $\sqrt{N}_{h}(\widehat{\tau}-\tauh)$ is
normally distributed with mean zero and asymptotic variance
\begin{equation*}
  \sigma^{2}_{\tau}=\sum_{g=1}^{G_{h}}\sigma^{2}_{g}\omega_{g},\qquad
  \omega_{g}=e_{1}'Q^{-1}Q_{g}Q^{-1}e_{1},
\end{equation*}
where $\sigma^{2}_{g}=\Var(Y_{i}\mid X_{i}=x_{g})$ denotes the conditional
variance of the outcome given the running variable,
$Q_{g}=\pi_{g}m(x_{g})m(x_{g})'$ and $\pi_{g}= P(X_i=x_g\mid \abs{X_{i}}\leq h)$
is the probability that the value of the running variable for a unit within the
estimation window equals $x_{g}$, the $g$th support point. The asymptotic
variance $\sigma^{2}_{\tau}$ is a weighted sum of the conditional variances of
the outcome given the running variable, with the weights $\omega_{g}$ reflecting
the influence of observations with $X_{i}=x_{g}$ on the estimate
$\widehat{\tau}$. It follows that in large samples,
$\sqrt{N}_{h}(\widehat{\tau}-\tau)$ is normally distributed with asymptotic bias
$b$ and variance $\sigma^{2}_{\tau}$. Lemma~\ref{lemma:auxilliary-lemma}
also shows that the EHW variance estimator is consistent for
$\sigma^{2}_{\tau}$.

As discussed in Section~\ref{sec:basic-model-infer}, CIs based on EHW standard errors
 generally undercover $\tau$, as they are not correctly centered unless
$b=0$. A CI based on the \CRV\ standard error $\CRVse/\sqrt{N_{h}}$ is centered
at the same point estimate. In order for a \CRV\ CI to have correct coverage,
the \CRV\ standard error thus has to be larger than the EHW standard error by
a suitable amount, at least on average. In the remainder of this section,
we show that this is \emph{not} the case, and that CIs based on \CRV\ standard
errors (i) undercover the ATE under correct specification and (ii) can either
over- or undercover the ATE under misspecification by essentially arbitrary
amounts. The asymptotic behavior of $\CRVse^{2}$ differs depending on whether
the number of support points inside the estimation window ${G}_{h}$ is treated
as fixed or as increasing with the sample size. We examine each of these two
cases in turn now.

\subsection{Fixed number of support points}
In Theorem~\ref{th1} in Appendix~\ref{sec:proofs-results-sect}, we study the
properties of $\CRVse^{2}$ when ${G}_{h}$ is held fixed as the sample size
increases. This should deliver useful approximations for settings with a small
to moderate number of
support points within the estimation window $[-h,h]$. We show that in this case
$\CRVse^{2}$ does not converge to a constant, but to a non-degenerate limit
denoted by $\limCRVse^{2}$. This means that it remains stochastic even in large
samples.\footnote{In general, clustered standard errors converge to a
  non-degenerate limiting distribution unless the number of clusters increases
  with the sample size at a suitable rate \citep[see for example,][]{hansen2007asymptotic}.} It follows
from Theorem~\ref{th1} that we can decompose the expected difference between
$\limCRVse^{2}$ and $\sigma_{\tau}^{2}$ as
\begin{equation}\label{eq:CRV-hetero-limit}
  \E(\limCRVse^{2})-\sigma_\tau^2=
  \sum_{g=1}^{G_{h}}
  d_g^2\pi_{g}\omega_{g} +\sum_{g=1}^{G_{h}}
  m(x_{g})'  Q^{-1}
  \left(
    \sum_{j=1}^{G_{h}}\sigma^{2}_{j}Q_{j}
    -  2\sigma^{2}_{g}Q\right) Q^{-1} m(x_{g})\cdot
  \pi_{g}\omega_{g}.
\end{equation}
The first term on the right-hand side of~\eqref{eq:CRV-hetero-limit} is
positive, and its magnitude depends on the degree of misspecification. The
second term does not depend on the degree of misspecification and vanishes if we
replace the estimated regression residuals $\widehat{u}_{i}$ in the \CRV\
formula with the true error $u_{i}$. Its sign is difficult to determine in general.
It is therefore helpful to consider the case in which the conditional variance
of the outcomes is homoskedastic, $\sigma^{2}_{g}=\sigma^{2}$, so that
$\sigma^{2}_{\tau}=\sigma^{2}\sum_{g=1}^{G_{h}}\omega_{g}$. The expected
difference between $\limCRVse^{2}$ and $\sigma_{\tau}^{2}$, relative to
$\sigma^{2}_{\tau}$, can then be written as,
\begin{equation}\label{eq:CRV-homo-limit}
  \frac{  \E(\limCRVse^{2})-\sigma_{\tau}^{2}}{\sigma^{2}_{\tau}}=
  \frac{  \sum_{g=1}^{G_{h}}
    d_g^2\pi_{g}\omega_{g}}{\sigma^{2}\sum_{g=1}^{G_{h}}\omega_{g}}
  -  \frac{ \sum_{g=1}^{G_{h}}  m(x_{g})'  Q^{-1} m(x_{g})
    \pi_{g}\omega_{g}}{\sum_{g=1}^{G_{h}}\omega_{g}}\equiv T_1 + T_2.
\end{equation}
This decomposition is key to understanding the properties of $\CRVse^2$ in
settings with a few support points $G_{h}$. Since the EHW variance
estimator is consistent for $\sigma^{2}_{\tau}$, the notation introduced
in~\eqref{eq:CRV-homo-limit} implies that average magnitude of $\CRVse^2$ is about
$(1 +T_1 + T_2)$ times that of $\widehat\sigma^2_\textnormal{EHW}$. In order for
the CRV CI to have correct coverage, the terms $T_1 + T_2$ have to be positive,
with a magnitude that suitably reflects the degree of misspecification. That is
generally not the case.

First, the term $T_1$ is indeed positive and increasing in the ``average''
misspecification, as measured by the average squared local specification error,
$d_{g}^{2}$, weighted by the cluster size $\pi_{g}$ and $\omega_{g}$.
Heuristically, this term arises because the CRV variance estimator treats the
specification errors as random effects, and tries to account for their
``variance''. The magnitude of $T_1$ depends on the magnitude of the average
misspecification, relative to $\sigma^{2}$. Second, the term $T_2$ is negative
under homoskedasticity because both the weights $\omega_{g}$ and
$m(x_{g})' Q^{-1} m(x_{g})$ are positive. We thus also expect the term to be
negative for small and moderate levels of heteroskedasticity. Third, the
magnitude of $T_2$ depends only on the marginal distribution of the running
variable, and not on $\sigma^{2}$. It is therefore the dominant term in the
decomposition~\eqref{eq:CRV-homo-limit} if the variance $\sigma^2$ of the error
terms is large. Fourth, the term $T_2$ is increasing in the polynomial order
$p$. To see this, observe that
$\sum_{g=1}^{G_{h}}m(x_{g})'Q^{-1}m(x_{g})\pi_{g}=
\trace(\sum_{g=1}^{G_{h}}Q_{g}Q^{-1})=2(p+1)$. The term $T_2$ can thus be
written as a weighted average of terms whose unweighted average is equal to
$2(p+1)$. Heuristically, this means that the expected difference between
$\limCRVse^{2}$ and $\sigma_{\tau}^{2}$ should be increasing with the order of
the polynomial. For example, it is generally larger for local quadratic
regression ($p=2$) than for local linear regression ($p=1$). Fifth, the term
$T_2$ is analogous to the downward bias of cluster-robust variance estimators in
settings with a few clusters \citep[e.g.][]{CaMi14}. While the existence of such
biases is well-known in principle, in the current setting the distortion can be
substantial even for moderate values of $G_{h}$, as shown by our simulations in
Section~\ref{sec:dist-crv-conf} and the supplemental material.

In summary, our theoretical results show if the degree of misspecification
relative to the variance of the residuals and the number of clusters are both
small to moderate, the \CRV\ standard error is on average \emph{smaller} than
the EHW standard error. Clustering the standard error by the running variable
will then {amplify}, rather than solve, the problems for inference caused by
specification bias. This problem is particularly severe under correct
specification. These results are exactly in line with the evidence presented in
Section~\ref{sec:dist-crv-conf}.

\subsection{Increasing number of support points}\label{sec:incr-numb-supp}

In Theorem~\ref{theorem:g-increasing} in Appendix~\ref{sec:proofs-results-sect},
we study the properties of $\CRVse^{2}$ when ${G}_{h}$ increases with the sample
size. This should deliver a useful approximation when there are many support
points within the estimation window $[-h, h]$. We show that if $G_{h}\to\infty$
as $N_{h}\to\infty$, the variance of $\CRVse^{2}$ vanishes, and we obtain that
\begin{equation*}
  \CRVse^{2}-\sigma^{2}_{\tau}=\sum_{g=1}^{G_{h}}d_{g}^{2}\pi_{g}\omega_{g}+o_{P}(1).
\end{equation*}
The right-hand side coincides with the first term in
Equation~\eqref{eq:CRV-hetero-limit}; the second term vanishes. This means that
if the number of support points $G_{h}$ is large, the \CRV\ standard error
indeed tends to be larger than the EHW standard error, so that clustering by the
running variable improves coverage of the corresponding CI\@.\footnote{A similar
  result could be obtained for the alternative clustered variance estimators
  described in Remark~\ref{remark:few-clusters-robust}, which involve a ``few
  clusters'' bias correction. Such corrections would shrink the second term in
  Equation~\eqref{eq:CRV-hetero-limit}, and possibly remove it under additional
  assumptions on the error term.}

This fact alone does not imply that CRV confidence intervals have correct
coverage for all possible CEFs $\mu(\cdot)$, but only that there exist a
set $\mathcal{M}_{\textnormal{\CRV}}$ of CEFs over the estimation window $[-h, h]$
for which correct asymptotic coverage is delivered. While
$\mathcal{M}_{\textnormal{\CRV}}$ is strictly larger than the set of all $p$th
order polynomial functions, this property by itself is not a sufficient argument
for justifying the use of \CRV\ CIs. This is because any inference method that
leads to CIs wider than those based on EHW standard errors---including
undesirable approaches such as simply adding an arbitrary constant to the EHW
standard errors---gives an improvement in coverage, and delivers correct
inference if the true CEF lies in some set that is strictly larger than the set
of all $p$th order polynomial functions.

Whether \CRV\ standard errors provide a meaningful degree of robustness to
misspecification depends on whether $\mathcal{M}_{\textnormal{\CRV}}$ contains
CEFs that are empirically relevant and easy to interpret, so that it is clear in
practice what types of restrictions on $\mu(x)$ the researcher is imposing. In
this respect, the set $\mathcal{M}_{\textnormal{\CRV}}$ has the undesirable
feature that it depends on the distribution of the running variable through the
weights $\omega_{g}$ and the relative cluster sizes $\pi_{g}$. This makes the
set difficult to characterize. Furthermore, it implies that two researchers
studying the same RDD may arrive to different conclusions if, say, one of them
oversampled observations closer to the cutoff (so that the relative cluster size
$\pi_{g}$ for those observations is larger). Finally, simulation results in the
supplemental material show that the set does not include mild departures from a
polynomial specification, which is in line with the intuition given in
Section~\ref{sec:motiv-using-clust} that the model that motivates the CRV
approach implicitly rules out smooth CEFs.

For these reasons, when the number of support points within the estimation
window is large, and the researcher is worried that $\widehat{\tau}$ is biased,
a better alternative to using CRV CIs is to choose a smaller bandwidth.
Alternatively, one can use one of the honest inference procedures that we
outline in Section~\ref{sec:honest-conf-interv}, which guarantee proper coverage
under easily interpretable restrictions on $\mu(x)$.

\section{Honest Confidence Intervals}\label{sec:honest-conf-interv}

If the CEF $\mu(x)$ is allowed to vary arbitrarily between the two support
points of the running variable closest to the threshold, no method for inference
on the ATE can be both valid and informative: without any restrictions on
$\mu(x)$, even in large samples, any point on the real line would remain a
feasible candidate for the value of $\tau$. To make progress, we need to place
constraints on $\mu(x)$. Formally, we impose the assumption that
$\mu\in\mathcal{M}$ for some class of functions $\mathcal{M}$, and then seek to
construct CIs $C^{1-\alpha}$ that satisfy
\begin{equation}\label{eq:uniform-coverage-M}
  \liminf_{N\to\infty}\inf_{\mu\in\mathcal{M}}
  P_\mu (\tau\in C^{1-\alpha}) \geq 1-\alpha,
\end{equation}
where the notation $P_{\mu}$ makes explicit that the coverage probability
depends on the CEF\@. Following the terminology in \citet{li89}, we refer to
such CIs as \emph{honest with respect to $\mathcal{M}$}. Honesty with respect to
a meaningful and interpretable function class is desirable for a CI, as it
guarantees good coverage properties even when facing the ``the worst possible'' CEF
that still satisfies the postulated constraints. In the following subsections,
we present honest CIs with respect to two different function classes, which correspond
to two different formalizations of the notion that $\mu(x)$ can be well-approximated by a
polynomial. The
implementation of these CIs is straightforward using the software package \texttt{RDHonest},
available at \texttt{https://github.com/kolesarm/RDHonest}. Proofs are relegated
to Appendix~\ref{sec:honest-conf-interv-proofs}.

\subsection{Bound on the second derivative}\label{sec:bound-second-deriv}

The first restriction on $\mu(x)$ that we consider is based on bounding the
magnitude of its second derivative, which is perhaps the most natural way of
imposing smoothness. Specifically, we assume that $\mu(x)$ is twice
differentiable on either side of the cutoff, with a second derivative that is
bounded in absolute value by a known constant $\SC$. By choosing a value of
$\SC$ close to zero, the researcher can thus formalize the notion that $\mu(x)$
is close to being linear, whereas choosing a large value of $\SC$ allows for
less smooth CEFs. For technical reasons we only require $\mu(x)$ to be twice
differentiable almost everywhere, which leads to the second-order Hölder class
\begin{equation*}
 \mathcal{M}_{\textnormal{H}}(\SC)=\{\mu\colon \abs{\mu'(a)-\mu'(b)}\leq \SC\abs{a-b}\textrm{ for all } a,b\in\mathbb{R}_{-} \textrm{ and all } a,b\in\mathbb{R}_{+}\}.
\end{equation*}
A local version of this assumption is used in \citet{cfm97}, whose results
formally justify using local polynomial regression to estimate the ATE parameter
when the running variable is continuous. Our goal is inference: we seek to
construct CIs that are honest with respect to
$\mathcal{M}_{\textnormal{H}}(\SC)$, and based on a local linear estimator
$\widehat{\tau}$ as defined in Section~\ref{sec:properties-lee-card} (with
$p=1$). This can be accomplished using results in \citet{ArKo16honest}, who
study the construction of honest CIs in general nonparametric regression
problems.\footnote{The related problem of honest testing for a jump in the
  density of the running variable at the threshold has been considered in
  \citet{frandsen16manipulation}, who uses a bound on the second derivative of
  the density that yields a class of densities similar to
  $\mathcal{M}_\textnormal{H}(\SC)$.}

It turns out that it is easier to construct CIs that are honest conditional on
the realizations of the running variable, which is a slightly stronger
condition than~\eqref{eq:uniform-coverage-M}. To explain the construction, let
$\widetilde{\tau}_{h}=\E(\widehat{\tau}\mid X_{1},\dotsc,X_{N_{h}})$ denote
conditional expectation of the treatment effect estimator, and let
$\widehat\sigma^2_{\textnormal{NN}}/N_{h}$ denote the nearest-neighbor estimator
\citep{AbIm06,AbImZh14} of $\Var(\widehat{\tau}\mid X_{1},\dotsc,X_{N_{h}})$,
the conditional variance of $\widehat{\tau}$. With discrete running variable,
this estimator can be written as
\begin{align*}
  \widehat\sigma^2_{\textnormal{NN}}
  = e_1'\widehat Q^{-1} \widehat\Omega_{\textnormal{NN}}\widehat
  Q^{-1}e_1,\qquad
  \widehat\Omega_{\textnormal{NN}} = \frac{1}{N_{h}}
  \sum_{g=1}^{G_{h}}n_{g}\widehat{\sigma}^{2}_{g}
  m(x_{g})m(x_{g})',
\end{align*}
where $n_{g}$ is the number of observations with $X_{i}=x_{g}$, and
$\widehat{\sigma}^{2}_{g}=\sum_{i\colon
  X_{i}=x_{g}}(Y_{i}-\overline{Y}_{g})^{2}/(n_{g}-1)$ is an unbiased estimator
of the conditional variance $\sigma^{2}_{g}=\Var(Y_{i}\mid X_{i}=x_{g})$, with
$\overline{Y}_{g}=n_{g}^{-1}\sum_{i\colon
  X_{i}=x_{g}}Y_{i}$.\footnote{\citet{AbImZh14} show that the EHW variance
  estimator $\EHWse^{2}/N_{h}$ overestimates the conditional variance of
  $\widehat\tau$, while the nearest-neighbor estimator is consistent under mild
  regularity conditions, in the sense that
  $\Var(\widehat{\tau}\mid X_{1},\dotsc,X_{N_{h}}) /(
  \widehat\sigma^2_\textnormal{NN}/N_{h})\stackrel{P}{\rightarrow} 1$. One could
  nevertheless use the EHW standard error for the construction of the CI
  described in this subsection, but the resulting CI would be conservative.} We
can decompose the $t$-statistic based on this variance estimator as
\begin{equation*}
  \frac{\widehat{\tau}-\tau}{\widehat{\sigma}_\textnormal{NN}/ \sqrt{N_{h}}}=
  \frac{\widehat{\tau}-\widetilde{\tau}_{h}}{\widehat{\sigma}_\textnormal{NN}/\sqrt{N_{h}}}
  +\frac{\widetilde{\tau}_{h}-\tau}{\widehat{\sigma}_\textnormal{NN}/\sqrt{N_{h}}}.
\end{equation*}
Under mild regularity conditions, a Central Limit Theorem ensures that the first
term has standard normal distribution in large samples, and the second term is
bounded in absolute value by
\begin{equation*}
  r_{\sup}=
  \frac{\sup_{\mu\in\mathcal{M}_{H}(\SC)}
    \abs{\widetilde{\tau}_{h}-\tau}}{\widehat{\sigma}_\textnormal{NN}/\sqrt{N_{h}}}.
\end{equation*}

In Appendix~\ref{sec:honest-conf-interv-proofs}, we show that the supremum on
the right-hand side of the last equation is attained by the function $\mu(x)$
that equals $-\SC x^{2}$ for $x\geq 0$ and $\SC x^{2}$ for $x<0$. This yields
the explicit expression
\begin{align}\label{eq:holder-bias-expression}
  r_{\sup}
  &=
    -\frac{\SC}{2}
    \frac{\sum_{i=1}^{N_{h}} w(X_{i}) X_{i}^{2} \sign(X_i)}{
    \widehat{\sigma}_\textnormal{NN}/\sqrt{N_{h}}},
  & w(X_{i})
  &= \frac{1}{N_h} \cdot e_1'\widehat{Q}^{-1} M_i,
\end{align}
for the upper bound $r_{\sup}$ on the second component of the $t$-statistic.
This approach of bounding the second derivative (BSD) thus leads to the
following CI\@:
\begin{proposition}\label{theorem:holder-honest}
  Let $\cv_{1-\alpha}(r)$ denote the $1-\alpha$ quantile of the
  $\abs{\mathcal{N}(r,1)}$ distribution (the distribution of the absolute value
  of a normal random variable with mean $r$ and variance 1). Then, under
  regularity conditions stated in Appendix~\ref{sec:honest-conf-interv-proofs},
  the CI
  \begin{equation}\label{eq:holder-CI}
    C_{\textnormal{BSD}}^{1-\alpha} =
    \left(\widehat{\tau}\pm \cv_{1-\alpha}(r_{\sup})\times
      \widehat{\sigma}_\textnormal{NN}/\sqrt{N_{h}}\right).
  \end{equation}
  is honest with respect to $\mathcal{M}_{\textnormal{H}}(\SC)$.
\end{proposition}
This CI has several attractive features. First, since the CI is valid
conditional on the realizations of the running variable, it is not necessary to
distinguish between the cases of a discrete and a continuous distribution of the
running variable. The CI is valid in both cases. Second,
$C_{\textnormal{BSD}}^{1-\alpha}$ takes into account the exact finite-sample bias of
the estimator, and is thus valid for \emph{any} choice of bandwidth. In
particular, it does not rely on asymptotic promises about what the bandwidth
would have been had the sample size been larger, and remains valid even if the
bandwidth is treated as fixed. In contrast, the usual methods of inference in
RDDs rely on the assumption that the bandwidth shrinks to zero sufficiently
quickly relative to the sample size, which can be hard to evaluate in practice.
Third, to achieve the tightest possible CI, one can simply choose a bandwidth
$h$ that minimizes its length
$2\cv_{1-\alpha}(r_{\sup})\times \widehat{\sigma}_\textnormal{NN}/\sqrt{N_{h}}$.
Since this quantity depends on the outcome data only through the variance
estimate $\widehat{\sigma}_\textnormal{NN}$, which does not depend on $\mu$, and
which can be shown to be consistent under mild regularity conditions, doing so
does not invalidate the honesty of the resulting CI\@.

An important implementation aspect of $C_{\textnormal{BSD}}^{1-\alpha}$ is that
it requires the researcher to choose a value for $\SC$. This choice reflects the
researcher's beliefs of what constitutes a plausible level of fluctuation in
the function $\mu(x)$ in the particular empirical context. In practice, we
recommend reporting results for a range of $\SC$ values in order to illustrate
the sensitivity of empirical conclusions to this choice. We find the following
heuristic useful to get an intuition for the restrictions implied by
particular choices of $\SC$. Consider the function $\mu$ over some interval
$[x^{*},x^{**}]$. One can show by simple algebra that if
$\mu\in\mathcal{M}_{\textnormal{H}}(\SC)$, then it differs by at most
$\SC\times(x^{**}-x^*)^2/8$ from a straight line between $\mu(x^*)$ and
$\mu(x^{**})$. Thus, if a researcher believes, for example, that the CEF differs
by no more than $S$ from a straight line between the CEF values at the endpoints
of any interval of length one in the support of the running variable, a
reasonable choice for the bound on the second derivative is $\SC= 8S$.

The use of subject knowledge for choosing $\SC$ is necessary, as it is not
possible to use a data-driven method without additional assumptions. In
particular, it follows from the results in \citet{low97} and
\citet{ArKo16optimal} that it is not possible to form honest CIs that are
tighter than those in Proposition~\ref{theorem:holder-honest} by using
data-dependent tuning parameters, and at the same time maintain coverage over
the whole function class $\mathcal{M}_{\textnormal{H}}(\SC)$, for some
conservative upper bound $\SC$. This is related to the fact that it is not
possible to upper bound $\SC$ from the data. It is possible, however, to lower
bound $\SC$ from the data. In the supplemental material, we construct an
estimate and a left-sided CI for this lower bound by adapting the construction
from \citet{ArKo16optimal} to our setup. We apply this method to the two
empirical applications in Section~\ref{sec:empir-appl} below. The left-sided CI
can be used as a specification test to check if the chosen value of $\SC$ is too
low to be compatible with the data.

\subsection{Bounds on specification errors at the
  threshold}\label{sec:bounds-spec-errors}

The second restriction on $\mu(x)$ that we consider formalizes the intuitive
notion that the chosen model fits no worse at the cutoff than anywhere else
in the support of the running variable within the estimation window $[-h,h]$. In
particular, we assume that the left limit of the specification bias at the
threshold, $\lim_{x\uparrow 0}\delta(x)$, is no larger than the specification
bias $\delta(x_{g})$ at any point $x_{g}$ in the support of the running variable
below the threshold, and similarly for the right limit. This leads to the class
\begin{equation*}
  \mathcal{M}_{\textnormal{BME}}(h) = \left\{\mu \colon
  |\lim_{x\uparrow 0}\delta(x)| \leq
  \max_{1\leq g\leq G_{h}^{-}} |\delta(x_g)| \textnormal{ and } |\lim_{x\downarrow
  0}\delta(x)| \leq \max_{ G_{h}^{-}<g\leq G_{h}} |\delta(x_g)|\right\},
\end{equation*}
where we assume that the support points $\{x_{1},\dotsc,x_{G_{h}}\}$ are ordered
so that the first $G_{h}^{-}$ points are below the cutoff, and
$\delta(x) \equiv \mu(x) - m(x)'\theta_{h}$ is the specification bias. Note that
the specification bias depends on the function $\mu(x)$ directly, but also
indirectly as the definition of the best polynomial predictor $m(x)'\theta_{h}$
depends on $\mu(x)$. We refer to this class as the bounded misspecification
error (BME) class.

The assumption that
$\mathcal{M}_{\textnormal{BME}}(h)$ contains the true CEF implies that we can
bound the bias $\tauh-\tau$ as follows:
\begin{equation*}
 |\tauh-\tau|\leq
  b_{\max},\qquad b_{\max}\equiv \max_{G_{h}^{-}<g\leq G_{h}} |\delta(x_g)| +
  \max_{1\leq g\leq G_{h}^{-}} |\delta(x_g)|.
\end{equation*}
It will be convenient to rewrite the bias bound $b_{\max}$ in a way that
avoids using the absolute value operator as
\begin{equation}\label{eq:bias-bound}
  b_{\max}=\max_{W\in\mathcal{W}}b(W),\qquad b(W)=s^+\delta(x_{g^+}) + s^-\delta(x_{g^-}),
\end{equation}
where $W=(g^{-},g^{+},s^{-},s^{+})$, and the maximum is taken over the set
$\mathcal{W}=\{1,\dotsc,G_{h}^{-}\}\times \{G_{h}^{-}+1,\dotsc,G_{h}\}\times
\{-1,1\}^{2}$.

As a first step towards constructing a two-sided CI, consider first the slightly
simpler problem of constructing an honest \emph{right-sided} CI for $\tau$.
Suppose we knew that~\eqref{eq:bias-bound} was maximized at
$W_{0}=(g_{0}^{-},g_{0}^{+},s_{0}^{-},s_{0}^{+})$. Then we could estimate the
bias bound by
\begin{equation*}
  \widehat{b}(W_{0})\equiv
  s_{0}^{+}\widehat{\delta}(x_{g_{0}^{+}})+s_{0}^{-}\widehat{\delta}(x_{g_{0}^{-}}).
\end{equation*}
Here $\widehat{\delta}(x_{g})=\widehat\mu(x_{g}) - m(x_{g})'\widehat{\theta}$ is
an estimator of $\delta(x_{g})$, where
$\widehat\mu(x_{g}) = \sum_{i=1}^{N_{h}}Y_{i}\1{X_i=x_g}/n_{g}$ is simply the
average outcome for the $n_{g}$ observations whose realization of the running
variable is equal to $x_{g}$. Since $\widehat{\tau}+\widehat{b}(W_{0})$ is
asymptotically normal, we could then construct a right-sided CI for $\tau$ with
confidence level $1-\alpha$ as
\begin{equation*}
  \hol{-\infty,c^{1-\alpha}_{R}(W_{0})},\qquad
  c^{1-\alpha}_{R}(W_{0})=\widehat{\tau}+\widehat{b}(W_{0})
  +z_{1-\alpha}\widehat{V}(W_{0})/\sqrt{N_{h}},
\end{equation*}
where $\widehat{V}(W_{0})$ is a consistent estimator of the asymptotic variance
of $\widehat{\tau}+\widehat{b}(W_{0})$ given in Equation~\eqref{eq:hat-VW} in
Appendix~\ref{sec:honest-conf-interv-proofs}, and $z_{1-\alpha}$ denotes the
$1-\alpha$ quantile of the standard normal distribution.

Although we do not know which $W\in\mathcal{W}$ maximizes~\eqref{eq:bias-bound},
we do know that the union of CIs,
\begin{equation*}
  C^{1-\alpha}_{\text{BME},R}\equiv
  \bigcup_{W\in\mathcal{W}}\hol{-\infty,c^{1-\alpha}_{R}(W)}
  =\hol{-\infty,\max_{W\in\mathcal{W}}c^{1-\alpha}_{R}(W)},
\end{equation*}
contains the CI corresponding to that value of $W$ that
maximizes~\eqref{eq:bias-bound}. Hence, $C^{1-\alpha}_{\text{BME},R}$ is a valid
CI for $\tau$. This type of construction is called the union-intersection
principle \citep[Chapter 8.2.3]{CaBe02}, and produces CIs with correct asymptotic
coverage. 
By analogous reasoning, a left-sided CI for $\tau$ can be constructed as
\begin{equation*}
  C^{1-\alpha}_{\text{BME},L}\equiv
  \bigcup_{W\in\mathcal{W}}\hor{c^{1-\alpha}_{L}(W),\infty}
  =\hor{\min_{W\in\mathcal{W}}c^{1-\alpha}_{L}(W),\infty},\;\;
  c^{1-\alpha}_{L}(W)=\widehat{\tau}
  +\widehat{b}(W)-z_{1-\alpha}\widehat{V}(W)/\sqrt{N_{h}}.
\end{equation*}
An intersection of the right- and left-sided CIs with a Bonferroni adjustment
for their confidence levels then yields a two-sided CI for $\tau$ with
confidence level $1-\alpha$:
\begin{proposition}\label{th:bounds-missp-errors}
  Under regularity conditions in Appendix~\ref{sec:honest-conf-interv-proofs},
  the CI
  \begin{equation*}
    C^{1-\alpha}_{\textnormal{BME}} \equiv
    \left(
      \min_{W\in\mathcal{W}}c^{1-\alpha/2}_{L}(W)\;,\;
      \max_{W\in\mathcal{W}}c^{1-\alpha/2}_{R}(W)\right).
  \end{equation*}
  is honest with respect to $\mathcal{M}_{\textnormal{BME}}(h)$.
\end{proposition}
In contrast to the CI described in the previous subsection, which is valid
regardless of whether the running variable $X_{i}$ is discrete or continuous,
$C^{1-\alpha}_{\textnormal{BME}}$ is explicitly tailored to settings in which
the running variable is discrete. It also requires many observations to be
available for every support point within the estimation window. This is needed
to justify joint asymptotic normality of the estimators
$\widehat\mu(x_{1}),\ldots,\widehat\mu(x_G)$, which is in turn needed to ensure
that $\widehat{\tau}+\widehat{b}(W)$ is asymptotically normal. We do not
recommend using the BME CI when the variance of $\widehat\mu(x_{g})$ is large
relative to the absolute magnitude of $\delta(x_{g})$, or when there are many
support points within the estimation window.

\begin{remark}
  In the supplemental material, we evaluate the performance of the honest CIs
  described in this section in the CPS placebo study from
  Section~\ref{sec:dist-crv-conf}, and also in an additional simulation study.
  BSD CIs are shown to have excellent coverage properties given an appropriate
  choice of the constant $\SC$. BME CIs perform well in designs with a few
  support points when many observations are available at each of them; they are
  conservative otherwise.
\end{remark}

\section{Empirical Applications}\label{sec:empir-appl}

In this section, we use data from \citet{oreopoulos2006estimating} and
\citet{lalive08} to illustrate the performance of CRV CIs and honest CIs from
Section~\ref{sec:honest-conf-interv} in empirical settings.

\subsection{\texorpdfstring{\citet{oreopoulos2006estimating}}{Oreopoulos (2006)}}\label{sec:oreopoulos-2006}

\begin{figure}[!t]
  \begin{center}
    \includegraphics[width=.9\textwidth]{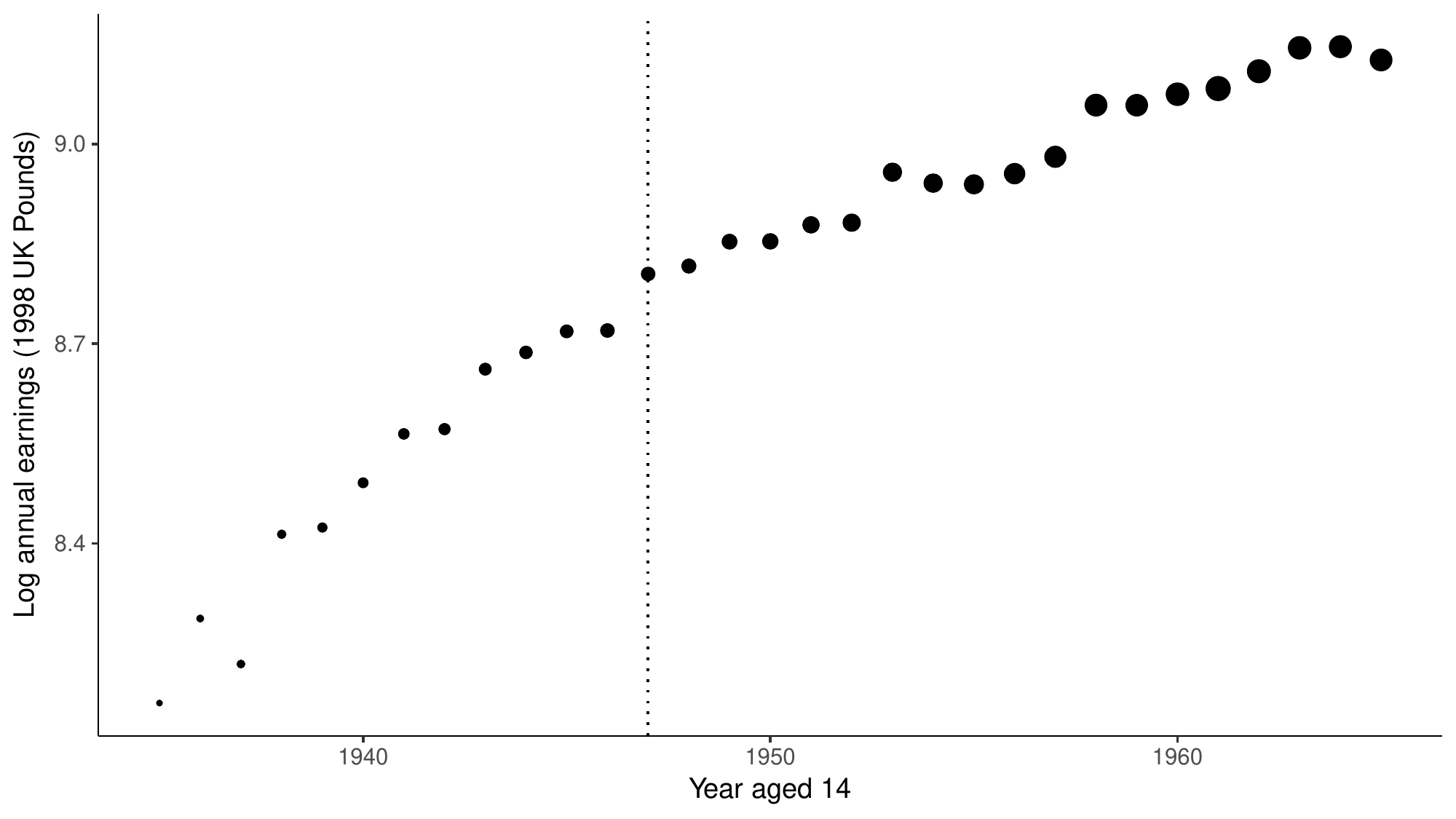}
    \caption{\label{fig_est} Average of natural logarithm of annual earnings by
      year aged 14. Vertical line indicates the year 1947, in which the minimum
      school-leaving age changed from 14 to 15. Volume of dots is proportional
      to share of workers in the full data with the corresponding age.}
  \end{center}
\end{figure}

\citet{oreopoulos2006estimating} studies the effect of a change in the
minimum school-leaving age in the United Kingdom from 14 to 15 on schooling
attainment and workers' earnings. The change occurred in 1947 in Great
Britain (England, Scotland and Wales), and in 1957 in Northern Ireland. The data
are a sample of UK workers  who turned 14 between 1935 and 1965, obtained by
combining the 1984--2006 waves of the U.K.\ General Household Survey; see
\citet{oreopoulos2006estimating,oreopoulos2006online} for details.

For simplicity, we focus on the sub-sample of British workers, and restrict
attention to the effect of facing a higher minimum school-leaving age on (the
natural logarithm of) annual earnings measured in 1998 U.K.\ Pounds. Oreopoulos
uses a sharp RDD to estimate this parameter. The running variable is the year in which
the worker turned 14, and the treatment threshold is 1947. The running variable
thus has $G=31$ support points, of which $G^+=19$ are above the threshold, and
$G^-=12$ ones are below.
\citet{oreopoulos2006estimating,oreopoulos2006online} uses the global specification
\begin{equation}
  \log(\text{EARN}_i) = \beta_0 + \tau\cdot\1{\text{YEAR14}_i\geq 1947}+
  \sum_{k=1}^4 \beta_k\cdot \text{YEAR14}_i^k + U_i.\label{originalspec}
\end{equation}
Table~\ref{tab:results4} reports the resulting treatment effect estimate along
with \CRV\ and conventional EHW standard errors and CIs for this specification
(column (1)). In addition, we consider linear and quadratic specifications
fitted separately on each side of the threshold using either the full data set
(columns (2)--(3)), or with the estimation window restricted to $h=6$ years
around the threshold (columns (4)--(5)), or $h=3$ years (columns
(6)--(7)).\footnote{\label{fn:oreop-online} Our analysis is based on the data
  distributed together with the online corrigendum \citet{oreopoulos2006online},
  which fixes some coding errors in the original paper and includes data from
  additional waves of the U.K.\ General Household Survey. The results in column
  (1) therefore differ from those reported \citet{oreopoulos2006estimating}, but
  they are identical to those given in
  \citet{oreopoulos2006online}.}\footnote{We consider different specifications
  in order to illustrate how these affect the relative magnitude of CRV and EHW
  standard errors. Whether the estimated effect is significantly different from
  zero is secondary for our purposes. See e.g.\ \citet{devereux2010forced} for a
  discussion of the sensitivity of the point estimates in
  \citet{oreopoulos2006estimating} with respect to model specification.}

\afterpage{
\begin{landscape}
\begin{table}  \caption{Effect of being subject to increases minimum school-leaving
    age on natural logarithm of annual earnings.}\label{tab:results4}\vspace{1ex}
  \begin{tabular*}{\linewidth}{@{\extracolsep{\fill}}lccccccc@{}}
    \toprule
    \multicolumn{8}{c}{\textbf{Panel A\@: Point estimates and EHW/CRV/CRV2/CRV-BM/BME inference}} \\[0.5ex]
    & (1) & (2) & (3) & (4) & (5) & (6) & (7)  \\\midrule
Estimate  &            .055 &          $-.011$&            .042 &            .021 &            .085 &            .065 &            .110 \\
EHW SE    &            .030 &            .023 &            .038 &            .033 &            .058 &            .049 &            .127 \\
EHW CI    & $(-.003, .113)$& $(-.056, .035)$& $(-.032, .115)$& $(-.043, .085)$& $(-.029, .199)$& $(-.031, .161)$& $(-.138, .359)$\\[0.2ex]
CRV SE    &            .015 &            .027 &            .019 &            .020 &            .016 &            .009 &            .004 \\
CRV CI    &  $(.026, .084)$& $(-.063, .042)$&  $(.005, .079)$& $(-.018, .060)$&  $(.053, .117)$&  $(.048, .082)$&  $(.102, .119)$\\[0.2ex]
CRV2 SE    &            .017 &            .032 &            .026 &            .028 &            .031 &            .019 &            .014 \\
CRV2 CI    &  $(.022, .088)$& $(-.074, .052)$& $(-.010, .093)$& $(-.033, .075)$&  $(.025, .146)$&  $(.028, .101)$&  $(.082, .138)$\\[0.3ex]
CRV-BM CI     &  $(.013, .096)$& $(-.094, .073)$& $(-.046, .129)$& $(-.063, .106)$& $(-.036, .207)$& $(-.043, .173)$& $(-.072, .293)$\\[0.2ex]
BME CI    & $(-.237, .344)$& $(-.334, .313)$&  $(-.217,.300)$& $(-.132, .175)$& $(-.107, .275)$& $(-.070, .202)$& $(-.156, .376)$\\
Polyn.\ order & 4 & 1 & 2 & 1 & 2 & 1 &2\\
Separate fit & No &Yes& Yes & Yes & Yes & Yes & Yes\\
Bandwidth $h$ & $\infty$ & $\infty$ & $\infty$ & $6$ & $6$ & $3$ & $3$\\
Eff.\ sample size & 73,954 & 73,954&  73,954&  20,883 & 20,883 & 10,533 & 10,533 \\
    \midrule
  \end{tabular*}
  \begin{tabular*}{\linewidth}{@{\extracolsep{\fill}}lcccc@{}}
    \multicolumn{5}{c}{\textbf{Panel B\@: BSD inference}} \\[0.5ex]
& (8) & (9) & (10) & (11)\\
\midrule
    $\SC$   & $.004$ & $.02$ & $.04$ & $.2$\\
    Estimate &  .037 & .065 & .079 & .079\\
    BSD CI & $(-.044, .118)$& $(-.060,.190)$ & $(-.081,.239)$ & $(-.269,.428)$ \\
    Polyn.\ order & 1 & 1 & 1 & 1\\
    Separate fit &  Yes &  Yes &Yes & Yes\\
    Implied bandwidth $h$ & $5$ & $3$ & $2$ & $2$\\
    Eff.\ sample size &17,240 & 10,533 &7,424 & 7,424 \\
    \bottomrule
  \end{tabular*}
  \vspace{1ex}

  {\small\emph{Note:} Estimates use data for Great Britain only. Specification
    (1) is from \citet{oreopoulos2006estimating}. Specifications (2) and (3) fit
    global linear and quadratic models on each side of the cutoff.
    Specifications (4)--(7) fit local linear and local quadratic
    models on each side of the cutoff, using bandwidths $h=6$ and $h=3$. For BSD
    CIs in columns (8)--(11), the implied bandwidth is the one that minimizes
    the length of the resulting CI for a given choice of $\SC$.}
\end{table}
\end{landscape}}

For Oreopoulos' original specification in column (1), the EHW standard error is
twice as large as \CRV\ standard error. The corresponding 95\% EHW CI covers
zero, whereas the 95\% \CRV\ CI does not. For the linear specification using the
full data in column (2), the point estimate is negative, and EHW standard errors
are slightly smaller than \CRV\ ones. Figure~\ref{fig_est} suggests that this
may be due to substantial misspecification of a global linear model below the
threshold.
For the remaining specifications in Table~\ref{tab:results4}, all EHW standard
errors are larger than the \CRV\ standard errors, by a factor between 1.6 and
28.9.
For both the linear and the quadratic specifications, the factor generally
increases as the bandwidth (and thus the number of support points used for
estimation) decreases. Moreover, the factor is larger for quadratic
specifications than it is for the linear specifications. This is in line with
the theoretical results presented in Section 4. While our analysis does not
imply that EHW CIs have correct coverage in this setting, it \emph{does} imply
that any CI with correct coverage must be at least as wide as the EHW CI\@. As a
result, since none of the EHW CIs in Table~\ref{tab:results4} exclude zero, for
the specifications in columns (1), (3) and (5)--(7), LC's approach leads to
incorrect claims about the statistical significance of the estimated treatment
effect.

Next, Table~\ref{tab:results4} reports results for a bias-reduced version of the
CRV estimator developed by \citet{bm02}, which we term CRV2, and results for CIs
that combine this bias-reduction with an adjusted critical value, also due to
\citet{bm02}, which we term CRV-BM.\footnote{We also considered the
  \citet{ImKo16} version of the critical value adjustment. The results, not
  reported here, are virtually identical to the CRV-BM results.} See the
supplemental material for formal definitions, and
Remark~\ref{remark:few-clusters-robust} in Section~\ref{sec:dist-crv-conf} for a
discussion of the performance of these CIs in simulations. While these
adjustments lead to larger CIs relative to CRV, CRV-BM CIs are still smaller
than EHW CIs for specifications (1) and (7), and the effect in Oreopoulos'
original specification (1) is still significant. CRV2 standard errors remain
smaller than EHW standard errors in all specifications except (2).

Finally, Table~\ref{tab:results4} reports the values of the honest CIs proposed
in the Section~\ref{sec:honest-conf-interv}. It shows that the BME procedure
leads to CIs that are wider than the EHW CIs. The difference is most pronounced
for Oreopoulos' original specification in column (1) and the specifications that
use the full data in columns (2) and (3), due a large number of support points
within the estimation window. For the estimation windows $h=6$ and $h=3$, the
BME CI is only slightly wider than the EHW CIs.

We also report BSD CIs for $\SC\in\{0.004,0.02,0.04,0.2\}$. We use the heuristic
described in Section~\ref{sec:bound-second-deriv} that fixing a value of $\SC$
corresponds to assuming that the true CEF does not deviate from a straight line
by more than $\SC/8$ over a one-year interval. Given that a typical increase in
log earnings per extra year in age is about 0.02, we consider $\SC=0.04$ and
$\SC=0.02$ to be reasonable choices, with the other values corresponding to a
very optimistic and a very conservative view, respectively. We check these
choices by estimating the lower bound for $\SC$ using the method described in
the supplemental material, which yields a point estimate equal to $0.012$, with
a 95\% CI given by $\hor{0.0013,\infty}$. Except for $\SC=0.2$, for which the
implied maximum bias is as large as $0.24$ (reflecting that this choice of $\SC$
is very conservative), the resulting CIs, reported in columns (8)--(11) of
Table~\ref{tab:results4}, are reasonably tight, and in line with the results
based on EHW and BME CIs.

\subsection{\texorpdfstring{\citet{lalive08}}{Lalive
    (2008)}}\label{sec:lalive-2008}

\citet{lalive08} exploits a sharp discontinuity in Austria's Regional Extended
Benefit Program that was in place in 1989--1991 to study how changes in
unemployment benefits affect the duration of unemployment. To mitigate labor
market problems in the steel sector that was undergoing restructuring, the
program extended the potential duration of unemployment benefits from 30 weeks
to 209 weeks for job seekers aged 50 or older, living in one of 28 labor market
districts of Austria with high steel industry concentrations. The sharp age and
location cutoffs in eligibility for this program allow \citet{lalive08} to
exploit two separate regression discontinuity designs.

\begin{figure}[!t]
\begin{center}
    \includegraphics[width=.9\textwidth]{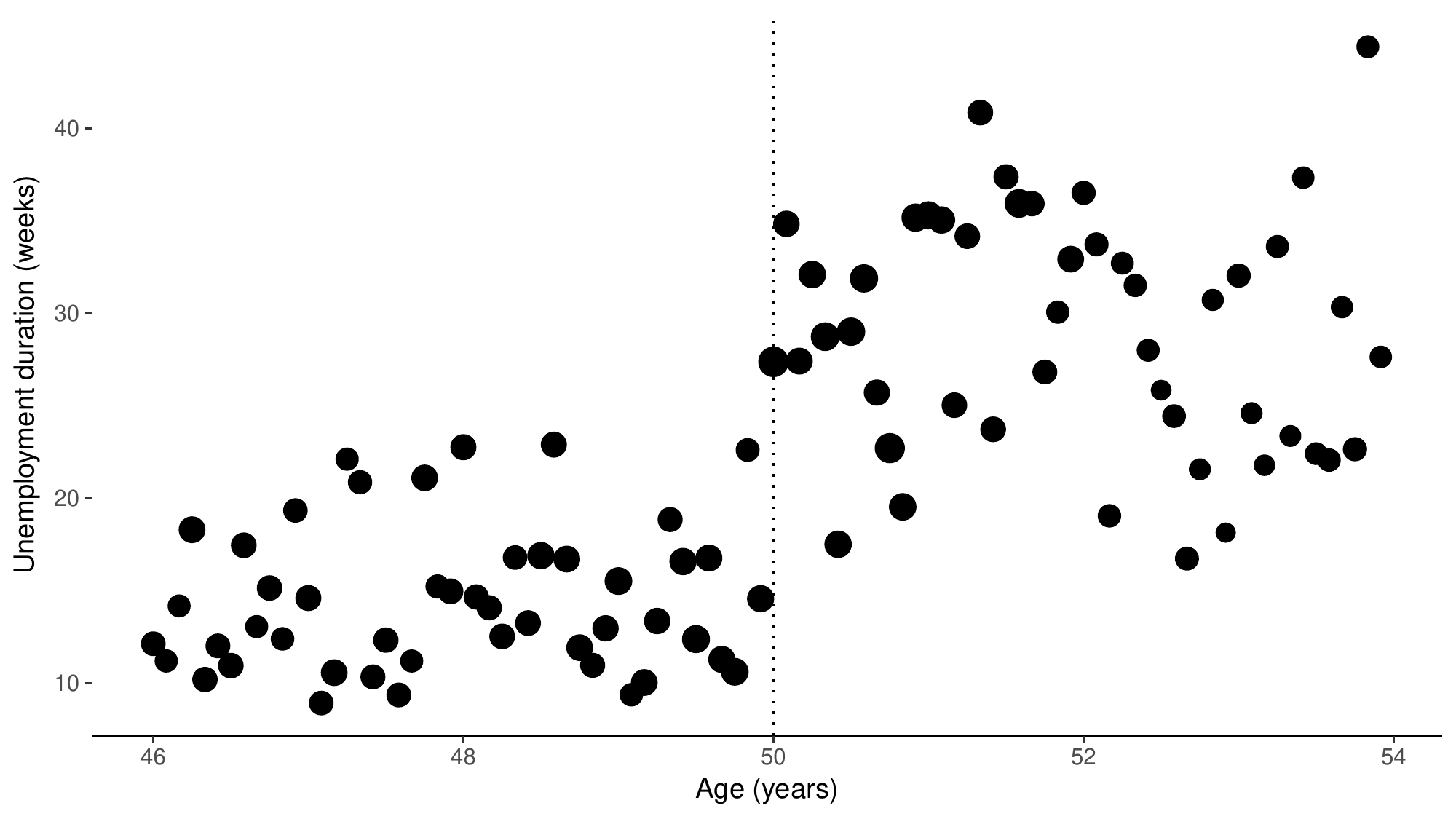}
    \caption{Average unemployment duration by age for men living in the treated
      districts. Vertical line indicates age 50, the cutoff for Regional
      Extended Benefit Program eligibility. Volume of dots is proportional to
      share of workers in the full data with the corresponding
      age.\label{fig_lalive}}
\end{center}
\end{figure}

We focus here on the design with age in months as the running variable, with 50
as the treatment threshold, and focus on the effect on unemployment duration for
men aged 46--54, previously employed in non-steel sectors, living in the treated
districts.\footnote{ As discussed in \citet{lalive08}, there may be a concern
  about the validity of this design if firms wait to lay off employees until
  they just satisfy the age requirement. If such manipulation was happening in
  the data, it may call into question the causal interpretation of the RD
  estimand. This concern about validity is secondary here, as our focus is on
  whether CRV CIs provide adequate coverage of the RDD estimand irrespective of
  whether it has causal interpretation. See \citet{gerard2016bounds} for bounds
  on causal parameters in RDDs with manipulation.} Because the data used in
\citet{lalive08} record age only up to a precision of one month, there are
$G=96$ support points, 48 above, and 48 below the cutoff.
Figure~\ref{fig_lalive} plots the data. \citet{lalive08} uses the global
polynomial regression
\begin{multline}
  \text{DURATION}_i = \beta_0 + \tau \cdot \1{\text{AGE}_i \geq 50} +
  \sum_{j=1}^p \beta_{j}^- \cdot (\text{AGE}_i-50)^p   \\
  + \sum_{j=1}^p \beta_{j}^+ \cdot \1{\text{AGE}_i \geq 50}\cdot
  (\text{AGE}_i-50)^p + U_i,
\end{multline}
with $p=0,1$ or $3$, as well as a local linear regression with bandwidth $h=2$
and the Epanechnikov kernel, to estimate the parameter of interest.
\citeauthor{lalive08} also reports CRV standard errors. Table~\ref{tab:lalive}
reproduces these estimates and standard errors, and also reports conventional
EHW standard errors for these specifications (columns (1)--(4)), except in
column (4) we use the uniform rather than the Epanechnikov kernel for
simplicity. Columns (5)--(7) report estimates for local linear and local cubic
regressions with bandwidth $h=2$ and $h=1$. Like in
Section~\ref{sec:oreopoulos-2006}, we also report results for CRV-BM, and CRV2
that correspond to the bias-reduction modification of the CRV estimator
developed by \citet{bm02}, with and without an additional critical value
adjustment.

The point estimate is relatively insensitive to changes of the specification,
indicating that men stay unemployed for about 13 additional weeks in response to
receiving up to 179 extra weeks of unemployment benefits. For all
specifications, the EHW standard errors are bigger than CRV standard errors, by
a factor of 1.1 to 2.0
that is increasing in the order of the polynomial and decreasing in the
bandwidth. This is in spite of the moderate number of support points on either
side of the threshold. Although the results remain significant, the CRV standard
errors overestimate the precision of the results. The CRV2 adjustment does not
change this conclusion. Similarly, the CRV-BM CIs are shorter than EHW CIs in
all specifications considered except (5) and (7).

Table~\ref{tab:lalive} also reports honest CIs proposed in
Section~\ref{sec:honest-conf-interv}. BME CIs turn out to be very wide for all
specifications, and are essentially uninformative. As discussed after
Proposition~\ref{th:bounds-missp-errors}, this is because there is substantial
uncertainty here about the magnitudes of the specification errors (only few
observations are available for each support points, and the number of support
points is relatively large). This is an instance of a setup where BME CIs are
very conservative. For BSD CIs, using the heuristic from
Section~\ref{sec:bound-second-deriv} that the CEF should not deviate by more than
$SC/8$ over a one-year interval, we set $\SC=8$ and $\SC=16$ as reasonable
choices. We also consider $\SC=1$ and $\SC=32$ as a robustness check. The
specification test described in the supplemental material cannot reject linear
CEF ($\SC=0$). Again, this is due to a relatively small number of observations
available for each support point. The resulting CIs for these choices of $\SC$,
reported in columns (8)-(11) of Table~\ref{tab:lalive}, are much tighter than
those based on the BME method, and quite close to the EHW CIs based local linear
regression and bandwidth $h=1$.

\afterpage{
\begin{landscape}
\begin{table}  \caption{Estimated effect of extended unemployment benefits on
    duration of unemployment.}\label{tab:lalive}
  \begin{tabular*}{\linewidth}{@{\extracolsep{\fill}}lccccccc@{}}
    \toprule
    \multicolumn{8}{c}{\textbf{Panel A\@: Point estimates and EHW/CRV/BME inference}} \\[0.5ex]
    & (1) & (2) & (3) & (4) & (5) & (6) & (7)  \\\midrule
Estimate & 14.6         & 14.8          & 11.2          & 13.4          & 14.5          & 12.5          & 12.2          \\[0.2ex]
EHW SE   & 1.19         & 2.23          & 4.57          & 3.13          & 6.54          & 4.45          & 8.89           \\
EHW CI   & (12.3, 16.9) & (10.4, 19.2)  & (2.2, 20.1)   & (7.2, 19.5)   & (1.7, 27.4)   & (3.8, 21.2)   & $(-5.2, 29.6)$\\[0.2ex]
CRV SE   & 1.07         & 1.93          & 3.47          & 2.45          & 4.46          & 3.29          & 4.35           \\
CRV CI   & (12.5, 16.7) & (11, 18.6)    & (4.3, 18)     & (8.6, 18.2)   & (5.8, 23.3)   & (6, 18.9)     & (3.7, 20.7)   \\[0.2ex]
CRV2 SE  & 1.08         & 1.97          & 3.75          & 2.55          & 5.36          & 3.56          & 6.25           \\
CRV2 CI  & (12.5, 16.7) & (10.9, 18.7)  & (3.8, 18.5)   & (8.4, 18.4)   & (4, 25)       & (5.5, 19.5)   & $(-0.1, 24.5)$\\[0.2ex]
CRV-BM CI& (12.5, 16.7) & (10.8, 18.8)  & (2.8, 19.5)   & (8, 18.7)     & (0.2, 28.9)  &(4.4, 20.6)   &$(-10.5, 34.9)$\\[0.2ex]
BME CI   &$(-31.6, 60.7)$&$(-31.8, 60.7)$&$(-34.7, 56.3)$&$(-27.2, 52.7)$ &$(-25.8, 54.1)$ &$(-23.1, 46.7)$&$(-28.5, 50.2)$ \\
    Polyn.\ order & 0 & 1 & 3 & 1 & 3 & 1 &3\\
    Separate fit & Yes &Yes& Yes & Yes & Yes & Yes & Yes\\
    Bandwidth $h$ & $\infty$ & $\infty$ & $\infty$ & 2 & 2 & 1 & 1\\
Eff.\ sample size &9,734              &9,734              &9,734              &5,582              &5,582              &3,030              &3,030\\
    \bottomrule
  \end{tabular*}
  \begin{tabular*}{\linewidth}{@{\extracolsep{\fill}}lcccc@{}}
    \multicolumn{5}{c}{\textbf{Panel B\@: BSD inference}} \\[0.5ex]
& (8) & (9) & (10) & (11)\\
\midrule
    $\SC$   & $1$ & $8$ & $16$ & $32$\\
    Estimate &  15.4 &12.1  & 12.9 & 15.5 \\
    BSD CI & (9.7, 21.0)& (3.3,20.8) & (2.9,22.9) & (3.5,27.5) \\\midrule
    Polyn.\ order & 1 & 1 & 1 & 1\\
    Separate fit &  Yes &  Yes &Yes & Yes\\
    Implied bandwidth $h$ & $2.83$ & $1.17$ & $0.92$ & $0.67$\\
    Eff.\ sample size &8,723 & 6,875  & 6,521 &  6,115\\
    \bottomrule
  \end{tabular*}
  \vspace{1ex}

  {\small\emph{Note:} Specification (1)--(3) are from \citet{lalive08}, and they
    fit global linear and cubic models on each side of the cutoff.
    Specifications (4)--(7) fit local linear and local cubic models on each
    side of the cutoff, using bandwidths $h=2$ and $h=1$. For BSD CIs in columns
    (8)--(11), the implied bandwidth is the one that minimizes the length of the
    resulting CI for a given choice of $\SC$.}
\end{table}
\end{landscape}
}

\section{Conclusions}\label{sec:conclusions}

RDDs with a discrete running variable are ubiquitous in empirical practice. In
this paper, we show that the commonly used confidence intervals based on
standard errors that are clustered by the running variable have poor coverage
properties. We therefore recommend that they should not be used in practice, and
that one should instead proceed as follows. First, there is no need to
distinguish sharply between the case of a discrete and a continuous running
variable. In particular, if the discrete running variable has rich support, one
can make the bandwidth smaller to reduce the bias of the treatment effect
estimate, and use EHW standard errors for inference. Discreteness of the running
variable only causes problems if the number of support points close to the
cutoff is so small that using a smaller bandwidth to make the bias of the
estimator negligible relative to its standard deviation is not feasible. Second,
if one wants to deal with bias issues explicitly, one can  use one of the two
honest CIs described in this paper. The BSD approach works whether the running
variable is discrete or continuous, whereas the BME approach is tailored to
settings in which the running variable only takes on a few distinct values.

\appendix
\allowdisplaybreaks

\section{Proofs of results in
  Section~\ref{sec:asymptotic-results}}\label{sec:proofs-results-sect}

The claims in Section~\ref{sec:asymptotic-results} follow directly from general
results on the properties of $\CRVse^{2}$ that are given in the following
subsection. The proofs of these results are given in turn in Sections A.2--A.4.
To state these results, we use the notation $\diag\{a_g\}$ to denote a diagonal
matrix with diagonal elements given by $a_1,\ldots,a_G$, and
$\vvec\{a_g\} =(a_1',\ldots,a_G')'$.

\subsection{Properties of \texorpdfstring{$\CRVse^{2}$}{CRV variance} under
  General Conditions}

In this subsection, we consider a setup that is slightly more general than that
in Section 3, in that it also allows the bandwidth $h$ to change with the sample
size. For convenience, the following assumption summarizes this more general
setup.
\begin{assumption}[Model]\label{assumption:model}
  For each $N$, the data $\{Y_{i},X_{i}\}_{i=1}^{N}$ are i.i.d., distributed
  according to a law $P_{N}$. Under $P_{N}$, the marginal distribution of
  $X_{i}$ is discrete with $G=G_{-}+G_{+}$ support points denoted
  $x_{1}<\dotsb<x_{G^{-}}<0\leq x_{G^{-}+1}<\dotsb <x_{G}$. Let
  $\mu(x)=\E_{N}(Y_{i}\mid X_{i}=x)$ denote the CEF under
  $P_{N}$. Let $\varepsilon_{i}=Y_{i}-\mu(X_{i})$, and let
  $\sigma^{2}_{g}=\Var_{N}(\varepsilon_{i}\mid X_{i}=x_{g})$ denote its
  conditional variance. Let $h=h_{N}$ denote a non-random bandwidth sequence,
  and let $\mathcal{G}_{h}\subseteq\{1,\dotsc,G\}$ denote the indices for which
  $\abs{x_{g}}\leq h$, with $G_{h}^{+}$ and $G_{h}^{-}$ denoting the number of
  elements in $\mathcal{G}_{h}$ above and below zero. Let
  $\pi_{g}=P_{N}(X_{i}=x_{g})$, $\pi=P_{N}(\abs{X_{i}}\leq h)$, and
  $N_{h}=\sum_{i=1}^{N}\1{\abs{X_{i}}\leq h}$. For a fixed integer $p\geq 0$,
  define
  \begin{equation*}
    m(x)
    = (\1{x\geq 0},1,x,\ldots,x^p,\1{x\geq 0}x,\ldots,\1{x\geq 0}x^p)',
  \end{equation*}
  $M_{i}=\1{\abs{X_{i}}\leq h}m(X_{i})$, and
  $m_{g}=\1{\abs{x_{g}}\leq h}m(x_{g})$. Let
  $\widehat{Q}=N_{h}^{-1}\sum_{i=1}^{n}M_{i}M_{i}'$ and
  $Q_{N}=\E_{N}(M_{i}M_{i}')/\pi$. Let
  $\thetah=Q_{N}^{-1}\E_{N}(M_{i}Y_{i})/\pi$, and denote its first element my
  $\tauh$. Let
  $\widehat{\theta}=\widehat{Q}^{-1}\frac{1}{N_{h}}\sum_{i=1}^{N}M_{i}Y_{i}$,
  and denote its first element by $\widehat\tau$. Define
  $\delta(x)=\mu(x)-m(x)'\thetah$, and
  $u_{i}=Y_{i}-m(X_{i})'\thetah=\delta(X_{i})+\epsilon_{i}$. Define
  $\Omega=\E_{N}(u_{i}^{2}M_{i}M_{i}')/\pi
  =\sum_{g=1}^{G}(\sigma^{2}_{g}+\delta^{2}(x_{g}))Q_{g}$, where
  $Q_{g}=\frac{\pi_{g}}{\pi}m_{g}m_{g}'$.
\end{assumption}
Note that the setup allows various quantities that depend on $P_{N}$ and $h$ to
change with $N$, such as the number of support points $G$, their locations
$x_{g}$, the conditional expectation function $\mu(x)$, or the specification
errors $\delta(X_{i})$.

\begin{assumption}[Regularity conditions]\label{assumption:regularity}
  (i)
  $\sup_{N}\max_{g\in\{1,\dotsc,G\}}\E_{N}(\epsilon_{i}^{4}\mid
  X_{i}=x_{g})<\infty$,
  $\det(H^{-1}Q_{N}H^{-1})=\det(\sum_{g\in\mathcal{G}_{h}}
  \frac{\pi_{g}}{\pi}m(x_{g}/h)m(x_{g}/h)')>C$ for some $C>0$ that does not
  depend on $N$, where $H=\diag\{m(h)\}$, $N\pi\to\infty$, and the limit
  $\lim_{N\to\infty}H^{-1}Q_{N}H^{-1}$ exists. (ii)
  $\sup_{N}\max_{g\in\mathcal{G}_{h}}\delta(x_{g})<\infty$; and the limit
  $\lim_{N\to\infty}H^{-1}\Omega H^{-1}$ exists.
\end{assumption}
The assumption ensures that that bandwidth shrinks to zero slowly enough so that
the number of effective observations $N\pi$ increases to infinity, and that the
number of effective support points ${G}_{h}=G_{h}^{+}+G_{h}^{-}$ is large enough
so that the parameter $\theta_{h}$ and the asymptotic variance of
$\widehat{\theta}$ remain well-defined with well-defined limits. We normalize
$Q_{N}$ and $\Omega$ by the inverse of $H$ since if $h\to 0$, their
elements converge at different rates.

Our first result is an asymptotic approximation in which $G_h^+$ and $G_h^-$ are
fixed as the sample size increases. Let $B_1,\dotsc ,B_G$ be a collection of
random vectors such that $\vvec\{B_g\} \sim \mathcal{N}(0,V)$, with
\begin{equation*}
  V=\frac{1}{\pi}\diag\{\pi_{g}(\sigma^{2}_{g}+\delta(x_{g})^{2})\}
  -\frac{1}{\pi}\vvec\{\pi_{g}\delta(x_{g})\} \vvec\{\pi_{g}\delta(x_{g})\}'.
\end{equation*}
Note that if $\abs{x_{g}}>h$, then $B_{g}=0$ and $Q_{g}=0$, and that the
limiting distribution of the statistic $\sqrt{N_{h}}(\widehat\tau-\tauh)$
coincides with the distribution of $e_{1}'Q_{N}^{-1}\sum_{g=1}^{G}m_{g}B_{g}$.
Finally, define
\begin{equation*}
  W_{g}=e_{1}'Q_{N}^{-1}m_{g}\left( B_{g} -
    \frac{\pi_{g}}{\pi}m_{g}'Q_{N}^{-1}\sum_{j=1}^{G}m_{j}B_{j}
    +(N/\pi)^{1/2}\pi_{g} \delta(x_{g}) \right).
\end{equation*}
With this notation, we obtain the following generic result.
\begin{theorem}\label{th1}
  Suppose that Assumptions~\ref{assumption:model}
  and~\ref{assumption:regularity} hold. Suppose also that, as $N\to \infty$, (i)
  $G_{h}^{+}$ and $G_{h}^{-}$ are fixed; and (ii) the limit of $V$ exists. Then
  \begin{align*}
    \CRVse^{2}&\stackrel{d}{=} (1+o_{P_{N}}(1))\sum_{g=1}^{G} W_{g}^{2}.
  \end{align*}
\end{theorem}

Our second result is an asymptotic approximation in which the number of support
points of the running variable (or, equivalently, the number of ``clusters'')
that are less than $h$ away from the threshold increases with the sample size.

\begin{theorem}\label{theorem:g-increasing}
  Suppose that Assumptions~\ref{assumption:model}
  and~\ref{assumption:regularity} hold. Suppose also that, as $N\to\infty$,
  $G_{h}\to\infty$ and $\max_{g\in\mathcal{G}_{h}}\pi_{g}/\pi\to 0$. Then
  \begin{equation*}
    \CRVse^{2}=    (1+o_{P_{N}}(1))
    e_{1}'Q_{N}^{-1}\left(\Omega +
      (N-1)\sum_{g=1}^{G}Q_{g}\cdot \pi_{g}\delta(x_{g})^{2}.
    \right)Q_{N}^{-1}e_{1}.
  \end{equation*}
\end{theorem}
The assumption that $\max_{g\in\mathcal{G}_{h}}\pi_{g}/\pi\to 0$ ensures that
each ``cluster'' comprises a vanishing fraction of the effective sample size.

\subsection{Auxiliary Lemma}

Here we state an intermediate result that is used
in the proofs of Theorem~1 and~2 below, and that shows that
$\EHWse^{2}$ is consistent for the asymptotic variance of $\widehat{\theta}$.

\begin{lemma}\label{lemma:auxilliary-lemma}
  Suppose that Assumptions~\ref{assumption:model}
  and~\ref{assumption:regularity}~(i) hold. Then
  \begin{align}
    \frac{N_{h}/N}{\pi}&= 1+o_{P_{N}}(1),\label{eq:Nh}\\
    H^{-1}\widehat{Q}H^{-1}-H^{-1}{Q}_{N}H^{-1}&=o_{P_{N}}(1).\label{eq:hatQ}
  \end{align}
  If, in addition, Assumption~\ref{assumption:regularity}~(ii) holds, then
  \begin{equation}\label{eq:theta-normal}
    \sqrt{N_{h}}    H(\widehat{\theta}-\thetah)\stackrel{d}{=}
    HQ_{N}^{-1}S+o_{P_{N}}(1),
  \end{equation}
  where $S\sim \mathcal{N}(0,\Omega)$. Let
  $n_{g}=\sum_{i=1}^{N}\1{X_{i}=x_{g}}$,
  $\widehat{q}_{g}=H\widehat{Q}^{-1}Hm(x_{g}/h)\1{\abs{x_{g}}\leq h}$, and
  $A_{g}=\frac{\1{\abs{x_{g}}\leq
      h}}{\sqrt{N\pi}}\sum_{i=1}^{N}\left(\1{X_{i}=x_{g}}\varepsilon_{i}
    +(\1{X_{i}=x_{g}}-\pi_{g})\delta(x_{g})\right)$. Then
  $H^{-1}\sum_{g=1}^{G}m_{g}A_{g}\stackrel{d}{=}H^{-1}S+o_{P_{N}}(1)$, and
  \begin{equation}\label{eq:crv-se}
    \CRVse^{2}=(1+o_{P_{N}}(1))\sum_{g=1}^{G}(e_{1}'\widehat{q}_{g})^{2}
    \left( A_{g}
      - \frac{n_{g}}{N_{h}} \widehat{q}_{g}'\sum_{j=1}^{G}
      m(x_{j}/h) A_{j}
      + \frac{\sqrt{N}\pi_{g}\delta(x_{g})}{\sqrt{\pi}}
    \right)^{2}.
  \end{equation}
  Furthermore, $\EHWse^{2}=e_{1}'Q_{N}^{-1}\Omega Q_{N}^{-1}e_{1}+o_{P_{N}}(1)$.
\end{lemma}
\begin{proof}
  We have $\Var_{N}(N_{h}/N)=\pi(1-\pi)/N\leq \pi/N$. Therefore, by Markov's
  inequality, $N\pi\to\infty$ implies
  $\frac {N_{h}/N}{\pi}=\E_{N}(N_{h}/(N\pi))+o_{P_{N}}(1)=1+o_{P_{N}}(1)$, which
  proves~\eqref{eq:Nh}. Secondly, since elements of $H^{-1}M_{i}$ are bounded by
  $\1{\abs{X_{i}}\leq h}$, the second moment of any element of
  $\frac{N_{h}}{N\pi}H^{-1}\widehat{Q}H^{-1}-H^{-1}Q_{N}H^{-1}
  =\frac{1}{N\pi}\sum_{i=1}^{n}H^{-1}(M_{i}M_{i}'-E[M_{i}M_{i}'])H^{-1}$ is
  bounded by $1/(N\pi)$, which converges to zero by assumption. Thus, by Markov's
  inequality,
  $\frac{N_{h}}{N\pi}H^{-1}\widehat{Q}H^{-1}-H^{-1}Q_{N}H^{-1}=o_{P_{N}}(1)$.
  Combining this result with~\eqref{eq:Nh} and the fact that $H^{-1}Q_{N}H^{-1}$
  is bounded then yields~\eqref{eq:hatQ}.

  Next note that since $\sum_{g=1}^{N}\pi_{g}m_{g}\delta(x_{g})=0$,
  $H^{-1}
  \sum_{g=1}^{G}m_{g}A_{g}=\frac{1}{\sqrt{N}}\sum_{i=1}^{N}\frac{1}{\sqrt{\pi}}H^{-1}M_{i}u_{i}$,
  and that by the central limit theorem,
  $\frac{1}{\sqrt{N}}\sum_{i=1}^{N}\frac{1}{\sqrt{\pi}}H^{-1}M_{i}u_{i}
  \stackrel{d}{=} H^{-1}S+o_{P_{N}}(1)$. Therefore,
  \begin{equation*}
    \sqrt{N_{h}}    H(\widehat{\theta}-\thetah)=\sqrt{\frac{\pi N}{N_{h}}}
    (H^{-1}\widehat{Q}H^{-1})^{-1}\frac{1}{\sqrt{N}}
    \sum_{i=1}^{N}\frac{1}{\sqrt{\pi}}H^{-1}M_{i}u_{i}\stackrel{d}{=}
    HQ_{N}^{-1}S+o_{P_{N}}(1),
  \end{equation*}
  as claimed. Next, we prove~\eqref{eq:crv-se}. Let
  $J_{g}= \sum_{i=1}^{N}\1{X_{i}=x_{g}}\widehat{u}_{i}M_{i}$. Then
  by~\eqref{eq:Nh}, the cluster-robust variance estimator can be written as
  \begin{equation*}
    \CRVse^{2} = e_1'\widehat{Q}^{-1}
    \frac{1}{N_{h}}\sum_{g=1}^{G} J_{g}J_{g}'\widehat{Q}^{-1}e_1=(1+o_{P_{N}}(1))
    \sum_{g=1}^G\left( \frac{1}{\sqrt{N\pi}}
      e_{1}'\widehat{Q}^{-1}J_{g}\right)^{2}.
  \end{equation*}
  The expression in parentheses can be decomposed as
  \begin{equation*}
  \begin{split}
    \frac{1}{\sqrt{N\pi}} e_{1}'\widehat{Q}^{-1}J_{g} &=
    e_{1}'\widehat{q}_{g}\left( A_{g} +N \pi_{g}\delta(x_{g})/\sqrt{N\pi}
      -n_{g}m_{g}'(\widehat{\theta}-\thetah)/\sqrt{N\pi}\right)\\
    &=e_{1}'\widehat{q}_{g}\left(A_{g} +N \pi_{g}\delta(x_{g})/\sqrt{N\pi}
      -(n_{g}/N_{h}) \widehat{q}_{g}' H^{-1}\sum_{j=1}^{G}m_{j}A_{j}\right),
  \end{split}
\end{equation*}
which yields the result.

It remains to prove consistency of $\EHWse^{2}$. To this end,
using~\eqref{eq:Nh}, decompose
\begin{equation*}
  H^{-1}\widehat\Omega_{\textnormal{EHW}}H^{-1}
  =(1+o_{P_{N}}(1))\frac{1}{N\pi}\sum_{i=1}^{N}\widehat{u}_{i}^{2}H^{-1}M_{i}
  M_{i}'H^{-1}=(1+o_{P_{N}}(1))(\mathcal{C}_{1}+\mathcal{C}_{2}+\mathcal{C}_{3}),
\end{equation*}
where
$\mathcal{C}_{1}= \frac{1}{N\pi} \sum_{i=1}^{N}{u}_{i}^{2}H^{-1}M_{i}
M_{i}'H^{-1}$,
$\mathcal{C}_{2}= \frac{1}{N\pi}
\sum_{i=1}^{N}(M_{i}'(\widehat{\theta}-\thetah))^{2}H^{-1}M_{i} M_{i}'H^{-1}$,
and
$\mathcal{C}_{3}=-\frac{2}{N\pi}
\sum_{i=1}^{N}u_{i}M_{i}'(\widehat{\theta}-\thetah)H^{-1}M_{i} M_{i}'H^{-1}$.
Since elements of $M_{i}$ are bounded by $\1{\abs{X_{i}}\leq h}$, variance of
$\mathcal{C}_{1}$ is bounded by
$\E[u_{i}^{4}\1{\abs{X}_{i}\leq h}]/N\pi^{2}=o_{P_{N}}(1)$, so that by Markov's
inequality,
$\mathcal{C}_{1}=E[\mathcal{C}_{1}]+o_{P_{N}}(1)= H^{-1}\Omega
H^{-1}+o_{P_{N}}(1)$. Similarly, all elements of $\mathcal{C}_{2}$ are bounded,
$\frac{1}{N\pi}
\sum_{i=1}^{N}(M_{i}'(\widehat{\theta}-\thetah))^{2}=o_{P_{N}}(1)$
by~\eqref{eq:Nh}--\eqref{eq:theta-normal}. Finally,
$\mathcal{C}_{3}=o_{P_{N}}(1)$ by Cauchy-Schwarz inequality. Thus,
\begin{equation}\label{eq:Omega-EHW-consistency}
H^{-1}\widehat\Omega_{\textnormal{EHW}}H^{-1}=H^{-1}\Omega H^{-1}+o_{P_N}(1),
\end{equation}
and consistency of $\EHWse^{2}$ then follows by combining this result
with~\eqref{eq:hatQ}.
\end{proof}

\subsection{Proof of Theorem~\ref{th1}}
Let ${q}_{g}=HQ_{N}^{-1}Hm(x_{g}/h)\1{\abs{x_{g}}\leq h}$, and define
$\widehat{q}_{g}$, $A_{g}$, and $n_{g}$ as in the statement of
Lemma~\ref{lemma:auxilliary-lemma}. By Lemma~\ref{lemma:auxilliary-lemma},
$\widehat{q}_{g}=q_{g}(1+o_{P_{N}}(1))$, and by Markov's inequality and
Equation~\eqref{eq:Nh}, $n_{g}/N_{h}=\pi_{g}/\pi+o_{P_{N}}(1)$ for
$g\in\mathcal{G}_{h}$. Combining these results with Equation~\eqref{eq:crv-se},
it follows that the cluster-robust variance estimator satisfies
\begin{equation*}
  \CRVse^{2}=(1+o_{P_{N}}(1))\sum_{g=1}^{G}
  (e_{1}'{q}_{g})^{2}\left(A_{g} + {\sqrt{N\pi}}
    \frac{\pi_{g}}{\pi}\delta(x_{g})- \frac{\pi_{g}}{\pi}
    {q}_{g}'\sum_{j=1}^{G}
    m(x_{j}/h) A_{j}\right)^{2},
\end{equation*}
To prove the theorem, it therefore suffices to show that
\begin{equation}\label{eq:rtp-theorem1}
  e_{1}'{q}_{g}\left(A_{g}- \frac{\pi_{g}}{\pi}
    {q}_{g}'\sum_{j=1}^{G}
    m(x_{j}/h) A_{j} + {\sqrt{N\pi}}
    \frac{\pi_{g}}{\pi}\delta(x_{g})\right)=W_{g}(1+o_{P_{N}}(1)).
\end{equation}
This follows from Slutsky's lemma and the fact that by the central limit
theorem,
\begin{align}
  \vvec\{A_{g}\}&\stackrel{d}{=} \vvec\{B_{g}\}(1+o_{P_{N}}(1)).\label{eq:rtp-3}
\end{align}

\subsection{Proof of Theorem~\ref{theorem:g-increasing}}
Throughout the proof, write $a\preceq b$ to denote $a<Cb$ for some constant $C$
that does not depend on $N$. By Equation~\eqref{eq:crv-se} in
Lemma~\ref{lemma:auxilliary-lemma}, we can write the cluster-robust estimator as
$\CRVse^{2}=(\mathcal{C}_{1}+\mathcal{C}_{2}+\mathcal{C}_{3})(1+o_{P_{N}}(1))$, with
\begin{align*}
  \mathcal{C}_{1}&=\sum_{g=1}^{G}(e_{1}'\widehat{q}_{g})^{2}\left( A_{g} +
                   (N/\pi)^{1/2} \pi_{g}\delta(x_{g})\right)^{2},\\
  \mathcal{C}_{2}&=\widehat{S}'H^{-1}\sum_{g=1}^{G}(\widehat{q}_{g}'e_{1})^{2}
                   \frac{n_{g}^{2}}{N_{h}^{2}}
                   \widehat{q}_{g}\widehat{q}_{g}'\cdot H^{-1}\widehat{S},\\
  \mathcal{C}_{3}&=-2\sum_{g=1}^{G} (e_{1}'\widehat{q}_{g})^{2}\left( A_{g} +
                   (N/\pi)^{1/2} \pi_{g}\delta(x_{g})\right)
                   \frac{n_{g}}{N_{h}}\widehat{q}_{g}' H^{-1}\widehat{S},
\end{align*}
where $\widehat{S}=\sum_{j=1}^{G}m_{j}A_{j}$, and $n_{g},A_{g}$ and
$\widehat{q}_{g}$ are defined in the statement of the Lemma.

We first show that $\mathcal{C}_{2}=o_{P_{N}}(1)$. Since
$H^{-1}\widehat{S}=O_{P_{N}}(1)$ by Lemma~\ref{lemma:auxilliary-lemma}, it
suffices to show that
\begin{equation*}
  \sum_{g=1}^{G}(\widehat{q}_{g}'e_{1})^{2} \frac{n_{g}^{2}}{N_{h}^{2}}
  \widehat{q}_{g}\widehat{q}_{g}'=o_{P_{N}}(1).
\end{equation*}
To this end, note that since elements of $m(x_{g}/h)$ are bounded by $1$, for
any $j$, by Cauchy-Schwarz inequality,
$\abs{\widehat{q}_{g}'e_{j}}\leq
\norm{e_{j}'H\widehat{Q}^{-1}H}_{2}\sqrt{2(p+1)}$,
where $\norm{v}_{2}$ denotes the Euclidean norm of a vector $v$. By
Lemma~\ref{lemma:auxilliary-lemma},
$\norm{e_{j}'H\widehat{Q}^{-1}H}_{2}=O_{P_{N}}(1)$ and
$N_{h}/\pi N=1+o_{P_{N}}(1)$ so that
\begin{equation*}
  \abs*{\sum_{g=1}^{G}(\widehat{q}_{g}'e_{1})^{2} \frac{n_{g}^{2}}{N_{h}^{2}}
    e_{j}\widehat{q}_{g}\widehat{q}_{g}'e_{k}}\leq
  O_{P_{N}}(1)\sum_{g\in\mathcal{G}_{h}} \frac{n_{g}^{2}}{N_{h}^{2}}
  =O_{P_{N}}(1)\sum_{g\in\mathcal{G}_{h}} \frac{n_{g}^{2}}{\pi^{2}N^{2}}.
\end{equation*}
Now, since $\E_{N}(n_{g}^{2})=N\pi_{g}(1-\pi_{g})+N^{2}\pi_{g}^{2}$, and
$\sum_{g\in\mathcal{G}_{h}}\pi_{g}=\pi$,
\begin{equation*}
  \E_{N}\sum_{g\in\mathcal{G}_{h}} \frac{n_{g}^{2}}{N^{2}\pi^{2}}
  =\sum_{g\in\mathcal{G}_{h}}\frac{\pi_{g}(1-\pi_{g})}{N\pi^{2}}+
  \sum_{g\in\mathcal{G}_{h}}\frac{\pi_{g}^{2}}{\pi^{2}}
  \leq \left(\frac{1}{N\pi}+\frac{\max_{g\in\mathcal{G}_{h}}\pi_{g}}{\pi}\right)
  \sum_{g\in\mathcal{G}_{h}}\frac{\pi_{g}}{\pi}\to 0.
\end{equation*}
Therefore, by Markov's inequality, $\sum_{g\in\mathcal{G}_{h}}
\frac{n_{g}^{2}}{\pi^{2}N^{2}}=o_{P_{N}}(1)$, so that
$\mathcal{C}_{2}=o_{P_{N}}(1)$ as claimed.

Next, consider $\mathcal{C}_{1}$. Let
${q}_{g}=HQ_{N}^{-1}Hm(x_{g}/h)\1{\abs{x_{g}}\leq h}$. We have
\begin{equation*}
  \begin{split}
    \mathcal{C}_{1}&
    =\frac{1}{N\pi}\sum_{i=1}^{N}\sum_{j=1}^{N}\sum_{g=1}^{G}(e_{1}'\widehat{q}_{g})^{2}
    \1{X_{i}=x_{g}}\1{X_{j}=x_{g}}(\varepsilon_{i}+\delta(x_{g}))
    (\varepsilon_{j}+\delta(x_{g}))\\
    &
    =(1+o_{P_{N}}(1))\frac{1}{N\pi}\sum_{i=1}^{N}\sum_{j=1}^{N}\sum_{g=1}^{G}(e_{1}'{q}_{g})^{2}
    \1{X_{i}=x_{g}}\1{X_{j}=x_{g}}(\varepsilon_{i}+\delta(x_{g}))
    (\varepsilon_{j}+\delta(x_{g}))\\
    &=(1+o_{P_{N}}(1))\left(\mathcal{C}_{11}+2(\mathcal{C}_{12}+\mathcal{C}_{13}
      +\mathcal{C}_{14}+\mathcal{C}_{15}+\mathcal{C}_{16})\right),
  \end{split}
\end{equation*}
where
\begin{align*}
  \mathcal{C}_{11}&=\frac{1}{N\pi}\sum_{i=1}^{N}\sum_{g=1}^{G}(e_{1}'{q}_{g})^{2}
  \1{X_{i}=x_{g}}(\varepsilon_{i}+\delta(x_{g}))^{2},\\
  \mathcal{C}_{12}&=\frac{1}{N\pi}
  \sum_{i=1}^{N}\sum_{j=1}^{i-1}\sum_{g=1}^{G}(e_{1}'{q}_{g})^{2}
  \1{X_{i}=x_{g}}\1{X_{j}=x_{g}}\varepsilon_{i}\varepsilon_{j},\\
  \mathcal{C}_{13}&=\frac{1}{N\pi}
  \sum_{i=1}^{N}\sum_{j=1}^{i-1}\sum_{g=1}^{G}(e_{1}'{q}_{g})^{2}
  \1{X_{i}=x_{g}}\1{X_{j}=x_{g}}
  \varepsilon_{j}\delta(x_{g}),\\
  \mathcal{C}_{14}&=\frac{1}{N\pi}
  \sum_{i=1}^{N}\sum_{j=1}^{i-1}\sum_{g=1}^{G}(e_{1}'{q}_{g})^{2}
  \1{X_{i}=x_{g}}\1{X_{j}=x_{g}}
  \varepsilon_{i}\delta(x_{g}),\\
  \mathcal{C}_{15}&=\frac{1}{N\pi}
  \sum_{i=1}^{N}\sum_{j=1}^{i-1}\sum_{g=1}^{G}(e_{1}'{q}_{g})^{2}
  \1{X_{i}=x_{g}}(\1{X_{j}=x_{g}}-\pi_{g})\delta(x_{g})^{2},\\
  \mathcal{C}_{16}&=\frac{1}{N\pi}
  \sum_{g=1}^{G}\sum_{i=1}^{N}(i-1)(e_{1}'{q}_{g})^{2}
  \1{X_{i}=x_{g}}\pi_{g}\delta(x_{g})^{2}.
\end{align*}
We have
\begin{equation*}
  \E_{N}(\mathcal{C}_{11})=\frac{1}{\pi}\sum_{g=1}^{G}(e_{1}'{q}_{g})^{2}
  \pi_{g}(\sigma^{2}_{g}+\delta(x_{g})^{2})=e_{1}'Q_{N}^{-1}\Omega Q_{N}^{-1}e_{1},
\end{equation*}
and
\begin{equation*}
  \begin{split}
    \Var_{N}(\mathcal{C}_{11})&\leq\frac{1}{N\pi^{2}}
    \sum_{g=1}^{G}(e_{1}'{q}_{g})^{4}
    \pi_{g}\E_{N}[(\varepsilon_{i}+\delta(x_{g}))^{4}\mid X_{i}=x_{g}]\preceq
    \frac{\sum_{g\in\mathcal{G}_{h}}\pi_{g}}{N\pi^{2}}=\frac{1}{N\pi}\to 0.
  \end{split}
\end{equation*}
Next, $\E_{N}(\mathcal{C}_{12})=0$, and
\begin{equation*}
  \Var_{N}(\mathcal{C}_{12})= \frac{N-1}{2N\pi^{2}}
  \sum_{g=1}^{G}(e_{1}'{q}_{g})^{4}
  \pi_{g}^{2}\sigma^{2}_{g}\sigma^{2}_{g}\preceq
  \frac{\max_{g}\pi_{g}\sum_{g=1}^{G}\pi_{g}}{\pi^{2}}=
  \frac{\max_{g}\pi_{g}}{\pi}\to 0.
\end{equation*}
The expectations for the remaining terms satisfy
$\E_{N}(\mathcal{C}_{13})=\E_{N}(\mathcal{C}_{14})=\E_{N}(\mathcal{C}_{15})=0$,
and
\begin{equation*}
  \E_{N}(\mathcal{C}_{16})=\frac{N-1}{2\pi}
  \sum_{g=1}^{G}(e_{1}'q_{g})^{2}\pi_{g}^{2}\delta(x_{g})^{2}.
\end{equation*}
The variances of $\mathcal{C}_{13},\dotsc,\mathcal{C}_{16}$ are all of smaller
order than this expectation:
\begin{align*}
  \Var_{N}(\mathcal{C}_{13})&=\frac{1}{N^{2}\pi^{2}}
  \sum_{g}^{G}\sum_{i,k=1}^{N}\sum_{j=1}^{\min\{i,k\}-1} (e_{1}'{q}_{g})^{4}
  \pi_{g}^{3}\sigma^{2}_{g}\delta(x_{g})^{2}\preceq
  \frac{N\max_{g}\pi_{g}}{\pi^{2}} \sum_{g}^{G}(e_{1}'{q}_{g})^{2}
  \pi_{g}^{2}\delta(x_{g})^{2}\\
  &=o(\E_{N}(\mathcal{C}_{16}))\\
  \Var_{N}(\mathcal{C}_{14})&=\frac{1}{N^{2}\pi^{2}}
  \sum_{i=1}^{N}\sum_{j=1}^{i-1}\sum_{k=1}^{i-1}
  \sum_{g=1}^{G}(e_{1}'{q}_{g})^{4} \pi_{g}^{3}\sigma^{2}_{g}\delta(x_{g})^{2}
  =o(\E_{N}(\mathcal{C}_{16}))\\
  \Var_{N}(\mathcal{C}_{15})&=\frac{1}{N^{2}\pi^{2}}
  \sum_{i=1}^{N}\sum_{k=1}^{N}\sum_{j=1}^{\min\{i,k\}-1} \sum_{g,f=1}^{G}
  \left(\1{g=f}\pi_{g}-\pi_{g}\pi_{f}\right) \pi_{g}\pi_{f}(e_{1}'{q}_{g})^{2}
  (e_{1}'{q}_{f})^{2} \delta(x_{f})^{2}\delta(x_{g})^{2}\\
  &\leq\frac{1}{N^{2}\pi^{2}}
  \sum_{i=1}^{N}\sum_{k=1}^{N}\sum_{j=1}^{\min\{i,k\}-1} \sum_{g=1}^{G}
  \pi_{g}^{3}(e_{1}'{q}_{g})^{4} \delta(x_{g})^{4}= o(\E_{N}(\mathcal{C}_{16})),
\end{align*}
and
\begin{equation*}
  \begin{split}
    \Var_{N}(\mathcal{C}_{16})& =\frac{1}{N^{2}\pi^{2}} \sum_{g=1}^{G}
    \sum_{f=1}^{G} \sum_{i=1}^{N}(i-1)^{2}
    (\1{g=f}\pi_{g}-\pi_{g}\pi_{f})\pi_{g}
    \pi_{f}\delta(x_{g})^{2} \delta(x_{f})^{2}(e_{1}'{q}_{g})^{2}(e_{1}'{q}_{f})^{2}\\
    &\leq \frac{N}{\pi^{2}} \sum_{g=1}^{G}
    \pi_{g}^{3}\delta(x_{g})^{4}(e_{1}'{q}_{g})^{4}=o(\E_{N}(\mathcal{C}_{16})).
  \end{split}
\end{equation*}
It therefore follows that
\begin{equation*}
  \mathcal{C}_{1}=(1+o_{P_{N}}(1))\E_{N}(\mathcal{C}_{1})=
  (1+o_{P_{N}}(1))\left(e_{1}'Q_{N}^{-1}\Omega Q_{N}^{-1}e_{1}+
    \frac{N-1}{\pi}\sum_{g=1}^{G}(e_{1}'q_{g})^{2}\pi_{g}^{2}\delta(x_{g})^{2}
  \right).
\end{equation*}
Finally, the cross-term $\mathcal{C}_{3}$ is
$o_{P_{N}}(\E_{N}(\mathcal{C}_{1})^{1/2})$ by Cauchy-Schwarz
inequality, so that $\CRVse^{2}=(1+o_{P_{N}}(1))\E_{N}(\mathcal{C}_{1})$, which
yields the result.

\section{Proofs of results in   Section~\ref{sec:honest-conf-interv}}\label{sec:honest-conf-interv-proofs}

For the proof of Propositions~\ref{theorem:holder-honest}, we suppose that
Assumptions~\ref{assumption:model} and~\ref{assumption:regularity}~(i) hold. We
denote the conditional variance of $\tau$ by
$\tilde{\sigma}^{2}_{\tau}=e_{1}'HQ_{N}^{-1}H\tilde{\Omega}HQ_{N}^{-1}He_{1}$,
where $\tilde{\Omega}=\E[\sigma^{2}(X_{i})\lambda(X_{i})]/\pi$, and
$\lambda(x)=m(x/h)\cdot m(x/h)'\1{\abs{x}\leq h}$. We assume that
$\tilde{\sigma}^{2}_{\tau}$ is bounded and bounded away from zero. To ensure
that $\widehat{\sigma}^{2}_{NN}$, as defined in
Section~\ref{sec:bound-second-deriv}, is consistent, we assume that as
$N\to\infty$, $P_{N}(\sum_{g=1}^{G_{h}}\1{n_{g}\leq 1})\to 0$, so that in large
samples there are at least two observations available for each support point. We
put $\widehat{\sigma}^{2}_{g}=0$ if $n_{g}\leq 1$. For simplicity, we also
assume that $h\to 0$; otherwise $r_{\sup}$ will diverge with the sample size.

For the proof of Proposition~\ref{th:bounds-missp-errors}, we suppose that
Assumptions~\ref{assumption:model} and~\ref{assumption:regularity} hold. For
simplicity, we also assume that as $N\to\infty$, $G_{h}^{+}$ and $G^{-}_{h}$ are
fixed, and that $\min_{g\in\mathcal{G}_{h}}\pi_{g}/\pi$ is bounded away from
zero. We also assume that the asymptotic variance of
$\widehat{\tau}+\widehat{b}(W)$ is bounded away from zero for some
$W\in\mathcal{W}$.

\subsection{Proof of Proposition~\ref{theorem:holder-honest}}

We first derive the expression for $r_{\sup}$, following the arguments in
Theorem B.1 in \citet{ArKo16honest}. Note first that the local linear estimator
$\widehat{\tau}$ can be written as a linear estimator,
$\widehat{\tau}=\sum_{i=1}^{N}w(X_{i})Y_{i}$, with the weights $w(x)$ given
in~\eqref{eq:holder-bias-expression}. Put $w_{+}(x)=w(x)\1{x\geq 0}$, and
$w_{-}(x)=w(x)\1{x<0}$, and put $\mu_{+}(x)=\mu(x)\1{x\geq 0}$ and
$\mu_{-}(x)=\mu(x)\1{x<0}$, with the convention that
$\mu_{-}(0)=\lim_{x\uparrow 0}\mu(0)$. Since
$\sum_{i=1}^{N}w_{+}(X_{i})=-\sum_{i=1}^{N}w_{-}(X_{i})=1$ and
$\tau=\mu_{+}(0)-\mu_{-}(0)$, the conditional bias has the form
\begin{equation*}
  \tilde{\tau}_{h}-\tau=
  \sum_{i}w_{+}(X_{i})(\mu_{+}(X_{i})-\mu_{+}(0))
  +\sum_{i}w_{-}(X_{i})(\mu_{-}(X_{i})-\mu_{-}(0)).
\end{equation*}
By assumption, the first derivatives of the functions $\mu_{+}$ and $\mu_{-}$
are Lipschitz, and hence absolutely continuous, so that, by the Fundamental
Theorem of Calculus and Fubini's theorem, we can write, for $x\geq 0$,
$\mu_{+}(x)=\mu_{+}(0)+\mu_{+}'(0)x+r(x)$, and for $x<0$,
$\mu_{-}(x)=\mu_{-}(0)+\mu_{-}'(0)x+r(x)$, where
$r(x)=\1{x\geq 0}\int_{0}^{x}\mu''(s)(x-s) \,ds+\1{x<0}\int_{x}^{0}\mu''(s)(x-s)
\,ds$. Since the weights satisfy $\sum_{i=1}^{N}X_{i}w_{+}(X_{i})=0$, and
$\sum_{i=1}^{N}X_{i}w_{-}(X_{i})=0$, it follows that
\begin{equation*}
  \begin{split}
    \tilde{\tau}_{h}-\tau
    &=\sum_{i\colon X_{i}\geq 0}w(X_{i})r(X_{i})
    +\sum_{i\colon X_{i}<0}w(X_{i})r(X_{i})\\
    &=\int_{0}^{\infty}\mu''(s) \sum_{i\colon X_{i}\geq s}w(X_{i})(X_{i}-s)\,ds
    +\int_{-\infty}^{0}\mu''(s)
    \sum_{i\colon X_{i}\leq -s}w(X_{i})(X_{i}-s)\,ds,
  \end{split}
\end{equation*}
where the second line uses Fubini's theorem to change the order of summation and
integration. Next, note that
$\bar{w}_{+}(s)=\sum_{i\colon X_{i}\geq s}w(X_{i})(X_{i}-s)$ is negative for all
$s\geq 0$, because $\bar{w}_{+}(0)=0$, $\bar{w}_{+}(s)=0$ for $s\geq h$, and
$\bar{w}_{+}'(s)=-\sum_{X_{i}\geq s}w(X_{i})$ is monotone on $[0,h]$ with
$\bar{w}_{+}'(0)=-1$. Similarly,
$\bar{w}_{-}(s)=\sum_{i\colon X_{i}\leq -s}w(X_{i})(X_{i}-s)$ is positive for
all $s\geq 0$. Therefore, the expression in the preceding display is maximized
by setting $\mu''(x)=-\SC\operatorname{sign}(x)$, and minimized by setting
$\mu''(x)=\SC\operatorname{sign}(x)$. Plugging these expressions into the
preceding display then gives $\abs{\tilde{\tau}_{h}-\tau}\leq B_{N}$, with
$B_{N}=-\SC\sum_{i=1}^{N}w(X_{i})X_{i}^{2}\sign(X_{i})/2$, which
yields~\eqref{eq:holder-bias-expression}.

Let $o_{P_{N}}(1)$ denote a term that's asymptotically negligible, uniformly
over $\mathcal{M}_{\textnormal{H}}(\SC)$. To complete the proof, we need to show
that (i) $\widehat{\sigma}_{NN}=\tilde{\sigma}^{2}_{\tau}+o_{P_{N}}(1)$, (ii)
$\sqrt{N}_{h}(\widehat{\tau}-\tilde{\tau}_{h})
=\mathcal{N}(0,\tilde{\sigma}^{2}_{\tau})+o_{P_{N}}(1)$, (iii)
$\sqrt{N_{h}}(\tilde{\tau}_{h}-\tau)=b(\mu)+o_{P_{N}}(1)$, and (iv)
$\sqrt{N_{h}}B_{N}=B_{\infty}+o_{P_{N}}(1)$, where
$b(\mu)=\sqrt{N/\pi}\cdot e_{1}'Q_{N}^{-1}\E[M_{i}r(X_{i})]$ and
$B_{\infty}=-(\SC/2\sqrt{N/\pi}) \cdot
e_{1}'Q_{N}^{-1}\E[M_{i}X_{i}^{2}\sign(X_{i})]$ are non-random, and by an
argument analogous to that in the preceding paragraph, satisfy
$\sup_{\mu\in\mathcal{M}_{\textnormal{H}}(\SC)}\abs{b(\mu)}\leq B_{\infty}$. It
then follows from uniform continuity of $\cv_{1-\alpha}(\cdot)$ that
\begin{equation*}
  P_{N}(\sqrt{N_{h}}\abs{\widehat{\tau}-\tau}\leq \cv_{1-\alpha}(r_{\sup})
  \widehat{\sigma}_{NN}^{2})
  =P_{N} (\abs{Z+b(\mu)/\tilde{\sigma}^{2}_{\tau}}
  \leq \cv_{1-\alpha}(B_{\infty}/\tilde{\sigma}^{2}_{\tau})
  \tilde{\sigma}^{2}_{\tau})+o_{P_{N}}(1),
\end{equation*}
where $Z\sim\mathcal{N}(0,1)$, from which honesty follows.

To show (i), note that by~\eqref{eq:Nh},~\eqref{eq:hatQ}, and the law of large
numbers, it suffices to show that
$H^{-1}\widehat{\Omega}_{NN}H^{-1}
-N_{h}^{-1}\sum_{i=1}^{N}\sigma^{2}(X_{i})\lambda(X_{i})=o_{P_{N}}(1)$. Note
that
$N_{h}\widehat{\Omega}_{NN}=\sum_{i}\1{n(X_{i})>1}\cdot
\epsilon_{i}^{2}\lambda(X_{i})-
\sum_{i>j}\epsilon_{i}\epsilon_{j}\1{X_{i}=X_{j},n(X_{i})>1}/(n(X_{i})-1)$,
where $n(x)=\sum_{j=1}^{N}\1{X_{j}=x}$ (so that $n(x_{g})=n_{g}$). This yields
the decomposition
\begin{multline*}
  H^{-1}\widehat{\Omega}_{NN}H^{-1}-
  \frac{1}{N_{h}}\sum_{i=1}^{N}\sigma^{2}(X_{i})\lambda(X_{i})
  =
  \frac{1}{N_{h}}\sum_{i=1}^{N}(\epsilon_{i}^{2}-\sigma^{2}(X_{i}))\lambda(X_{i})\\
+  \frac{(1+o_{P_{N}}(1))\cdot 2}{\sqrt{N\pi N_{h}}}\sum_{i>j}\epsilon_{i}\epsilon_{j}
  \frac{\1{n(X_{i})>1}\1{X_{i}=X_{j}}\lambda(X_{i})}{n(X_{i})-1}
-\frac{1}{N_{h}}\sum_{i=1}^{N}\lambda(X_{i})\epsilon_{i}^{2}\1{n(X_{i})=1}.
\end{multline*}
Since elements of $\lambda(X_{i})$ are bounded by $\1{\abs{X_{i}}\leq h}$, the
first term on the right-hand side of the preceding display is of the order
$o_{P_{N}}(1)$ by the law of large numbers. Conditional on $X_{1},\dotsc,X_{N}$,
the second term has mean zero and variance bounded by
$\frac{8}{N\pi} \max_{i}\sigma^{4}(X_{i})$, which implies that unconditionally,
it also has mean zero, and variance that converges to zero. Therefore, by Markov's
inequality, it is also of the order $o_{P_{N}}(1)$. Finally, by assumption of
the proposition, the probability that the third term is zero converges to one.
Thus, $\widehat{\sigma}_{NN}=\tilde{\sigma}^{2}_{\tau}+o_{P_{N}}(1)$ as
required.

Next, (ii) holds by~\eqref{eq:Nh},~\eqref{eq:hatQ}, and a central limit theorem.
To show (iii), note that
\begin{equation*}
\sqrt{N}_{h}(\tau_{h}-\tau)
=(1+o_{P_{N}}(1))e_{1}'{Q}_{N}^{-1}H
\frac{1}{\sqrt{N\pi}}\sum_{i}H^{-1}M_{i}r(X_{i}).
\end{equation*}
Since elements of $H^{-1}M_{i}r(X_{i})$ are bounded by
$\1{\abs{X_{i}}\leq h}\SC X_{i}^{2}/2\leq \1{\abs{X_{i}}\leq h}\SC h^{2}/2$, it
follows that elements of the variance matrix of
$(N\pi)^{-1/2}\sum_{i}H^{-1}M_{i}r(X_{i})$ are bounded by $\SC^{2}h^{4}/4$.
Thus, (iii) follows by Markov's inequality. Finally, the proof of (iv) is
analogous.

\subsection{Proof of Proposition~\ref{th:bounds-missp-errors}}
It suffices to show that for each $W\in\mathcal{W}$, the left- and right-sided
CIs $\hor{c_{L}^{1-\alpha}(W),\infty}$ and $\hol{-\infty,c_{R}^{1-\alpha}(W)}$
are asymptotically valid CIs for $\tau_{h}+b(W)$, for any sequence of
probability laws $P_{N}$ satisfying the assumptions stated at the beginning of
Appendix~\ref{sec:honest-conf-interv-proofs}, and satisfying
$\mu\in\mathcal{M}_{\text{BME}}(h)$. Honesty will then follow by the
union-intersection principle and the definition of
$\mathcal{M}_{\text{BME}}(h)$.

Note first that by the central limit theorem and the delta method,
\begin{equation*}
  \sqrt{N_{h}}
  \begin{pmatrix}
    \vvec(\widehat\mu_{g}-\mu_{g})\\
    N_{h}^{-1}\sum_{i=1}^{N}H^{-1}M_{i}u_{i}
  \end{pmatrix}
  =_{d}\mathcal{N}\left(0,\; \begin{pmatrix} \pi\diag(\sigma^{2}_{g}/\pi_{g})&
      \vvec(\sigma^{2}_{g}m_{g}'H^{-1})\\
      \vvec(\sigma^{2}_{g}m_{g}'H^{-1})' & H^{-1}\Omega H^{-1}
    \end{pmatrix}\right)+o_{P_{N}}(1).
\end{equation*}
Applying the delta method again, along with Lemma~\ref{lemma:auxilliary-lemma},
yields
\begin{equation*}\label{eq:delta-limit}
  \sqrt{N_{h}}  \begin{pmatrix}
    \vvec(\widehat{\delta}(x_{g})-\delta(x_{g}))\\
    \widehat{\tau}_{h}-\tauh
  \end{pmatrix}
  =_{d} \mathcal{N}\left(0,\Sigma \right)+o_{P_{N}}(1),
\end{equation*}
where the variance matrix $\Sigma$ is given by
\begin{equation*}
  \Sigma=  \begin{pmatrix}
    \diag(\sigma^{2}_{g}\cdot \pi/\pi_{g}) +\mathcal{V}
    &
    \vvec(\sigma^{2}_{g}m_{g}'Q_{N}^{-1}e_{1}-m_{g}'Q_{N}^{-1}\Omega Q_{N}^{-1}e_{1})
    \\
    \vvec(\sigma^{2}_{g}m_{g}'Q_{N}^{-1}e_{1} - m_{g}'Q_{N}^{-1}\Omega Q_{N}^{-1}e_{1})'
    & e_{1}'Q_{N}^{-1}\Omega Q_{N}^{-1}e_{1}
  \end{pmatrix},
\end{equation*}
and $\mathcal{V}$ is a $G_{h}\times G_{h}$ matrix with $(g,g^*)$ element equal
to
$m_{g}'Q^{-1}\Omega Q^{-1}m_{g^*}- (\sigma^{2}_{g}+\sigma^{2}_{g^*})
m_{g}'Q^{-1}m_{g^*}$.

Fix $W=(g^{-},g^{+},s^{-},s^{+})$, and let $a(W)\in\mathbb{R}^{G_{h}+1}$ denote
a vector with the $g_{-}$th element equal to $s^{-}$, $(G_{h}^{-}+g_{+})$th
element equal to $s^{+}$, the last element equal to one, and the remaining
elements equal to zero. It follows that
$\sqrt{N_{h}}(\widehat{\tau}+\widehat{b}(W)-\tau_{h}-b(W))$ is asymptotically
normal with variance $a(W)'\Sigma a(W)$. To construct the left- and right-sided
CIs, we use the variance estimator
\begin{equation}\label{eq:hat-VW}
  \widehat{V}(W)=a(W)'\widehat{\Sigma}a(W),
\end{equation}
where $\widehat{\Sigma}$ is a plug-in estimator of $\Sigma$ that replaces
$Q_{N}$ by $\widehat{Q}$, $\Omega$ by $\widehat{\Omega}_{\text{EHW}}$,
$\pi/\pi_{g}$ by $N_{h}/n_{g}$, and $\sigma^{2}_{g}$ by
$\widehat{\sigma}_{g}^{2}$ (given in Section~\ref{sec:bound-second-deriv}).
Since by standard arguments $n_{g}/N_{h}=\pi_{g}/\pi+o_{P_{N}}(1)$, and
$\widehat{\sigma}_{g}^{2}=\sigma^{2}_{g}+o_{P_{N}}(1)$, it follows
from~\eqref{eq:hatQ} and~\eqref{eq:Omega-EHW-consistency} that
$\widehat{V}(W)=a(W)'{\Sigma}a(W)+o_{P_{N}}(1)$, which, together with the
asymptotic normality of
$\sqrt{N_{h}}(\widehat{\tau}+\widehat{b}(W)-\tau_{h}-b(W))$, implies asymptotic
validity of $\hor{c_{L}^{1-\alpha}(W),\infty}$ and
$\hol{-\infty,c_{R}^{1-\alpha}(W)}$, as required.

\section{Additional Figures}

This appendix shows the fit of the specifications considered in
Section~\ref{sec:dist-crv-conf}. Specifically, Figure~\ref{fig_data1a} shows the
fit of a linear specification ($p=1$) for the four values of the bandwidth $h$
considered; Figure~\ref{fig_data1b} shows the analogous results for a quadratic
fit ($p=2$). In each case, the value of the parameter $\tauh$ is equal to height
of the jump in the fitted line at the 40-year cutoff.

\afterpage{
\begin{landscape}
\begin{figure}
\begin{center}
\includegraphics[width=\textwidth]{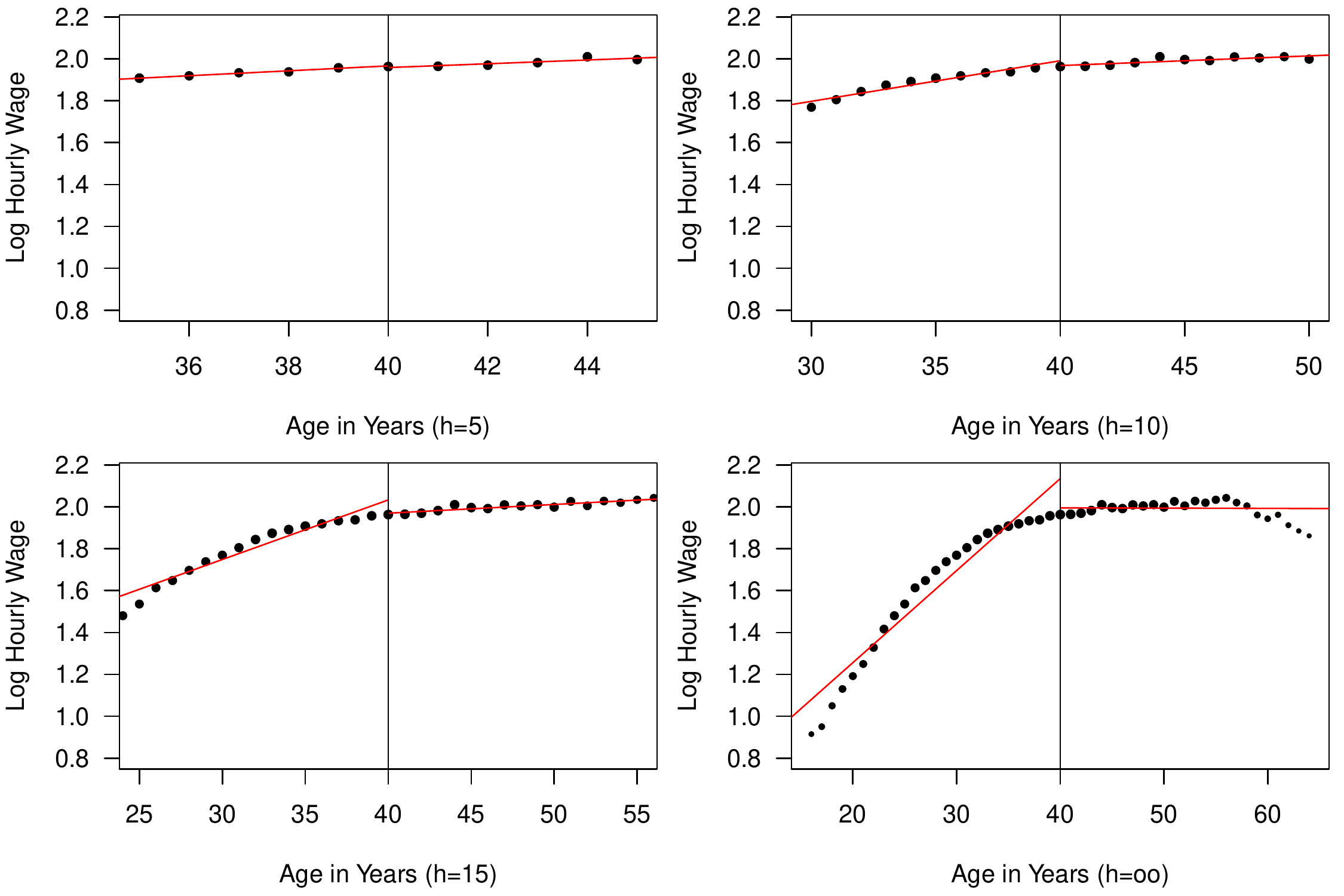}
\caption{Fit of specification~\eqref{eq:cps-regression} for $p=1$ (linear, red
  line) in the full CPS data for $h=5$ (top-left panel), $h=10$ (top-right
  panel), $h=15$ (bottom-left panel), and $h=\infty$ (bottom-right
  panel).\label{fig_data1a}}
\end{center}
\end{figure}
\end{landscape}
}

\afterpage{
\begin{landscape}
\begin{figure}
\begin{center}
\includegraphics[width=\textwidth]{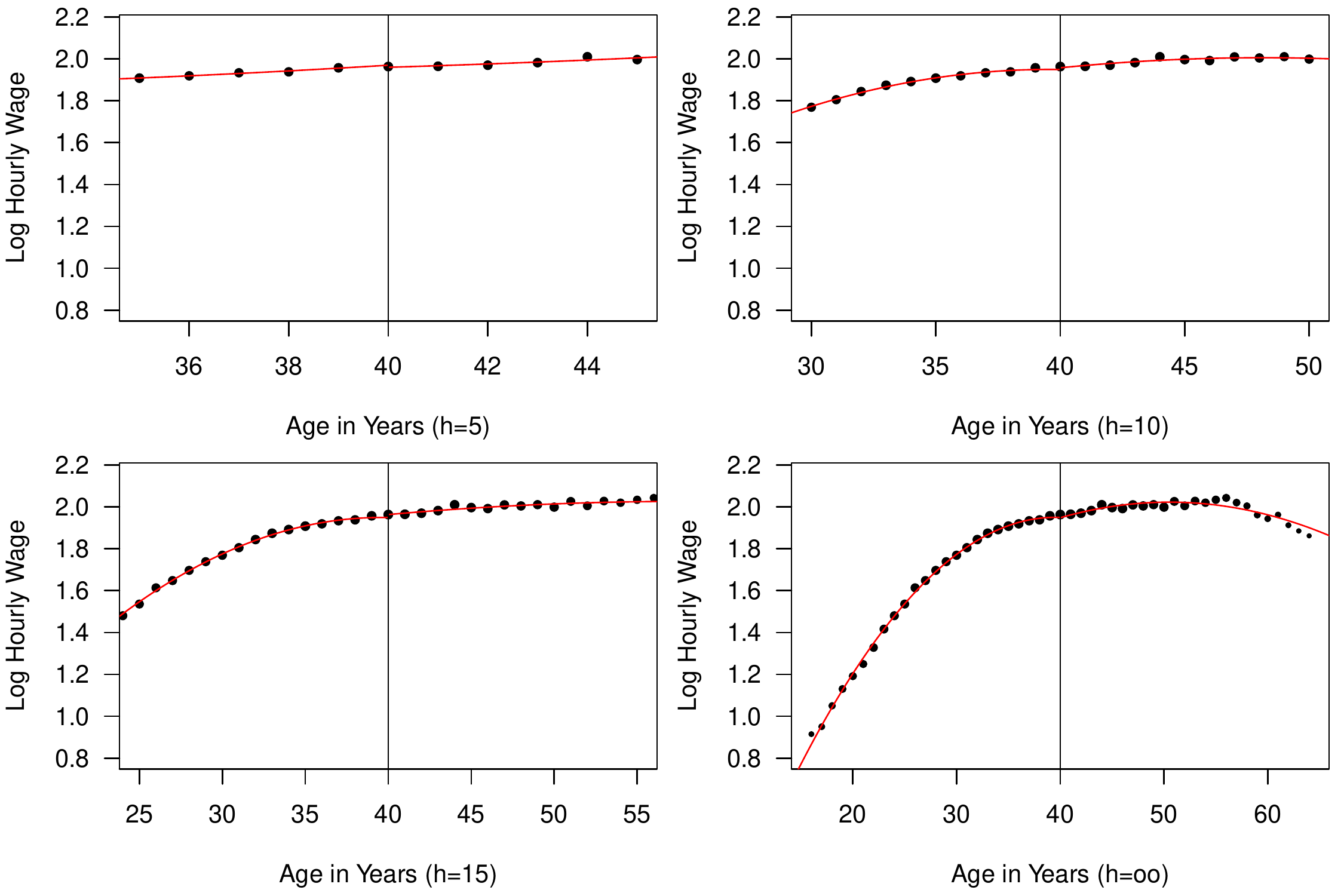}
\caption{Fit of specification~\eqref{eq:cps-regression} for $p=2$ (quadratic,
  red line) in the full CPS data for $h=5$ (top-left panel), $h=10$ (top-right
  panel), $h=15$ (bottom-left panel), and $h=\infty$ (bottom-right
  panel).\label{fig_data1b}}
\end{center}
\end{figure}
\end{landscape}
}

\newpage

\singlespacing
\bibliography{bibl}

\begin{thebibliography}{37}
\newcommand{\enquote}[1]{``#1''}
\expandafter\ifx\csname natexlab\endcsname\relax\def\natexlab#1{#1}\fi

\bibitem[\protect\citeauthoryear{Abadie and Imbens}{Abadie and
  Imbens}{2006}]{AbIm06}
\textsc{Abadie, A. and G.~W. Imbens} (2006): \enquote{Large Sample Properties
  of Matching Estimators for Average Treatment Effects,} \emph{Econometrica},
  74, 235--267.

\bibitem[\protect\citeauthoryear{Abadie, Imbens, and Zheng}{Abadie
  et~al.}{2014}]{AbImZh14}
\textsc{Abadie, A., G.~W. Imbens, and F.~Zheng} (2014): \enquote{Inference for
  Misspecified Models with Fixed Regressors,} \emph{Journal of the American
  Statistical Association}, 109, 1601--1614.

\bibitem[\protect\citeauthoryear{Armstrong and Kolesár}{Armstrong and
  Kolesár}{2016}]{ArKo16honest}
\textsc{Armstrong, T.~B. and M.~Kolesár} (2016): \enquote{Simple and honest
  confidence intervals in nonparametric regression,} ArXiv:1606.01200.

\bibitem[\protect\citeauthoryear{Armstrong and Kolesár}{Armstrong and
  Kolesár}{2017}]{ArKo16optimal}
---\hspace{-.1pt}---\hspace{-.1pt}--- (2017): \enquote{Optimal inference in a
  class of regression models,} ArXiv:1511.06028.

\bibitem[\protect\citeauthoryear{Bell and McCaffrey}{Bell and
  McCaffrey}{2002}]{bm02}
\textsc{Bell, R.~M. and D.~F. McCaffrey} (2002): \enquote{Bias Reduction in
  Standard Errors for Linear Regression with Multi-Stage Smples,} \emph{Survey
  Methodology}, 28, 169--181.

\bibitem[\protect\citeauthoryear{Calonico, Cattaneo, and Titiunik}{Calonico
  et~al.}{2014}]{calonico2014robust}
\textsc{Calonico, S., M.~D. Cattaneo, and R.~Titiunik} (2014): \enquote{Robust
  Nonparametric Confidence Intervals for Regression-Discontinuity Designs,}
  \emph{Econometrica}, 82, 2295--2326.

\bibitem[\protect\citeauthoryear{Cameron and Miller}{Cameron and
  Miller}{2014}]{CaMi14}
\textsc{Cameron, C.~A. and D.~L. Miller} (2014): \enquote{A Practitioner's
  Guide to Cluster-Robust Inference,} \emph{Journal of Human Resources}, 50,
  317--372.

\bibitem[\protect\citeauthoryear{Card, Dobkin, and Maestas}{Card
  et~al.}{2008}]{10.1257/aer.98.5.2242}
\textsc{Card, D., C.~Dobkin, and N.~Maestas} (2008): \enquote{The Impact of
  Nearly Universal Insurance Coverage on Health Care Utilization: Evidence from
  Medicare,} \emph{American Economic Review}, 98, 2242--58.

\bibitem[\protect\citeauthoryear{Card, Lee, Pei, and Weber}{Card
  et~al.}{2015}]{clpw15}
\textsc{Card, D., D.~S. Lee, Z.~Pei, and A.~Weber} (2015): \enquote{Inference
  on Causal Effects in a Generalized Regression Kink Design,}
  \emph{Econometrica}, 83, 2453--2483.

\bibitem[\protect\citeauthoryear{Casella and Berger}{Casella and
  Berger}{2002}]{CaBe02}
\textsc{Casella, G. and R.~L. Berger} (2002): \emph{Statistical Inference},
  Pacific Grove, CA: Duxbury/Thomson Learning, 2nd ed.

\bibitem[\protect\citeauthoryear{Cheng, Fan, and Marron}{Cheng
  et~al.}{1997}]{cfm97}
\textsc{Cheng, M.-Y., J.~Fan, and J.~S. Marron} (1997): \enquote{On automatic
  boundary corrections,} \emph{Annals of Statistics}, 25, 1691--1708.

\bibitem[\protect\citeauthoryear{Chetty, Friedman, and Saez}{Chetty
  et~al.}{2013}]{10.1257/aer.103.7.2683}
\textsc{Chetty, R., J.~N. Friedman, and E.~Saez} (2013): \enquote{Using
  Differences in Knowledge across Neighborhoods to Uncover the Impacts of the
  EITC on Earnings,} \emph{American Economic Review}, 103, 2683--2721.

\bibitem[\protect\citeauthoryear{Clark and Royer}{Clark and
  Royer}{2013}]{10.1257/aer.103.6.2087}
\textsc{Clark, D. and H.~Royer} (2013): \enquote{The Effect of Education on
  Adult Mortality and Health: Evidence from Britain,} \emph{American Economic
  Review}, 103, 2087--2120.

\bibitem[\protect\citeauthoryear{Devereux and Hart}{Devereux and
  Hart}{2010}]{devereux2010forced}
\textsc{Devereux, P.~J. and R.~A. Hart} (2010): \enquote{Forced to be Rich?
  Returns to Compulsory Schooling in Britain,} \emph{Economic Journal}, 120,
  1345--1364.

\bibitem[\protect\citeauthoryear{Dong}{Dong}{2015}]{dong2015regression}
\textsc{Dong, Y.} (2015): \enquote{Regression discontinuity applications with
  rounding errors in the running variable,} \emph{Journal of Applied
  Econometrics}, 30, 422--446.

\bibitem[\protect\citeauthoryear{Frandsen}{Frandsen}{2016}]{frandsen16manipulation}
\textsc{Frandsen, B.~R.} (2016): \enquote{Party Bias in Union Representation
  Elections: Testing for Manipulation in the Regression Discontinuity Design
  When the Running Variable is Discrete,} Working Paper.

\bibitem[\protect\citeauthoryear{Fredriksson, {\"O}ckert, and
  Oosterbeek}{Fredriksson et~al.}{2013}]{fredriksson2013long}
\textsc{Fredriksson, P., B.~{\"O}ckert, and H.~Oosterbeek} (2013):
  \enquote{Long-term effects of class size,} \emph{Quarterly Journal of
  Economics}, 128, 249--285.

\bibitem[\protect\citeauthoryear{Gelman and Imbens}{Gelman and
  Imbens}{2014}]{gelman2014high}
\textsc{Gelman, A. and G.~Imbens} (2014): \enquote{Why high-order polynomials
  should not be used in regression discontinuity designs,} NBER Working Paper.

\bibitem[\protect\citeauthoryear{Gerard, Rokkanen, and Rothe}{Gerard
  et~al.}{2016}]{gerard2016bounds}
\textsc{Gerard, F., M.~Rokkanen, and C.~Rothe} (2016): \enquote{Bounds on
  Treatment Effects in Regression Discontinuity Designs under Manipulation of
  the Running Variable, with an Application to Unemployment Insurance in
  Brazil,} NBER Working Paper.

\bibitem[\protect\citeauthoryear{Hahn, Todd, and {van der Klaauw}}{Hahn
  et~al.}{2001}]{hahn2001identification}
\textsc{Hahn, J., P.~Todd, and W.~{van der Klaauw}} (2001):
  \enquote{Identification and Estimation of Treatment Effects with a
  Regression-Discontinuity Design,} \emph{Econometrica}, 69, 201--209.

\bibitem[\protect\citeauthoryear{Hansen}{Hansen}{2007}]{hansen2007asymptotic}
\textsc{Hansen, C.~B.} (2007): \enquote{Asymptotic properties of a robust
  variance matrix estimator for panel data when T is large,} \emph{Journal of
  Econometrics}, 141, 597--620.

\bibitem[\protect\citeauthoryear{Hinnerich and Pettersson-Lidbom}{Hinnerich and
  Pettersson-Lidbom}{2014}]{hinnerich2014democracy}
\textsc{Hinnerich, B.~T. and P.~Pettersson-Lidbom} (2014): \enquote{Democracy,
  redistribution, and political participation: Evidence from Sweden
  1919--1938,} \emph{Econometrica}, 82, 961--993.

\bibitem[\protect\citeauthoryear{Imbens and Koles{\'{a}}r}{Imbens and
  Koles{\'{a}}r}{2016}]{ImKo16}
\textsc{Imbens, G.~W. and M.~Koles{\'{a}}r} (2016): \enquote{Robust Standard
  Errors in Small Samples: Some Practical Advice,} \emph{Review of Economics
  and Statistics}, 98, 701--712.

\bibitem[\protect\citeauthoryear{Lalive}{Lalive}{2008}]{lalive08}
\textsc{Lalive, R.} (2008): \enquote{How Do Extended Benefits Affect
  Unemployment Duration? A Regression Discontinuity Approach,} \emph{Journal of
  Econometrics}, 142, 785--806.

\bibitem[\protect\citeauthoryear{Lee and Card}{Lee and
  Card}{2008}]{lee2008regression}
\textsc{Lee, D.~S. and D.~Card} (2008): \enquote{Regression discontinuity
  inference with specification error,} \emph{Journal of Econometrics}, 142,
  655--674.

\bibitem[\protect\citeauthoryear{Lee and Lemieux}{Lee and Lemieux}{2010}]{ll10}
\textsc{Lee, D.~S. and T.~Lemieux} (2010): \enquote{Regression Discontinuity
  Designs in Economics,} \emph{Journal of Economic Literature}, 48, 281--355.

\bibitem[\protect\citeauthoryear{Lemieux}{Lemieux}{2006}]{lemieux2006increasing}
\textsc{Lemieux, T.} (2006): \enquote{Increasing residual wage inequality:
  Composition effects, noisy data, or rising demand for skill?} \emph{American
  Economic Review}, 96, 461--498.

\bibitem[\protect\citeauthoryear{Li}{Li}{1989}]{li89}
\textsc{Li, K.-C.} (1989): \enquote{{Honest confidence regions for
  nonparametric regression},} \emph{Annals of Statistics}, 17, 1001--1008.

\bibitem[\protect\citeauthoryear{Liang and Zeger}{Liang and
  Zeger}{1986}]{liang1986longitudinal}
\textsc{Liang, K.-Y. and S.~L. Zeger} (1986): \enquote{Longitudinal data
  analysis using generalized linear models,} \emph{Biometrika}, 73, 13--22.

\bibitem[\protect\citeauthoryear{Low}{Low}{1997}]{low97}
\textsc{Low, M.~G.} (1997): \enquote{On nonparametric confidence intervals,}
  \emph{Annals of Statistics}, 25, 2547--2554.

\bibitem[\protect\citeauthoryear{MacKinnon and White}{MacKinnon and
  White}{1985}]{MaWh85}
\textsc{MacKinnon, J.~G. and H.~White} (1985): \enquote{Some
  Heteroskedasticity-Consistent Covariance Matrix Estimators with Improved
  Finite Sample Properties,} \emph{Journal of Econometrics}, 29, 305--325.

\bibitem[\protect\citeauthoryear{Martorell and McFarlin}{Martorell and
  McFarlin}{2011}]{martorell2011help}
\textsc{Martorell, P. and I.~McFarlin} (2011): \enquote{Help or hindrance? The
  effects of college remediation on academic and labor market outcomes,}
  \emph{Review of Economics and Statistics}, 93, 436--454.

\bibitem[\protect\citeauthoryear{Oreopoulos}{Oreopoulos}{2006}]{oreopoulos2006estimating}
\textsc{Oreopoulos, P.} (2006): \enquote{Estimating average and local average
  treatment effects of education when compulsory schooling laws really matter,}
  \emph{American Economic Review}, 152--175.

\bibitem[\protect\citeauthoryear{Oreopoulos}{Oreopoulos}{2008}]{oreopoulos2006online}
---\hspace{-.1pt}---\hspace{-.1pt}--- (2008): \enquote{Estimating Average and
  Local Average Treatment Effects of Education When Compulsory Schooling Laws
  Really Matter: Corrigendum,} Available at
  \url{http://www.aeaweb.org/articles.php?doi=10.1257/000282806776157641}.

\bibitem[\protect\citeauthoryear{Sacks and Ylvisaker}{Sacks and
  Ylvisaker}{1978}]{sacks1978linear}
\textsc{Sacks, J. and D.~Ylvisaker} (1978): \enquote{Linear estimation for
  approximately linear models,} \emph{Annals of Statistics}, 1122--1137.

\bibitem[\protect\citeauthoryear{Schochet, Cook, Deke, Imbens, Lockwood,
  Porter, and Smith}{Schochet et~al.}{2010}]{schochet2010standards}
\textsc{Schochet, P., T.~Cook, J.~Deke, G.~Imbens, J.~Lockwood, J.~Porter, and
  J.~Smith} (2010): \enquote{Standards for Regression Discontinuity Designs.}
  \emph{What Works Clearinghouse, Institute of Education Sciences, U.S.
  Department of Education}.

\bibitem[\protect\citeauthoryear{Urquiola and Verhoogen}{Urquiola and
  Verhoogen}{2009}]{urquiola2009class}
\textsc{Urquiola, M. and E.~Verhoogen} (2009): \enquote{Class-size caps,
  sorting, and the regression-discontinuity design,} \emph{American Economic
  Review}, 99, 179--215.

\end{thebibliography}


\begin{thebibliography}{4}
\newcommand{\enquote}[1]{``#1''}
\expandafter\ifx\csname natexlab\endcsname\relax\def\natexlab#1{#1}\fi

\bibitem[\protect\citeauthoryear{Armstrong and Kolesár}{Armstrong and
  Kolesár}{2017}]{ArKo16optimal}
\textsc{Armstrong, T.~B. and M.~Kolesár} (2017): \enquote{Optimal inference in
  a class of regression models,} ArXiv:1511.06028.

\bibitem[\protect\citeauthoryear{Bell and McCaffrey}{Bell and
  McCaffrey}{2002}]{bm02}
\textsc{Bell, R.~M. and D.~F. McCaffrey} (2002): \enquote{Bias Reduction in
  Standard Errors for Linear Regression with Multi-Stage Smples,} \emph{Survey
  Methodology}, 28, 169--181.

\bibitem[\protect\citeauthoryear{Chernozhukov, Lee, and Rosen}{Chernozhukov
  et~al.}{2013}]{ChLeRo2013}
\textsc{Chernozhukov, V., S.~Lee, and A.~Rosen} (2013): \enquote{Intersection
  Bounds: estimation and inference,} \emph{Econometrica}, 81, 667--737.

\bibitem[\protect\citeauthoryear{Imbens and Koles{\'{a}}r}{Imbens and
  Koles{\'{a}}r}{2016}]{ImKo16}
\textsc{Imbens, G.~W. and M.~Koles{\'{a}}r} (2016): \enquote{Robust Standard
  Errors in Small Samples: Some Practical Advice,} \emph{Review of Economics
  and Statistics}, 98, 701--712.

\end{thebibliography}

\end{document}


\bibliographystyle{ecta}
\singlespacing

\maketitle

\section{Performance of Honest CIs and alternative CRV CIs in CPS simulation study}\label{sec:s1}

In this section, we again consider the CPS placebo study from
Section~\ref{sec:dist-crv-conf} in the main text to study the performance of the
honest CIs proposed in the Section~\ref{sec:honest-conf-interv}. Given that the
typical increase in log wages is about $0.017$ per extra year of age, guided by
the heuristic in Section~\ref{sec:bound-second-deriv}, we set $\SC=0.045$ for
the BSD CIs. In addition, we also consider two modifications of the CRV CIs that
have been shown to perform better in settings with a few
clusters. 
The first modification, which we term CRV2, is a bias-reduction modification
analogous to the HC2 modification of the EHW variance estimator developed by
\citet{bm02}. This modification makes the variance estimator unbiased when the
errors $u_{i}=Y_{i}-M_{i}'\theta_{h}$ are homoscedastic and independent. To
describe it, let $\textbf{M}$ denote the $N_{h}\times (2p+2)$ design matrix with
$i$th row given by $M_{i}'$, and let $\mathbf{M}_{g}$ denote the $n_{g}$ rows of
the design matrix $\mathbf{M}$ for which $M_{i}'=m(x_{g})'$ (that is, those rows
that correspond to ``cluster'' $g$). Here $n_{g}$ is the number of observations
with $X_{i}=x_{g}$. The CRV2 variance estimator replaces
$\widehat\Omega_{\textnormal{\CRV}}$ in the definition of $\CRVse^2$ given in
Section~\ref{sec:basic-model-infer} with
\begin{equation*}
\widehat\Omega_{\textnormal{CRV2}}=
\frac{1}{N_{h}}\sum_{g=1}^{G_{h}}
\mathbf{M}_{g}'\mathbf{A}_{g}\widehat{\mathbf{u}}_{g}
\widehat{\mathbf{u}}_{g}'\mathbf{A}_{g}\mathbf{M}_{g},
\end{equation*}
where $\widehat{\mathbf{u}}_{g}$ is an $n_{g}$-vector of the regression
residuals corresponding to ``cluster'' $g$, and $\mathbf{A}_{g}$ is the inverse
of the symmetric square root of the $n_{g}\times n_{g}$ block of the annihilator
matrix $\mathbf{I}_{N_{h}}-\mathbf{M}(\mathbf{M}'\mathbf{M})^{-1}\mathbf{M}'$
corresponding to ``cluster'' $g$,
$\mathbf{I}_{n_{g}} -\mathbf{M}_{g}(\mathbf{M}'\mathbf{M})^{-1}\mathbf{M}_{g}'$.
Thus,
$\mathbf{A}_{g}[\mathbf{I}_{n_{g}}
-\mathbf{M}_{g}(\mathbf{M}'\mathbf{M})^{-1}\mathbf{M}_{g}']\mathbf{A}_{g}=\mathbf{I}_{n_{g}}$.
In contrast, the CRV estimator, as implemented in STATA, sets
$\mathbf{A}_{g}=\sqrt{G_{h}/(G_{h}-1) \times
  (N_{h}-1)/(N_{h}-2(p+1))}\mathbf{I}_{n_{g}}$. The second modification that we
consider also uses the CRV2 variance estimator, but replaces the usual
$1.96$ critical value in the construction of 95\% confidence intervals with a
critical value based on a $t$-distribution with $a$ degrees of freedom, where
$a$ is chosen so that the first two moments of
$\widehat{\sigma}_{\textnormal{CRV2}}/\mathbb{V}(\widehat{\tau}\mid\mathbf{M})$ match
that of $\chi^{2}_{a}/a$, assuming correct specification and independent
homoscedastic errors. This modification has also been proposed by \citet{bm02},
and, as discussed in \citet{ImKo16}, it generalizes the classic
Welch-Sattherthwaite solution to the Behrens-Fisher problem.

The results are reported in Table~\ref{tab:cps-linear} for the linear
specification ($p=1$), and in Table~\ref{tab:cps-quadratic} for the quadratic
specification ($p=2$). For convenience, the tables also reproduce the results
for inference based on EHW and CRV standard errors, reported in
Table~\ref{tab:empmc_results1} in the main text. To compare the length of honest
CIs and CRV-BM CIs to those of EHW, CRV, and CRV2 CIs, the tables report average
normalized standard errors. We define normalized standard error of a CI $[a,b]$
with nominal level 95\% as $(b-a)/(2\times 1.96)$, so that the CI is given by
adding and subtracting the normalized standard error times the usual $1.96$
critical value from its midpoint. Finally, Table~\ref{tab:cps-rates} compares
magnitudes of the various clustered standard errors relative to that of the
conventional EHW estimator.

The coverage properties of honest CIs are excellent: they achieve at least 95\%
coverage in all specifications. The length of BME CIs is moderately larger than
that of EHW CIs in specifications in which EHW CIs achieve proper coverage. The
BME CIs are much longer, especially for large $h$ or small $N_{h}$. This is in
line with the discussion in Section~\ref{sec:bounds-spec-errors}.

In designs in which the fitted model is close to being correctly specified (all
quadratic specifications and linear specifications with $h< 15$) CRV-BM CIs do
about as well as EHW CIs in delivering approximately correct coverage, while the
CRV2 adjustment alone is not sufficient to bring coverage close to 95\%. On the
other hand, in designs in which the fitted model is misspecified (linear
specifications with $h=15$ or $h=\infty$), the coverage of CRV-BM CIs is only
slightly better than that of EHW CIs; the improvement in coverage is not
sufficient to bring coverage close to 95\%. This contrasts with the coverage of
the honest CIs, which remains close to, or above, 95\%. Furthermore, the CRV-BM
CIs are much more variable than EHW, and thus the average length of the CRV-BM
CI, as measured by the normalized SE, is greater than in the case of EHW\@: in
the designs considered, they are about 40\% longer on average, sometimes more
than twice as long. In part due to this variability, the probability that CRV2
and CRV-BM CIs are longer than EHW CIs in a particular sample is not very close
to 1, and falls between 0.7 and 0.9 for most of the designs that we consider, as
shown in Table~\ref{tab:cps-rates}. This suggests, in line with the results in
Section~\ref{sec:empir-appl}, that in any particular empirical application,
these CIs may not be larger than EHW CIs.

\section{Additional simulation evidence}

In this section, we consider a second Monte Carlo exercise in which we simulate
realizations of an outcome variable $Y_i$ and a running variable $X_i$ from
several data generating processes (DGPs) with different conditional expectation
functions and different numbers of support points, and also several sample
sizes. This allows us to disentangle the effect of model misspecification and
the number of the support points on the performance of CRV-based inference;
something that was not possible in the CPS placebo study from
Section~\ref{sec:dist-crv-conf}.

Each of our DGPs is such that the support of the running variable is the union
of an equally spaced grid of $G^-$ points $-1,-(G^{-}-1)/G^{-},\dotsc,-1/G^{-}$
and an equally spaced grid of $G^+$ points $1/G^{+},2/G^{+},\dotsc,1$. We
consider values $G^-,G^+\in\{ 5,25,50\}$. The distribution of $X_i$ then has
probability mass 1/2 spread equally across the support points above and below zero,
so that $P(X_{i}=x_{g})=1/G^{+}$ for $x_{g}>0$ and
$P(X_{i}=x_{g})=1/G^{-}$ for $x_{g}<0$.

The outcome variable is generated as $Y_i = \mu(X_i)+ \varepsilon_i$, where
$\varepsilon_i$ and $X_i$ are independent, $\varepsilon_i\sim \mathcal{N}(0, 0.1)$,
and
\begin{equation*}
  \mu(x)= x +\lambda_1\cdot\sin(\pi\cdot x)+ \lambda_2\cdot\cos(\pi\cdot x).
\end{equation*}
We also generate a treatment indicator that is equal to one if $X_i\geq 0$, and
equal to zero otherwise. Since $\mu(x)$ is continuous at $x=0$ for every
$(\lambda_1,\lambda_2)$, the causal effect of our treatment at the cutoff is zero in our all our
DGPs. We consider $(\lambda_1,\lambda_2) \in\{ (0,0), (0.05,0), (0,0.05)\}$ and
the sample sizes $N_{h}\in \{ 100,1000,10000\}$, and estimate the treatment effect
by fitting the linear model
\begin{equation}\label{rdreg2}
  Y_i = \beta_0 + \tau \cdot \1{X_i \geq 0} + \beta^- \cdot X_i +
  \beta^+ \cdot \1{X_i \geq 0} \cdot X_i + U_i.
\end{equation}
Note that this specification is analogous to the model~\eqref{eq:cps-regression}
in the main text with $p=1$ and $h=1$. We do not consider alternative values of
$h$ and $p$ in this simulation exercise because variation in the accuracy of the
fitted model is achieved by varying the DGP\@.

\begin{figure}[!t]
  \begin{center}
    \includegraphics[width=.99\textwidth]{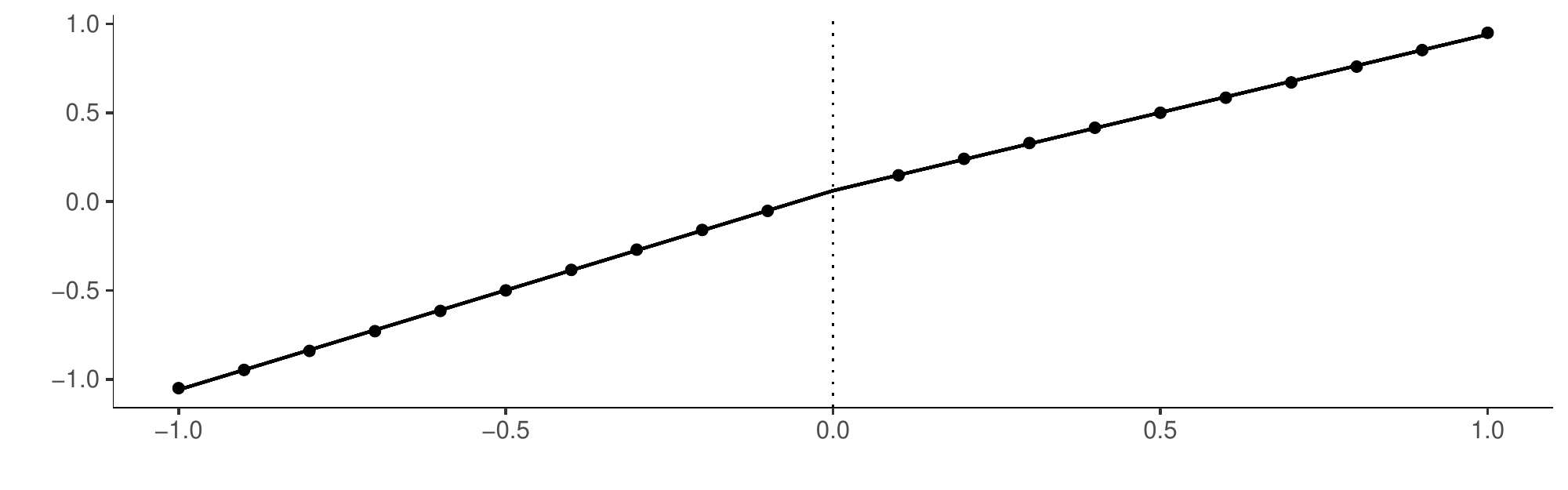}
    \includegraphics[width=.99\textwidth]{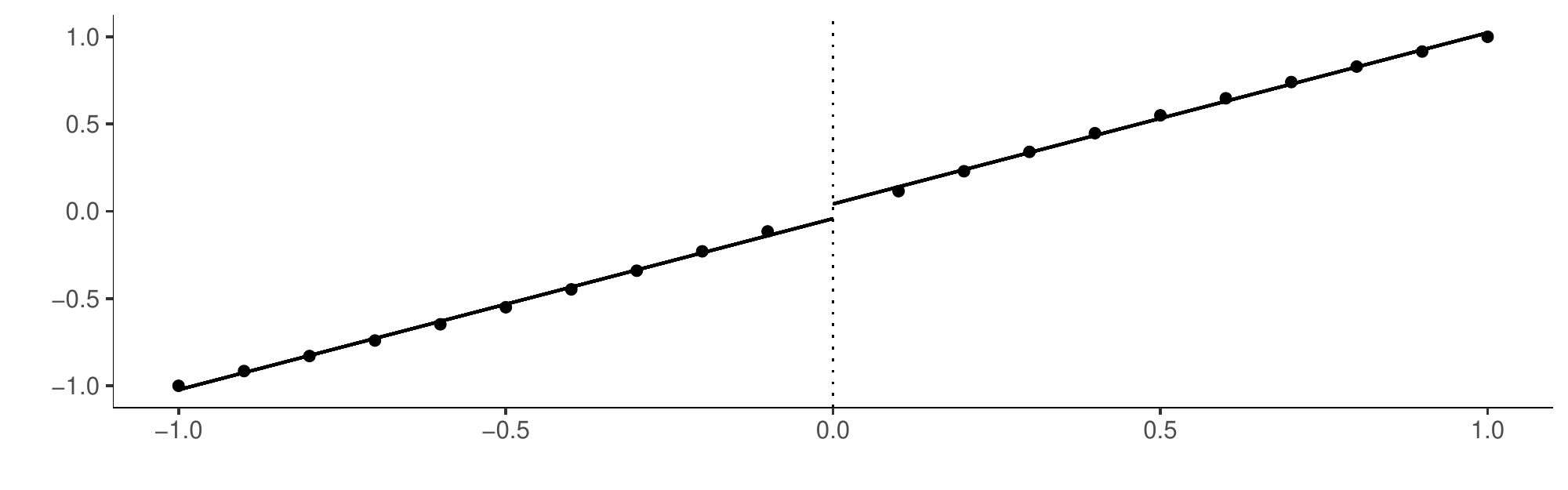}
    \caption{\label{fig_sim} Plot of $\mu(x)=x +.05\cdot\cos(\pi\cdot x)$ (top
      panel) and $\mu(x)=x +.05\cdot\sin(\pi\cdot x)$ (bottom panel) for
      $G^-=G^+=10$. Dots indicate the value of the function at the support
      points of the running variable; solid lines correspond to linear fit above
      and below the threshold.}
  \end{center}
\end{figure}

To asses the accuracy of model~\eqref{rdreg2}, we plot the versions of $\mu(x)$
that we consider together with the corresponding linear fit in
Figure~\ref{fig_sim} for the case that $G^-=G^+=10$. As one can see, the
departure from linearity is rather modest for
$(\lambda_1,\lambda_2) \in\{ (0.05,0), (0,0.05)\}$. In
Tables~\ref{tab:results1}--\ref{tab:results3} we then report the empirical
standard deviation of $\widehat\tau$, the empirical coverage probabilities of
the EHW and CRV CIs with nominal level 95\%, as well as the coverage
probabilities of the honest CIs, and the CRV2 and CRV-BM modifications. For BSD
CIs, we consider the values $\SC=\pi^{2}/20$, which corresponds to the largest
value of the second derivative of $\mu$ for the designs considered, as well as
the more conservative choice $\SC=\pi^{2}/10$.\footnote{As pointed out in the main
  text, the choice of $\SC$ requires subject knowledge, as it implies certain
  restrictions on the shape of $\mu(x)$. This is generally difficult to mimic in
  the context of a simulation study based on an artificial DGP\@. } To compare
length, the tables report normalized standard errors, defined in
Section~\ref{sec:s1}.

Table~\ref{tab:results1} reports results for the case
$(\lambda_1,\lambda_2) = (0,0)$, in which the true conditional expectation
function is linear and thus our fitted model is correctly specified. We see that
the \CRV\ standard error is a downward-biased estimate of the standard deviation
of $\widehat\tau$, and therefore the CRV confidence interval under-covers the
treatment effect. The distortion is most severe for the case with the least
number of points on either side of the threshold ($G^-=G^+=5$), where it amounts
to a deviation of 20 percentage points from the nominal level. With more support
points the distortion becomes less pronounced, but it is still noticeable even
for $G^-=G^+=50$. These findings are the same for all the sample sizes we
consider. The distortion of CRV2 CIs is slightly smaller, and the coverage of
CRV-BM is close to 95\%, at the cost of a loss in power (as measured by the
average normalized standard error). The honest CIs perform well in terms of
coverage, although they are quite conservative: this is the price for
maintaining good coverage over CEFs that are less smooth than $\mu(x)=x$ (see
\citet{ArKo16optimal} for theoretical results on impossibility of adaptation to
smooth functions).

Table~\ref{tab:results2} reports results for the case
$(\lambda_1,\lambda_2) = (0,.05)$. Here $\mu(x)$ is nonlinear, but due to the
symmetry properties of the cosine function $\tauh= 0$. This setup mimics
applications in which the bias of $\widehat\tau$ is small even though the
functional form of $\mu(x)$ is misspecified. In line with our asymptotic
approximations, the \CRV\ standard error is downward biased for smaller values
of $N$, and upward biased for larger sample sizes. Simulation results for the
case that $N=10^7$, which are not reported here, also confirm that the \CRV\
standard error does not converge to zero. Correspondingly, the CRV CI
under-covers the treatment effect for smaller values of $N$, and over-covers for
larger values. The distortions are again more pronounced for smaller values of
$G^+$ and $G^-$. The CRV-BM CIs correct for these size distortions, while the
CRV2 modification alone is not sufficient. The honest CIs perform well in terms
of coverage, but they are again quite conservative: again this is due to the
possible bias adjustment built into the CIs.

Table~\ref{tab:results3} reports results for the case
$(\lambda_1,\lambda_2) = (.05,0)$. Here the linear model is misspecified as
well, but in such a way that $\tauh$ is substantially different from zero; with
its exact value depending on $G^+$ and $G^-$. As with the previous sets of
results, the \CRV\ standard error is downward biased for smaller values of $N$,
and upward biased for larger sample sizes. However, since $\tauh\neq 0$ here,
the coverage probability of the CRV confidence interval is below the nominal
level for all $N$, and tends to zero as the sample size increases. For smaller
values of $N$, the coverage properties of the CRV confidence interval are also
worse than those of the standard EHW confidence interval. The \CRV\ confidence
interval only performs better in a relative sense than EHW confidence interval
when $N$ is large, but for these cases both CIs are heavily distorted and have
coverage probability very close to zero. So in absolute terms the performance of
the CRV confidence interval is still poor. The same conclusion applies to the
CRV2 and CRV-BM modifications. The assumption underlying the construction of BME
CIs is violated for $G_{+}=G_{-}=5$, since in this case, the specification bias
of a linear approximation to the CEF is actually worse at zero than at other
support points. Consequently, BME CIs undercover once $N_{h}$ is sufficiently
large. In contrast, the coverage of BSD CIs remains excellent for all
specifications, as predicted by the theory.

In summary, the simulation results for EHW and CRV CIs are in line with the
theory presented in Sections~\ref{sec:properties-lee-card}
and~\ref{sec:asymptotic-results} in the main text, and the results for CRV2 and
CRV-BM CIs are in line with the results presented in Section~\ref{sec:s1}. The
honest CIs perform well in terms of coverage, although BME CIs are overly
conservative if $N_{h}$ is small or the number of support points is large. As
discussed in Section~\ref{sec:bounds-spec-errors}, they are best suited to
applications with a small number of support points, in with a sufficient number
of observations available for each support point. BSD CIs, combined with
appropriate choice of the smoothness constant $\SC$, perform very well.

\section{Lower bound for \texorpdfstring{$\SC$}{K}}

In this section, we use ideas in \citet{ArKo16optimal} to derive and estimate
a left-sided CI for a lower bound on the smoothness constant $\SC$ under the
setup of Section~\ref{sec:bound-second-deriv} in the main text.

To motivate our procedure, consider measuring the curvature of the CEF $\mu$
over some interval $[x_{1},x_{3}]$ by measuring how much it differs from a
straight line at some point $x_{2}\in (x_{1},x_{3})$. If the function were
linear, then its value at $x_{2}$ would equal
$\lambda \mu(x_{1})+(1-\lambda)\mu(x_{3})$, where
$\lambda=(x_{3}-x_{2})/(x_{3}-x_{1})$ is the distance of $x_{3}$ to $x_{2}$
relative to the distance to $x_{1}$. The next lemma gives a lower bound on $\SC$
based on the curvature estimate
$\lambda \mu(x_{1})+(1-\lambda)\mu(x_{3})-\mu(x_{2})$.
\begin{lemma}\label{lemma:sc-lower-bound}
  Suppose $\mu\in\mathcal{M}_{\text{H}}(\SC)$ for some $\SC\geq 0$. Then
  $\SC\geq \abs{\Delta(x_{1},x_{2},x_{3})}$, where
  \begin{equation}\label{eq:delta-x1x3}
    \Delta(x_{1},x_{2},x_{3})=
    2\frac{ (1-\lambda)\mu(x_{3})+\lambda \mu(x_{1})-\mu(x_{2})
    }{(1-\lambda)x_{3}^{2}+\lambda x_{1}^{2}
      -x_{2}^{2}}=
    2\frac{ (1-\lambda)\mu(x_{3})+\lambda \mu(x_{1})-\mu(x_{2})
    }{(1-\lambda)\lambda (x_{3}-x_{1})^{2}}.
  \end{equation}
\end{lemma}
\begin{proof}
  We can write any $\mu\in\mathcal{M}_{\text{H}}(\SC)$ as
  \begin{equation}\label{eq:rx}
    \mu(x)=\mu(0)+\mu'(0)x+r(x)\,\qquad r(x)=\int_{0}^{x}(x-u)\mu''(u)\,\dd u.
  \end{equation}
  This follows by an argument given in the proof of
  Proposition~\ref{theorem:holder-honest} in
  Appendix~\ref{sec:honest-conf-interv-proofs}. Hence,
  \begin{multline*}
    \lambda \mu(x_{1})+(1-\lambda)\mu(x_{3})-\mu(x_{2})= (1-\lambda)r(x_{3})+\lambda
    r(x_{1})-r(x_{2})\\= (1-\lambda)\int_{x_{2}}^{x_{3}}(x_{3}-u)\mu''(u)\,\dd u
    +\lambda\int_{x_{1}}^{x_{2}}(u-x_{1})\mu''(u)\,\dd u.
  \end{multline*}
  The absolute value of the right-hand side is bounded by $\SC$ times
  $(1-\lambda)\int_{x_{2}}^{x_{3}}(x_{3}-u)\,\dd u+
  +\lambda\int_{x_{1}}^{x_{2}}(u-x_{1})\,\dd
  u=(1-\lambda)(x_{3}-x_{2})^{2}/2+\lambda (x_{2}-x_{1})^{2}/2$, which, combined
  with the definition of $\lambda$, yields the result.
\end{proof}
\begin{remark}
  The maximum departure of $\mu$ from a straight line over $[x_{1},x_{3}]$ is
  given by
  $\max_{\lambda\in[0,1]}\abs{(1-\lambda)\mu(x_{3})+\lambda\mu(x_{1})-\mu(\lambda
    x_{1}+(1-\lambda)x_{3})}$. Lemma~\ref{lemma:sc-lower-bound} implies that
  this quantity is bounded by
  $\max_{\lambda\in[0,1]}(1-\lambda)\lambda(x_{3}-x_{1})^{2}
  \SC/2=\SC(x_{3}-x_{1})^{2}/8$, with the maximum achieved at $\lambda=1/2$. We
  use this bound for the heuristic for the choice of $\SC$ described in
  Section~\ref{sec:bound-second-deriv}.
\end{remark}

While it is possible to estimate a lower bound on $\SC$ using
Lemma~\ref{lemma:sc-lower-bound} by fixing points $x_{1},x_{2},x_{3}$ in the
support of the running variable and estimating $\Delta(x_{1},x_{2},x_{3})$ by
replacing $\mu(x_{j})$ in \eqref{eq:delta-x1x3} with the sample average of
$Y_{i}$ for observations with $X_{i}=x_{j}$, such a lower bound estimate may be
too noisy if only a few observations are available at each support
point. To overcome this problem, we consider curvature estimates that average
the value of $\mu$ over $s$ neighboring support points. To that end, the
following generalization of Lemma~\ref{lemma:sc-lower-bound} will be useful.
\begin{lemma}\label{lemma:sc-lower-bound-pooled}
  Let $I_{k}=[\ubar{x}_{k},\overbar{x}_{k}]$, $k=1,2,3$ denote three intervals
  on the same side of cutoff such that $\overbar{x}_{k}\leq \ubar{x}_{k+1}$. For
  any function $g$, let
  $E_{n}[g(x_{k})]=\sum_{i=1}^{N}g(x_{i})\1{x_{i}\in I_{k}}/n_{k}$,
  $n_{k}=\sum_{i=1}^{N}\1{x_{i}\in I_{k}}$ denote the average value in interval
  $k$. Let $\lambda_{n}=E_{n}(x_{3}-x_{2})/E_{n}(x_{3}-x_{1})$. Then
  $\SC\geq \abs{\mu(I_{1},I_{2},I_{3})}$, where
  \begin{equation}\label{eq:delta-I}
    \Delta(I_{1},I_{2},I_{3})=
    2\frac{    \lambda_{n}E_{n} \mu(x_{1})+(1-\lambda_{n})E_{n}\mu(x_{3})-E_{n}\mu(x_{2})
    }{E_{n}\left[ (1-\lambda_{n})x_{3}^{2}+\lambda_{n} x_{1}^{2}
        -x_{2}^{2}\right]}.
  \end{equation}
\end{lemma}
\begin{proof}
  By Equation~\eqref{eq:rx}, for any $\lambda\in[0,1]$ and any three points
  $x_{1},x_{2},x_{3}$, we have
  \begin{multline*}
    \lambda \mu(x_{1})+(1-\lambda)\mu(x_{3})-\mu(x_{2}) =\delta\mu'(0)+\lambda
    r(x_{1})+(1-\lambda)r(x_{3})-r(x_{2})\\
    =(1-\lambda)\int_{x_{2}}^{x_{3}}(x_{3}-u)\mu''(u)\,\dd u
    +\lambda\int_{x_{1}}^{x_{2}}(u-x_{1})\mu''(u)\,\dd u +\delta\left(\mu'(0)+
      \int_{0}^{x_{2}}\mu''(u)\,\dd u \right)
  \end{multline*}
  where $\delta=(1-\lambda)x_{3}+\lambda x_{1}-x_{2}$. Setting
  $\lambda=\lambda_{n}$, taking expectations with respect to the empirical
  distribution, observing that $E_{n}\delta=0$ and
  $E_{n}x_{2}=(1-\lambda_{n})E_{n}x_{3}+\lambda_{n}E_{n}x_{1}$, and using iterated
  expectations yields
  \begin{multline}\label{eq:ed}
    \lambda_{n}E_{n}\mu(x_{1})+(1-\lambda_{n})E_{n}\mu(x_{3})-E_{n}\mu(x_{2})\\
    =(1-\lambda_{n})E_{n}\int_{\overbar{x}_{2}}^{x_{3}}(x_{3}-u)\mu''(u)\,\dd u
    +\lambda_{n}E_{n}\int_{x_{1}}^{\ubar{x}_{2}}(u-x_{1})\mu''(u)\,\dd u+S,
  \end{multline}
where
\begin{equation*}
  \begin{split}
    S&=(1-\lambda_{n})E_{n}\int_{x_{2}}^{\overbar{x}_{2}}(E_{n}x_{3}-u)\mu''(u)\,\dd
    u +E_{n}\int_{\ubar{x}_{2}}^{x_{2}}(\lambda_{n} u-\lambda_{n} E_{n}x_{1}+
    E_{n}x_{2}-x_{2})\mu''(u)\,\dd u \\
    &= \int_{\ubar{x}_{2}}^{\overbar{x}_{2}}q(u)\mu''(u)\,\dd u ,\qquad q(u)=
    (1-\lambda_{n})(E_{n}x_{3}-u) +E_{n}\1{u\leq x_{2}}( u-x_{2}).
  \end{split}
\end{equation*}
Observe that the function $q(x)$ is weakly positive on $I_{2}$, since
$q(\ubar{x}_{2})=(1-\lambda_{n})(E_{n}x_{3}-\ubar{x}_{2}) +
\ubar{x}_{2}-E_{n}x_{2} =\lambda_{n}(\ubar{x}_{2}- E_{n}x_{1})\geq 0$,
$q(\overbar{x}_{2})= (1-\lambda_{n})(E_{n}x_{3}-\overbar{x}_{2})\geq 0$, and
$q'(u)=\lambda_{n}-E_{n}\1{x_{2}\leq u}$, so that the first derivative is
positive at first and changes sign only once. Therefore, the right-hand side
of~\eqref{eq:ed} is maximized over $\mu\in\mathcal{F}_{\text{M}}(\SC)$ by taking
$\mu''(u)=\SC$ and minimized by taking $\mu''(u)=-\SC$, which yields
\begin{equation*}
\abs{    \lambda_{n}E_{n} \mu(x_{1})+(1-\lambda_{n})E_{n}\mu(x_{3})-E_{n}\mu(x_{2})}
\leq
\frac{\SC}{2} E_{n}\left[ (1-\lambda_{n})x_{3}^{2}+\lambda_{n} x_{1}^{2}
  -x_{2}^{2}\right].
\end{equation*}
The result follows.
\end{proof}

To estimate a lower bound on $\SC$ for three fixed intervals $I_{1},I_{2}$ and
$I_{3}$, we simply replace $E_{n}\mu(x_{k})$ in~\eqref{eq:delta-I} with the
sample average of $Y_{i}$ over the interval $I_{k}$. We do this over multiple
interval choices. To describe our choice of intervals, denote the support points
of the running variable by
$x_{G_{-}}^{-}<\dotsb<x_{1}^{-}<0\leq x_{1}^{+}<\dotsb <x_{G_{+}}^{+}$, and let
$J_{ms}^{+}=[x_{ms+1-s}^{+},x_{ms}^{+}]$ denote the $m$th interval containing
$s$ support points closest to the threshold, and similarly let
$J_{ms}^{-}=[x_{ms}^{-},x_{ms+1-s}^{-}]$. To estimate the curvature based on the
intervals $J^{+}_{3k-2,s}$, $J^{+}_{3k-1,s}$ and $J^{+}_{3k,s}$, we use the
plug-in estimate
\begin{equation*}
  \hat{\Delta}^{+}_{ks}=
  2\frac{\lambda_{ks}\overbar{y}(J_{3k-2,s}^{+})
    +(1-\lambda_{ks})\overbar{y}(J_{3k-1,s}^{+})-\overbar{y}(J_{3k,s}^{+})
  }{ (1-\lambda_{ks})\overbar{x}^{2}(J_{3k,s}^{+})+\lambda_{ks}
      \overbar{x}^{2}(J_{3k-2,s})-\overbar{x}^{2}(J_{3k-1,s}^{+})},
\end{equation*}
where we use the notation
$\overbar{y}(I)=\sum_{i=1}^{N}Y_{i}\1{X_{i}\in I}/n(I)$,
$n(I)=\sum_{i=1}^{N}\1{X_{i}\in I}$, and
$\overbar{x}^{2}(I)=\sum_{i=1}^{N}X_{i}\1{X_{i}\in I}/n(I)$, and
$\lambda_{ks}=(\overbar{x}(J_{3k,s}^{+})-\overbar{x}(J^{+}_{3k-1,s}))/
(\overbar{x}(J^{+}_{3k,s})-\overbar{x}(J^{+}_{3k-2,s}))$.

For simplicity, we assume that the regression errors
$\epsilon_{i}=Y_{i}-\mu(X_{i})$ are normally distributed with conditional
variance $\sigma^{2}(X_{i})$, conditionally on the running variable. Then
$\hat{\Delta}^{+}_{ks}$ is normally distributed with variance
\begin{equation*}
  \Var(\hat{\Delta}^{+}_{ks})=
  2\frac{\lambda_{ks}^{2}\overbar{\sigma}^{2}(J_{3k-2,s}^{+})/n(J_{3k-2,s}^{+})
    +(1-\lambda_{ks})\overbar{\sigma}^{2}(J_{3k-1,s}^{+})/
    n(J_{3k-1,s}^{+})+\overbar{\sigma}^{2}(J_{3k,s}^{+})/n(J_{3k,s}^{+})
  }{\left[ (1-\lambda_{ks})\overbar{x}^{2}(J_{3k,s}^{+})+\lambda_{ks}
      \overbar{x}^{2}(J_{3k-2,s})-\overbar{x}^{2}(J_{3k-1,s}^{+})\right]^{2}}.
\end{equation*}
By Lemma~\ref{lemma:sc-lower-bound-pooled} the mean of this curvature estimate
is bounded by $\abs{E[\hat{\Delta}^{+}_{ks}]}\leq \SC$. Define
$\hat{\Delta}^{-}_{ks}$ analogously.

Our construction of a lower bound for $\SC$ is based on the vector of estimates
$\hat{\Delta}_{s}=(\hat{\Delta}^{-}_{1s},\dotsc,\hat{\Delta}^{-}_{M_{-}s},
\hat{\Delta}^{+}_{1s},\dotsc,\hat{\Delta}^{+}_{M_{+}s})$, where $M_{+}$ and
$M_{-}$ are the largest values of $m$ such that $x^{+}_{3m,s}\leq x_{G_{+}}$ and
$x^{-}_{3m,s}\geq x_{G_{-}}$.

Since elements of $\hat{\Delta}_{s}$ are independent with means bounded in
absolute value by $\SC$, we can construct a left-sided CI for $\SC$ by inverting
tests of the hypotheses $H_{0}\colon \SC\leq \SC_{0}$ vs
$H_{1}\colon \SC>\SC_{0}$ based on the sup-$t$ statistic
$\max_{k,q\in\{-,+\}}
\abs{\hat{\Delta}^{q}_{ks}/\Var(\hat{\Delta}^{q}_{ks})^{1/2}}$. The critical
value $q_{1-\alpha}(\SC_{0})$ for these tests is increasing in $\SC_{0}$ and
corresponds to the $1-\alpha$ quantile of
$\max_{k,q\in\{-,+\}}
\abs{Z^{q}_{ks}+\SC_{0}/\Var(\hat{\Delta}^{q}_{ks})^{1/2}}$, where $Z^{q}_{ks}$
are standard normal random variables, which can easily be simulated. Inverting
these tests leads to the left-sided CI of the form
$\hor{\hat{K}_{1-\alpha},\infty}$, where $\hat{K}_{1-\alpha}$ solves
$\max_{k,q\in\{-,+\}}
\abs{\hat{\Delta}^{q}_{ks}/\Var(\hat{\Delta}^{q}_{ks})^{1/2}}=
q_{1-\alpha}(\hat{K}_{1-\alpha})$. Following \citet{ChLeRo2013}, we use
$\hat{\SC}_{1/2}$ as a point estimate, which is half-median unbiased in the sense
that $P(\hat{K}_{1/2}\leq K)\geq 1/2$.

An attractive feature of this method is that the same construction can be used
when the running variable is continuously distributed. To implement this method,
one has to choose the number of support points $s$ to average over. We leave the
optimal choice of $s$ to future research, and simply use $s=2$ in the empirical
applications in Section~\ref{sec:empir-appl} of the main text.

\singlespacing
\bibliography{bibl}

\clearpage

\begin{landscape}
  \begin{table}[htbp]
    \caption{Inference in CPS simulation study for placebo
      treatment: Linear specification.\label{tab:cps-linear}}
  \small
    \vspace{1ex}
    \begin{tabular}{@{}rrrc rrrrrr rrrrrr r@{}}
    \toprule
    &&&&\multicolumn{6}{c}{Avg.~normalized SE}&
      \multicolumn{6}{c}{CI coverage rate (\%)}
      \\
    \cmidrule(rl){5-10}\cmidrule(rl){11-16}
    $h$ & $\tauh$ & $N_{h}$ &$SD(\widehat\tau)$ & EHW & CRV  & CRV2 & BM & BME & BSD & EHW &CRV&CRV2&BM&BME&BSD&\\
    \midrule
 5&$-0.008$&   100&   0.239 & 0.234 & 0.166 & 0.218 & 0.364 & 0.606 &        0.309 &  93.8 & 77.3 & 86.3 & 96.9 & 100 &        97.2 \\
    &      &   500&   0.104 & 0.104 & 0.073 & 0.097 & 0.160 & 0.253 &        0.170 &  94.7 & 78.0 & 86.8 & 96.7 & 100 &        97.9 \\
    &      &  2000&   0.052 & 0.052 & 0.036 & 0.048 & 0.079 & 0.125 &        0.109 &  94.9 & 77.3 & 86.6 & 96.6 & 100 &        99.7 \\
    &      & 10000&   0.021 & 0.023 & 0.015 & 0.020 & 0.033 & 0.055 &        0.064 &  96.1 & 77.2 & 86.6 & 96.9 & 100 &        99.9 \\[1ex]
10&$-0.023$&   100&   0.227 & 0.223 & 0.193 & 0.218 & 0.270 & 0.946 &        0.394 &  93.9 & 87.3 & 90.8 & 95.2 & 100 &        97.0 \\
    &      &   500&   0.099 & 0.099 & 0.086 & 0.097 & 0.117 & 0.385 &        0.215 &  94.4 & 87.6 & 91.1 & 95.2 & 100 &        98.0 \\
    &      &  2000&   0.049 & 0.050 & 0.044 & 0.049 & 0.059 & 0.187 &        0.131 &  93.0 & 86.0 & 89.8 & 94.3 & 100 &        99.1 \\
    &      & 10000&   0.021 & 0.022 & 0.021 & 0.024 & 0.029 & 0.086 &        0.077 &  82.9 & 78.1 & 83.8 & 91.0 & 100 &        99.1 \\[1ex]
15&$-0.063$&   100&   0.222 & 0.216 & 0.197 & 0.214 & 0.246 & 1.137 &        0.453 &  92.7 & 88.4 & 90.7 & 94.4 & 100 &        96.8 \\
    &      &   500&   0.095 & 0.096 & 0.089 & 0.096 & 0.108 & 0.507 &        0.242 &  89.9 & 85.3 & 87.9 & 91.7 & 100 &        98.5 \\
    &      &  2000&   0.048 & 0.048 & 0.047 & 0.052 & 0.058 & 0.243 &        0.146 &  73.0 & 71.2 & 75.6 & 81.6 & 100 &        98.7 \\
    &      & 10000&   0.020 & 0.021 & 0.028 & 0.030 & 0.034 & 0.118 &        0.087 &  15.3 & 34.8 & 43.9 & 55.8 & 100 &        98.7 \\[1ex]
$\infty$&$-0.140$&100&0.208 & 0.205 & 0.196 & 0.208 & 0.229 & 1.281 &        0.513 &  88.6 & 85.6 & 87.8 & 90.9 & 100 &        96.5 \\
    &      &   500&   0.091 & 0.091 & 0.094 & 0.099 & 0.107 & 0.822 &        0.269 &  66.7 & 67.3 & 70.8 & 75.6 & 100 &        97.8 \\
    &      &  2000&   0.045 & 0.046 & 0.058 & 0.061 & 0.066 & 0.460 &        0.162 &  13.4 & 29.2 & 34.5 & 41.1 & 100 &        98.0 \\
    &      & 10000&   0.019 & 0.020 & 0.043 & 0.046 & 0.049 & 0.271 &        0.097 &  0.0  &  0.6 &  1.3 &  2.9 & 100 &        98.1 \\
    \bottomrule
  \end{tabular}

  \vspace{1ex}{\small Note: Results are based on 10,000 simulation runs. BM
    refers to CRV-BM CIs. For BSD, $M=0.045$. For CRV-BM, BME, and BSD,
    Avg.~normalized SE refers to average normalized standard error, described in
    the text. For EHW, CRV, and CRV2, it corresponds to average standard error,
    averaged over the simulation runs.}
\end{table}
\end{landscape}

\begin{landscape}
  \begin{table}[htbp]
    \caption{Inference in CPS simulation study for placebo
      treatment: Quadratic specification.\label{tab:cps-quadratic}}
  \small
    \vspace{1ex}
    \begin{tabular}{@{}rrrc rrrrrr rrrrrr r@{}}
    \toprule
    &&&&\multicolumn{6}{c}{Avg.~normalized SE}&
      \multicolumn{6}{c}{CI coverage rate}\\
    \cmidrule(rl){5-10}\cmidrule(rl){11-16}
    $h$ & $\tauh$ & $N_{h}$ &$SD(\widehat\tau)$ & EHW & CRV  & CRV2 & BM & BME & BSD & EHW &CRV&CRV2&BM&BME&BSD&\\
    \midrule
 5&$-0.100$&   100&   0.438 & 0.427 & 0.206 & 0.394 & 0.897 & 0.705 &        0.309 &      93.2 & 60.7 & 84.0 & 97.2 & 99.5 & 97.3 \\
    &      &   500&   0.189 & 0.190 & 0.086 & 0.172 & 0.382 & 0.301 &        0.170 &      94.7 & 59.9 & 83.8 & 97.4 & 99.7 & 97.9 \\
    &      &  2000&   0.093 & 0.095 & 0.042 & 0.084 & 0.187 & 0.149 &        0.109 &      95.1 & 58.7 & 83.2 & 97.4 & 99.8 & 99.7 \\
    &      & 10000&   0.038 & 0.042 & 0.018 & 0.036 & 0.079 & 0.067 &        0.064 &      96.4 & 59.5 & 84.4 & 97.6 & 99.9 & 99.9 \\[1ex]
10&$ 0.008$&   100&   0.361 & 0.349 & 0.258 & 0.342 & 0.540 & 1.000 &        0.394 &      93.3 & 79.5 & 88.3 & 97.2 &    100 & 97.0\\
    &      &   500&   0.157 & 0.156 & 0.110 & 0.147 & 0.231 & 0.405 &        0.215 &      94.8 & 79.0 & 88.1 & 96.8 &    100 & 98.0\\
    &      &  2000&   0.077 & 0.078 & 0.055 & 0.073 & 0.115 & 0.196 &        0.131 &      94.7 & 79.4 & 88.1 & 96.8 &    100 & 99.1\\
    &      & 10000&   0.033 & 0.035 & 0.025 & 0.034 & 0.053 & 0.087 &        0.077 &      95.6 & 80.8 & 89.4 & 97.6 &    100 & 99.1\\[1ex]
15&$ 0.014$&   100&   0.349 & 0.329 & 0.270 & 0.325 & 0.443 & 1.189 &        0.453 &      92.3 & 83.6 & 89.1 & 96.0 &    100 & 96.8\\
    &      &   500&   0.146 & 0.147 & 0.117 & 0.141 & 0.187 & 0.516 &        0.242 &      94.6 & 85.1 & 89.9 & 95.8 &    100 & 98.5\\
    &      &  2000&   0.073 & 0.073 & 0.058 & 0.070 & 0.092 & 0.243 &        0.146 &      94.6 & 83.9 & 89.7 & 95.7 &    100 & 98.7\\
    &      & 10000&   0.031 & 0.033 & 0.026 & 0.031 & 0.041 & 0.107 &        0.087 &      93.7 & 82.8 & 88.7 & 95.1 &    100 & 98.7\\[1ex]
$\infty$&$-0.001$&100&0.316 & 0.303 & 0.267 & 0.302 & 0.371 & 1.327 &        0.513 &      93.0 & 87.6 & 90.9 & 95.4 &    100 & 96.5\\
    &      &   500&   0.134 & 0.135 & 0.117 & 0.132 & 0.155 & 0.804 &        0.269 &      94.9 & 89.0 & 92.2 & 95.5 &    100 & 97.8\\
    &      &  2000&   0.068 & 0.067 & 0.058 & 0.065 & 0.077 & 0.402 &        0.162 &      94.7 & 88.7 & 91.6 & 95.4 &    100 & 98.0\\
    &      & 10000&   0.029 & 0.030 & 0.027 & 0.030 & 0.035 & 0.178 &        0.097 &      96.0 & 91.0 & 93.7 & 96.7 &    100 & 98.1\\
    \bottomrule
  \end{tabular}

  \vspace{1ex}{\small Note: Results are based on 10,000 simulation runs. BM
    refers to CRV-BM CIs. For BSD, $M=0.045$. For CRV-BM, BME, and BSD,
    Avg.~normalized SE refers to average normalized standard error, described in
    the text. For EHW, CRV, and CRV2, it corresponds to average standard error,
    averaged over the simulation runs.}
\end{table}
\end{landscape}

  \begin{table}[htbp]
  \caption{Inference in CPS simulation study for placebo treatment: Comparison
    of relative magnitude of EHW standard errors relative to CRV, CRV2 and CRV-BM
    standard errors under linear and quadratic specification.\label{tab:cps-rates}}
  \vspace{1ex}
  {\centering
    \begin{tabular}{@{}rrr rrr rrr@{}}
    \toprule
    &&&\multicolumn{3}{c@{\hskip 1ex}}{Linear}&
      \multicolumn{3}{c@{}}{Quadratic}\\
    \cmidrule(rl){4-6}\cmidrule(rl){7-9}
   &&& \multicolumn{3}{l}{Rate EHW SE $<$}&\multicolumn{3}{l}{Rate EHW SE $<$}\\
    $h$ & $\tauh$ & $N_{h}$ & CRV  & CRV2 & BM &CRV  & CRV2 & BM  \\
    \midrule
 5&$-0.100$&   100&   0.14 &     0.38 &    0.82 &0.03 &     0.38 &    0.87\\
    &      &   500&   0.13 &     0.37 &    0.81 &0.01 &     0.36 &    0.87\\
    &      &  2000&   0.13 &     0.36 &    0.80 &0.01 &     0.34 &    0.86\\
    &      & 10000&   0.09 &     0.30 &    0.77 &0.00 &     0.30 &    0.85\\[1ex]
10&$ 0.008$&   100&   0.26 &     0.44 &    0.74 &0.15 &     0.43 &    0.85\\
    &      &   500&   0.25 &     0.43 &    0.71 &0.12 &     0.38 &    0.81\\
    &      &  2000&   0.27 &     0.44 &    0.72 &0.11 &     0.37 &    0.81\\
    &      & 10000&   0.39 &     0.60 &    0.84 &0.13 &     0.40 &    0.82\\[1ex]
15&$ 0.014$&   100&   0.31 &     0.47 &    0.71 &0.20 &     0.45 &    0.81\\
    &      &   500&   0.34 &     0.47 &    0.69 &0.18 &     0.39 &    0.74\\
    &      &  2000&   0.45 &     0.61 &    0.79 &0.17 &     0.38 &    0.72\\
    &      & 10000&   0.92 &     0.96 &    0.99 &0.18 &     0.39 &    0.72\\[1ex]
$\infty$&$-0.001$&100&0.38 &     0.52 &    0.73 &0.26 &     0.48 &    0.78\\
    &      &   500&   0.54 &     0.66 &    0.80 &0.24 &     0.42 &    0.67\\
    &      &  2000&   0.93 &     0.96 &    0.99 &0.23 &     0.41 &    0.66\\
    &      & 10000&   1.00 &     1.00 &    1.00 &0.26 &     0.44 &    0.70\\
    \bottomrule
  \end{tabular}\par}

\vspace{1ex}{\small Note: Results are based on 10,000 simulation runs. BM refers
  to CRV-BM CIs. For CRV-BM CIs, SE refers to average normalized standard error,
  described in the text.}
\end{table}

\begin{landscape}
  \setlength{\tabcolsep}{3.2pt}
  \begin{table}[htbp]
    \caption{Simulation results in second Monte Carlo exercise for
      $\mu(x) = x$.}\label{tab:results1}
  \small
    \vspace{1ex}
    \begin{tabular}{@{}rrrcc rrrrrrr rrrrrrr@{}}
    \toprule
    &&&&&\multicolumn{7}{c}{Avg.~normalized SE}&
      \multicolumn{7}{c}{CI coverage rate (\%)}\\
    \cmidrule(rl){6-12}\cmidrule(rl){13-19}
    $G_{+}$ & $G_{-}$ & $N_{h}$ & $\tauh$ &$sd(\widehat\tau)$
            & EHW & CRV  & CRV2 & BM & BME & $\text{BSD}_{.49}$ & $\text{BSD}_{.99}$
            & EHW & CRV  & CRV2 & BM & BME & $\text{BSD}_{.49}$ & $\text{BSD}_{.99}$ \\
    \midrule
5&  5&   100&   0&0.148 & 0.149 & 0.105 & 0.141 & 0.221 & 0.332 &  0.197 &  0.249  & 94.6 & 79.1 & 88.0 & 96.6 & 100.0&  98.5 &  99.0 \\
&    &  1000&    &0.046 & 0.047 & 0.032 & 0.044 & 0.068 & 0.097 &  0.091 &  0.124  & 95.7 & 77.5 & 88.0 & 96.6 & 100.0&  99.0 &  99.5 \\
 &   & 10000&    &0.015 & 0.015 & 0.010 & 0.014 & 0.022 & 0.031 &  0.047 &  0.067  & 94.8 & 77.3 & 87.4 & 96.2 & 100.0&  99.5 &  100.0  \\[1ex]
5& 25&   100&   0&0.143 & 0.141 & 0.114 & 0.136 & 0.186 & 0.686 &  0.178 &  0.219  & 94.5 & 85.4 & 90.3 & 96.9 & 100.0&  97.8 &  97.8 \\
 &   &  1000&    &0.044 & 0.044 & 0.036 & 0.043 & 0.057 & 0.248 &  0.078 &  0.102  & 95.5 & 87.2 & 91.1 & 96.9 & 100.0&  98.6 &  98.1 \\
 &   & 10000&    &0.014 & 0.014 & 0.011 & 0.013 & 0.018 & 0.074 &  0.037 &  0.052  & 95.5 & 86.7 & 91.1 & 97.2 & 100.0&  99.4 &  100.0  \\[1ex]
5& 50&   100&   0&0.143 & 0.140 & 0.115 & 0.135 & 0.183 & 0.754 &  0.176 &  0.216  & 93.7 & 86.0 & 90.3 & 96.3 & 100.0&  97.4 &  97.8 \\
 &   &  1000&    &0.043 & 0.044 & 0.036 & 0.042 & 0.056 & 0.397 &  0.077 &  0.100  & 95.3 & 87.3 & 91.3 & 97.1 & 100.0&  99.0 &  98.1 \\
 &   & 10000&    &0.014 & 0.014 & 0.011 & 0.013 & 0.018 & 0.111 &  0.036 &  0.050  & 95.2 & 87.3 & 90.9 & 96.6 & 100.0&  99.2 &  99.9 \\[1ex]
25&25&   100&   0&0.136 & 0.131 & 0.124 & 0.131 & 0.143 & 0.791 &  0.159 &  0.186  & 93.8 & 91.3 & 92.9 & 94.8 & 100.0&  96.6 &  96.9 \\
 &   &100000&    &0.040 & 0.041 & 0.039 & 0.041 & 0.044 & 0.280 &  0.064 &  0.076  & 95.5 & 93.2 & 94.5 & 95.7 & 100.0&  97.9 &  98.0 \\
 &   & 10000&    &0.013 & 0.013 & 0.012 & 0.013 & 0.014 & 0.083 &  0.027 &  0.032  & 95.4 & 93.1 & 94.3 & 95.5 & 100.0&  97.8 &  98.3 \\[1ex]
25&50&   100&   0&0.132 & 0.130 & 0.125 & 0.130 & 0.140 & 0.843 &  0.156 &  0.183  & 94.2 & 92.8 & 93.8 & 95.4 & 100.0&  96.9 &  96.7 \\
 &   &  1000&    &0.042 & 0.041 & 0.039 & 0.041 & 0.043 & 0.430 &  0.063 &  0.074  & 94.7 & 93.0 & 93.9 & 95.0 & 100.0&  97.4 &  97.6 \\
 &   & 10000&    &0.013 & 0.013 & 0.012 & 0.013 & 0.013 & 0.119 &  0.026 &  0.031  & 94.9 & 93.2 & 93.9 & 94.8 & 100.0&  97.9 &  97.8 \\[1ex]
50&50&   100&   0&0.133 & 0.129 & 0.126 & 0.130 & 0.138 & 0.879 &  0.154 &  0.180  & 93.5 & 92.6 & 93.2 & 94.7 & 100.0&  96.6 &  96.7 \\
 &   &  1000&    &0.040 & 0.041 & 0.040 & 0.041 & 0.042 & 0.472 &  0.062 &  0.072  & 95.3 & 94.1 & 94.8 & 95.4 & 100.0&  97.6 &  97.7 \\
 &   & 10000&    &0.013 & 0.013 & 0.012 & 0.013 & 0.013 & 0.127 &  0.025 &  0.029  & 94.9 & 93.9 & 94.3 & 94.9 & 100.0&  97.5 &  97.7 \\
    \bottomrule
  \end{tabular}

  \vspace{1ex} {\small Note: Results are based on 5,000 simulation runs.
    $\text{BSD}_{.49}$ and $\text{BSD}_{.99}$ refer to BSD CIs with $\SC=0.493$
    and $\SC=0.986$, respectively, and BM refers to CRV-BM CIs. For CRV-BM, BME and BSD, Avg.~normalized SE refers to
    average normalized standard error, described in the text. For EHW, CRV and CRV2,
    it corresponds to average standard error, averaged over the simulation
    runs.}
\end{table}
\end{landscape}

\begin{landscape}
  \setlength{\tabcolsep}{3.2pt}
  \begin{table}[htbp]
    \caption{Simulation results in second Monte Carlo exercise for
      $\mu(x) = x +.05\cdot\cos(\pi\cdot x)$.}\label{tab:results2}
  \small
    \vspace{1ex}
    \begin{tabular}{@{}rrrcc rrrrrrr rrrrrrr@{}}
    \toprule
    &&&&&\multicolumn{7}{c}{Avg.~normalized SE}&
      \multicolumn{7}{c}{CI coverage rate (\%)}\\
    \cmidrule(rl){6-12}\cmidrule(rl){13-19}
    $G_{+}$ & $G_{-}$ & $N_{h}$ & $\tauh$ &$sd(\widehat\tau)$
            & EHW & CRV  & CRV2 & BM & BME & $\text{BSD}_{.49}$ & $\text{BSD}_{.99}$
            & EHW & CRV  & CRV2 & BM & BME & $\text{BSD}_{.49}$ & $\text{BSD}_{.99}$ \\
    \midrule
5&  5&   100&   0&0.148 & 0.149 & 0.105 & 0.142 & 0.222 & 0.332 & 0.197 & 0.249  & 94.7 & 78.9 & 88.1 & 96.6 & 100.0&   98.4 &  99.0 \\
&    &  1000&    &0.046 & 0.047 & 0.033 & 0.044 & 0.069 & 0.098 & 0.091 & 0.124  & 95.7 & 78.4 & 89.0 & 96.9 & 100.0&   99.0 &  99.5 \\
 &   & 10000&    &0.015 & 0.015 & 0.012 & 0.017 & 0.026 & 0.033 & 0.047 & 0.067  & 94.8 & 84.1 & 92.1 & 98.2 & 100.0&   99.5 &  100.0\\[1ex]
5& 25&   100&   0&0.143 & 0.141 & 0.115 & 0.136 & 0.186 & 0.686 & 0.178 & 0.219  & 94.5 & 85.4 & 90.2 & 96.8 & 100.0&   97.8 &  97.8 \\
 &   &  1000&    &0.044 & 0.044 & 0.036 & 0.043 & 0.058 & 0.248 & 0.078 & 0.102  & 95.5 & 87.3 & 91.4 & 97.0 & 100.0&   98.5 &  97.8 \\
 &   & 10000&    &0.014 & 0.014 & 0.012 & 0.015 & 0.020 & 0.075 & 0.037 & 0.052  & 95.4 & 88.8 & 93.2 & 98.1 & 100.0&   99.1 &  100.0\\[1ex]
5& 50&   100&   0&0.143 & 0.140 & 0.115 & 0.135 & 0.183 & 0.754 & 0.176 & 0.216  & 93.7 & 86.3 & 90.1 & 96.1 & 100.0&   97.4 &  97.9 \\
 &   &  1000&    &0.043 & 0.044 & 0.036 & 0.043 & 0.057 & 0.397 & 0.077 & 0.100  & 95.3 & 87.6 & 91.6 & 97.3 & 100.0&   98.9 &  98.0 \\
 &   & 10000&    &0.014 & 0.014 & 0.012 & 0.015 & 0.020 & 0.112 & 0.036 & 0.050  & 95.3 & 89.4 & 93.5 & 97.5 & 100.0&   99.0 &  99.9 \\[1ex]
25&25&   100&   0&0.136 & 0.131 & 0.124 & 0.131 & 0.143 & 0.791 & 0.159 & 0.186  & 93.8 & 91.5 & 92.9 & 94.7 & 100.0&   96.6 &  96.9 \\
 &   &100000&    &0.040 & 0.041 & 0.039 & 0.041 & 0.044 & 0.280 & 0.064 & 0.076  & 95.4 & 93.1 & 94.5 & 95.8 & 100.0&   97.9 &  98.0 \\
 &   & 10000&    &0.013 & 0.013 & 0.013 & 0.013 & 0.014 & 0.084 & 0.027 & 0.032  & 95.4 & 93.9 & 95.0 & 95.9 & 100.0&   97.8 &  98.3 \\[1ex]
25&50&   100&   0&0.132 & 0.130 & 0.125 & 0.130 & 0.140 & 0.843 & 0.156 & 0.183  & 94.2 & 92.9 & 93.7 & 95.4 & 100.0&   96.9 &  96.7 \\
 &   &  1000&    &0.042 & 0.041 & 0.039 & 0.041 & 0.043 & 0.430 & 0.063 & 0.074  & 94.7 & 92.9 & 93.9 & 95.1 & 100.0&   97.4 &  97.7 \\
 &   & 10000&    &0.013 & 0.013 & 0.013 & 0.013 & 0.014 & 0.119 & 0.026 & 0.031  & 94.8 & 93.5 & 94.1 & 95.1 & 100.0&   97.9 &  97.8 \\[1ex]
50&50&   100&   0&0.133 & 0.129 & 0.126 & 0.130 & 0.138 & 0.880 & 0.154 & 0.180  & 93.6 & 92.7 & 93.3 & 94.6 & 100.0&   96.6 &  96.7 \\
 &   &  1000&    &0.040 & 0.041 & 0.040 & 0.041 & 0.042 & 0.472 & 0.062 & 0.072  & 95.3 & 94.1 & 94.8 & 95.5 & 100.0&   97.6 &  97.7 \\
 &   & 10000&    &0.013 & 0.013 & 0.013 & 0.013 & 0.013 & 0.128 & 0.025 & 0.029  & 94.9 & 94.1 & 94.6 & 95.1 & 100.0&   97.5 &  97.7 \\
    \bottomrule
  \end{tabular}

  \vspace{1ex} {\small Note: Results are based on 5,000 simulation runs.
    $\text{BSD}_{.49}$ and $\text{BSD}_{.99}$ refer to BSD CIs with $\SC=0.493$
    and $\SC=0.986$, respectively, and BM refers to CRV-BM CIs. For CRV-BM, BME and BSD, Avg.~normalized SE refers to
    average normalized standard error, described in the text. For EHW, CRV and CRV2,
    it corresponds to average standard error, averaged over the simulation
    runs.}
\end{table}
\end{landscape}

\begin{landscape}
  \setlength{\tabcolsep}{3.2pt}
  \begin{table}[htbp]
    \caption{Simulation results in second Monte Carlo exercise for
      $\mu(x) = x +.05\cdot\sin(\pi\cdot x)$.}\label{tab:results3}
  \small
    \vspace{1ex}
    \begin{tabular}{@{}rrrcc rrrrrrr rrrrrrr@{}}
    \toprule
    &&&&&\multicolumn{7}{c}{Avg.~normalized SE}&
      \multicolumn{7}{c}{CI coverage rate (\%)}\\
    \cmidrule(rl){6-12}\cmidrule(rl){13-19}
    $G_{+}$ & $G_{-}$ & $N_{h}$ & $\tauh$ &$sd(\widehat\tau)$
            & EHW & CRV  & CRV2 & BM & BME & $\text{BSD}_{.49}$ & $\text{BSD}_{.99}$
            & EHW & CRV  & CRV2 & BM & BME & $\text{BSD}_{.49}$ & $\text{BSD}_{.99}$ \\
    \midrule
5&  5&   100&   0&0.148 & 0.150 & 0.107 & 0.145 & 0.227 & 0.334 & 0.197 & 0.249  & 88.6 & 68.9 & 81.6 & 94.0 & 100.0& 96.3 &  98.5 \\
&    &  1000&    &0.046 & 0.047 & 0.039 & 0.054 & 0.084 & 0.102 & 0.091 & 0.124  & 37.3 & 27.7 & 47.5 & 76.2 &96.0 & 97.2 &  99.2 \\
 &   & 10000&    &0.015 & 0.015 & 0.025 & 0.036 & 0.057 & 0.041 & 0.047 & 0.067  &  0.0 &  0.1 &  5.8 & 54.3 & 5.9 & 98.4 &  99.9 \\[1ex]
5& 25&   100&   0&0.143 & 0.141 & 0.116 & 0.138 & 0.188 & 0.687 & 0.178 & 0.219  & 89.4 & 78.7 & 84.5 & 94.2 & 100.0& 96.0 &  97.5 \\
 &   &  1000&    &0.044 & 0.044 & 0.040 & 0.049 & 0.065 & 0.252 & 0.078 & 0.102  & 47.7 & 39.9 & 53.0 & 72.6 & 100.0& 97.0 &  97.7 \\
 &   & 10000&    &0.014 & 0.014 & 0.021 & 0.028 & 0.038 & 0.085 & 0.037 & 0.052  &  0.0 &  0.2 &  5.1 & 25.0 & 100.0& 98.2 &  99.9 \\[1ex]
5& 50&   100&   0&0.143 & 0.140 & 0.116 & 0.137 & 0.185 & 0.755 & 0.176 & 0.216  & 89.3 & 80.2 & 85.5 & 94.0 & 100.0& 95.6 &  97.6 \\
 &   &  1000&    &0.043 & 0.044 & 0.039 & 0.048 & 0.064 & 0.400 & 0.077 & 0.100  & 49.4 & 41.3 & 53.9 & 72.9 & 100.0& 97.1 &  98.0 \\
 &   & 10000&    &0.014 & 0.014 & 0.020 & 0.028 & 0.037 & 0.121 & 0.036 & 0.050  &  0.0 &  0.3 &  5.6 & 25.9 & 100.0& 98.4 &  99.9 \\[1ex]
25&25&   100&   0&0.135 & 0.131 & 0.125 & 0.132 & 0.144 & 0.792 & 0.159 & 0.186  & 90.5 & 87.9 & 89.5 & 92.1 & 100.0& 95.1 &  96.4 \\
 &   &100000&    &0.040 & 0.041 & 0.041 & 0.042 & 0.045 & 0.283 & 0.064 & 0.076  & 59.4 & 57.8 & 61.4 & 65.9 & 100.0& 97.0 &  97.7 \\
 &   & 10000&    &0.013 & 0.013 & 0.016 & 0.017 & 0.018 & 0.092 & 0.027 & 0.032  &  0.0 &  0.2 &  0.4 &  0.6 & 100.0& 97.2 &  98.2 \\[1ex]
25&50&   100&   0&0.132 & 0.130 & 0.125 & 0.131 & 0.141 & 0.844 & 0.156 & 0.183  & 91.0 & 88.7 & 90.2 & 92.6 & 100.0& 95.5 &  96.4 \\
 &   &  1000&    &0.042 & 0.041 & 0.040 & 0.042 & 0.044 & 0.431 & 0.063 & 0.074  & 60.0 & 58.3 & 61.1 & 64.5 & 100.0& 96.3 &  97.5 \\
 &   & 10000&    &0.013 & 0.013 & 0.015 & 0.016 & 0.017 & 0.126 & 0.026 & 0.031  &  0.1 &  0.1 &  0.2 &  0.2 & 100.0& 97.1 &  97.8 \\[1ex]
50&50&   100&   0&0.132 & 0.130 & 0.126 & 0.130 & 0.138 & 0.881 & 0.154 & 0.180  & 91.0 & 89.6 & 90.5 & 92.2 & 100.0& 95.1 &  96.4 \\
 &   &  1000&    &0.040 & 0.041 & 0.040 & 0.041 & 0.043 & 0.474 & 0.062 & 0.072  & 62.2 & 61.0 & 62.8 & 64.9 & 100.0& 96.4 &  97.5 \\
 &   & 10000&    &0.013 & 0.013 & 0.014 & 0.015 & 0.015 & 0.135 & 0.025 & 0.029  &  0.1 &  0.2 &  0.2 &  0.2 & 100.0& 97.0 &  97.5 \\
    \bottomrule
  \end{tabular}

  \vspace{1ex} {\small Note: Results are based on 5,000 simulation runs.
    $\text{BSD}_{.49}$ and $\text{BSD}_{.99}$ refer to BSD CIs with $\SC=0.493$
    and $\SC=0.986$, respectively, and BM refers to CRV-BM CIs. For CRV-BM, BME and BSD, Avg.~normalized SE refers to
    average normalized standard error, described in the text. For EHW, CRV and CRV2,
    it corresponds to average standard error, averaged over the simulation
    runs.}
\end{table}
\end{landscape}

